\documentclass[12pt]{article}
\usepackage{authblk}

\usepackage{graphicx}
\usepackage{amsmath}
\usepackage{amssymb}
\usepackage{amsbsy}
\usepackage{mathtools}
\usepackage[left, running, displaymath, mathlines]{lineno}
 \usepackage{mathptmx}           
\usepackage{bm}   
\usepackage{framed}  
\usepackage{hyperref}
\numberwithin{equation}{section}

\newcommand{\bit}{\begin{itemize}}
\newcommand{\eit}{\end{itemize}}

\def\benu{\begin{enumerate}}
\def\eenu{\end{enumerate}}
\def\noi{\noindent}
\def\btab{\begin{tabbing}}
\def\etab{\end{tabbing}}

\def\bit{\begin{itemize}}
\def\eit{\end{itemize}}
\def\beq{\begin{equation}}
\def\eeq{\end{equation}}
\def\bec{\begin{center}}
\def\eec{\end{center}}
\def\btable{\begin{tabular}}
\def\etable{\end{tabular}}
\def\beqr{\begin{eqnarray}}
\def\eeqr{\end{eqnarray}}
\def\rarw{\rightarrow}
\def\Rarw{\Rightarrow}

\def\gm{\gamma}

\def\lm{\lambda}
\def\eps{\epsilon}

\def\bt{\beta}

\def\Dl{\Delta}
\def\sg{\sigma}
\def\Om{\Omega}
\def\rarw{\rightarrow}
\def\del{\partial}

\def\half{\frac{1}{2}}

\def\btab{\begin{tabbing}}
\def\etab{\end{tabbing}}
\def\beqrs{\begin{eqnarray*}}
\def\eeqrs{\end{eqnarray*}}
\def\noi{\noindent}

\def\bibi{\bibitem}
\def\bfig{\begin{figure}}
\def\efig{\end{figure}}
\def\fr{\frac}
\def\non{\nonumber}
\def\barr{\begin{array}}
\def\earr{\end{array}}

 \fontfamily{ptm}\selectfont

\title{\mbox{} \hfill {\normalsize FERMILAB-PUB-19-421-AD} \\
  \mbox{} \hfill {\normalsize Accepted} \\
\mbox{}  \\
  \bf Fields and Characteristic Impedances of  Dipole and Quadrupole Cylindrical Stripline Kickers}
\author[1]{T. Sen  \footnote{tsen@fnal.gov}}
\author[2]{Y. Tu} 
\author[1]{J.-F. Ostiguy}

\affil[1]{Fermi National Accelerator Laboratory, Batavia, IL 60510 }
\affil[2]{ University of Rochester, Rochester, NY 14627 }
  
\date{}

\begin{document}

\maketitle

\begin{abstract}
   We present semi-analytical methods for calculating the electromagnetic field in dipole and quadrupole stripline
   kickers with curved plates of infinitesimal thickness. Two different methods are used to solve Laplace's equation by reducing it either to 
   a single or to two coupled matrix equations; they are shown to yield equivalent results.
   Approximate analytic solutions for the lowest order fields (dipole or quadrupole) are presented and their useful range of
   validity are shown. 
   The kickers plates define a set of coupled transmission lines and the characteristic impedances of modes relevant
   to each configuration  are calculated.
 The solutions are compared with those obtained from a finite element solver  and found to be in good 
 agreement. Mode matching to an external impedance determines
 the kicker geometry and this is discussed for both kicker types. We show that a heuristic scaling law can be used to determine
the dependence of the characteristic impedance on plate thickness. The solutions found by  
 semi-analytical methods can be used as a starting point for a more detailed kicker design. 
\end{abstract}


\section{Introduction}

  Transverse dipole kickers change the transverse momentum of a beam in an accelerator and have multiple applications, e.g. in
 systems for  injection and extraction, feedback, tune measurement etc \cite{Handb, Gold_Lamb}. Transverse quadrupole kickers are  used
  in exciting coherent quadrupole oscillations in space charge dominated beams.
Stripline kickers (dipole and quadrupole)  are often preferred because of their relative simplicity and fast response time. 
  An application of interest where both types of kickers are needed
  is the generation of beam echoes \cite{Stupakov,  Stup_Kauf, Fischer_05, Sen_17, Sen_18}.
  Detailed  designs are usually done with electromagnetic codes
which solve for  the fields using a variety of numerical schemes, see e.g. \cite{Alesini_10, Belver-Aguilar_14}. In this paper, we
focus instead on   analytical methods   to solve Laplace's equation.
After incorporating   the boundary conditions, the solution is expressed as a series whose coefficients are obtained from
  matrix equation(s) of infinite dimension. The latter are then truncated and solved numerically. This approach 
leads to general insights about how the fields and characteristic impedances
  depend on kicker geometry. 

  This study was motivated by the need of these kickers for generating beam echoes in the IOTA ring at Fermilab
  \cite{Antipov_IOTA}. The small size of this ring (40 m circumference) calls for compact kickers. 
   Therefore, a design objective is to maximize the electric field (dipole) or field gradient (quadrupole) subject to the
   constraint of proper matching to external loads. A few simplifying assumptions are made in the analysis, 
   an important one being that the electrode plates are of infinitesimal thickness. We also assume circular symmetry so 
   the electrodes are arc shaped. 
   The two analytical methods presented here were originally applied to the design of striplines for microwave devices
   \cite{Wang_78}; one of them was later used in the analysis of a pickup with a single stripline \cite{Barry_Liu_90}. 
   In Section \ref{sec: Dipole} we introduce both methods and use them to analyze a dipole kicker; the quadrupole kicker
  is discussed in Section \ref{sec: Quad}. We compare our semi-analytical results with those from a finite element based
code (FEMM) in Section \ref{sec: Numerics}. In Section \ref{sec: thickness} we consider the influence of
  plate thickness and derive a heuristic scaling law. Our conclusions are presented in Section \ref{sec: conclus}.
  Appendix A contains approximate formulas to estimate the potentials in both kickers, Appendix B contains expressions for
  the mode capacitances which are related to the mode characteristic impedances while Appendix C shows how to
  match all modes with a properly chosen termination network.

\section{Dipole Kicker with circular symmetry} \label{sec: Dipole}

 In this section and the following two others, we make the following assumptions: a) the electrode plates are arc shaped and
 infinitesimally  thin, b) the electrode plates and beam pipe are perfect conductors and c) all plates have exactly the same
 shape and coverage angle, and they are placed at exactly symmetric locations inside the beam pipe.
 We first consider the two plate dipole kicker configuration. There are two modes to consider:
 the operational mode or odd mode where the plates are at equal and opposite voltages resulting -- to lowest order-- in a dipole
 field 
 and the even mode where both plates are at the same voltage. The even mode is relevant because it is excited by the circulating 
 beam which, assuming it is centered, induces identical charges (a fraction of the beam charge) and voltages on all plates.
 A more complete discussion of the odd and even modes can be found for example in \cite{Collin}.  
 If the beam current is high enough for beam instabilities to become a concern, the characteristic impedance of the even 
 mode should be optimized to prevent the field in the resonator formed by the plates from acting back on the
 circulating beam \cite{Belver-Aguilar_14}. As is discussed later in Section \ref{subsec: Zc_dip}, in general matching the
geometric means of both modal impedances to that of the external lines may be the best compromise.

\subsection{Potential for the Odd Mode}

We consider two arc shaped electrodes held at a constant voltage $\pm V_p$, a schematic is shown in
Fig. \ref{fig: dipole_polar}. The rods supporting the plates are omitted in this sketch. 
 For typical external voltages, the relative contribution of the beam induced voltage to the total 
 voltage is negligible so the potential can be assumed to obey Laplace's equation. In two dimensional polar coordinates,
 $(r, \theta)$ we have
\beqr
\fr{1}{r}\fr{\del }{\del r}(r \fr{\del \Phi}{\del r}) + \fr{1}{r^2}\fr{\del^2 \Phi}{\del \theta^2} &  = &  0  
\eeqr
For a well-posed problem, the specified boundary conditions ensure a unique solution.   
\bfig[h]
\centering
\includegraphics[scale=0.4]{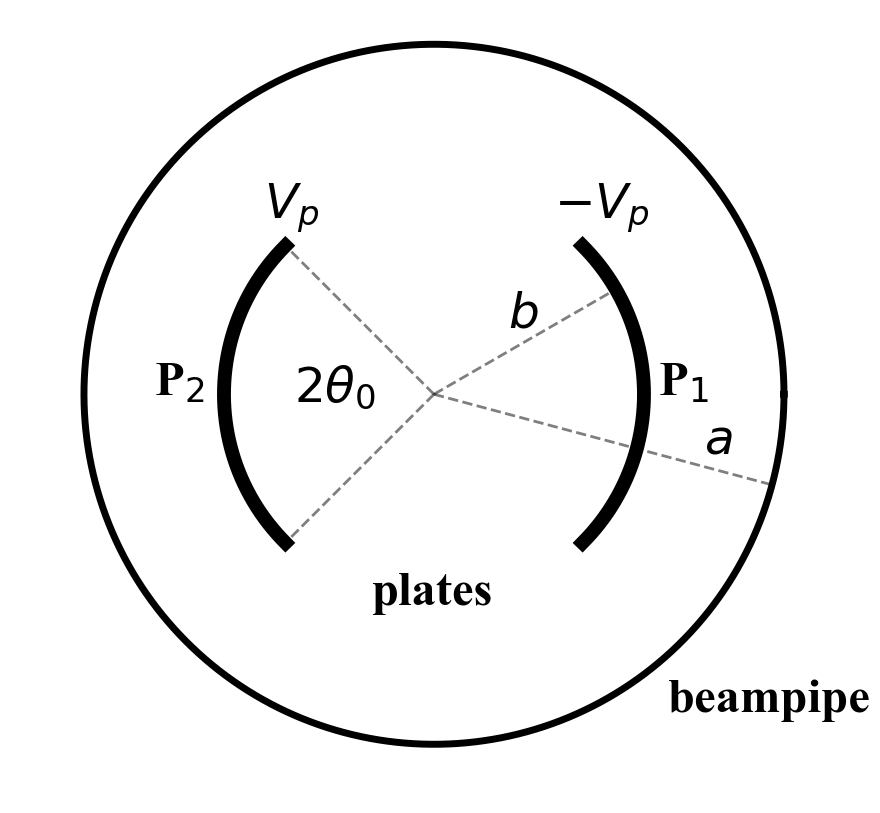}
\caption{Dipole kicker with arc shaped plates.}
\label{fig: dipole_polar}
\efig
Assuming a separation of variables, the potential can be written as  $ \Phi(r, \theta) = \Theta(\theta) R(r) $,  where $\Theta, R$ are as yet arbitrary functions of their
arguments. 
Using the fact that the potential is periodic in $\theta$ over $2\pi$, it can be shown that the general 
solution, a superposition of the separable solutions, is of the form
\beq
\Phi(r, \theta) = c_0 + a_0 \ln r + \sum_{m= 1}^{\infty} [a_m r^m + b_m r^{-m}][c_m \cos m\theta + d_m \sin m\theta]
\eeq
In the analytic expressions to follow, the upper limit is infinity while in the numerical evaluations, the upper limit is a suitably large integer. 
The plates are located at a radius $b$, and the beam pipe at a radius $a$.  
Each plate subtends an angle of magnitude $2\theta_0$.
We assume  the left and right plates are respectively at voltages $V_p$ and $-V_p$ so that a positively charged particle is kicked in the positive $x$ direction i.e. to the right. Thus $\Phi(r = b, \theta) = -V_p$
on plate $P_1$  which extends over the angles $ - \theta_0 \le \theta \le \theta_0$, and
$\Phi(r = b, \theta) = V_p$ over plate $P_2: \pi - \theta_0 \le \pi + \theta_0$.
The potential must be continuous across the entire boundary $r=b$.
In the region exterior to the plates, the potential vanishes on the beampipe 
$\Phi(a, \theta) = 0$. 
Across the interface between the interior and exterior regions $(r=b)$, the radial component of the electric field is continuous on the intervals where no electrode is present: 
\beqr
\fr{\del \Phi_{in}(r=b,\theta \in (G_1, G_2))}{\del r} & = & \fr{\del \Phi_{ex}(r=b,\theta \in (G_1, G_2))}{\del r}
\eeqr
where the gaps $G_1, G_2$ have the domains: $G_1: \theta_0 \le \theta \le \pi - \theta_0$ and
$G_2: \pi + \theta_0 \le \theta \le 2\pi - \theta_0$ respectively. 
Due to the presence of charge on the electrode surfaces, it is necessary to consider the interior and exterior regions separately.    

The potential within the interior of the plates must be be well behaved as $r \rarw 0$. This eliminates the coefficients $a_0, b_m$.
\beq
\Phi_{in}(r \le b, \theta) = c_0  +  \sum_{m=1} r^m (c_m \cos m \theta + d_m \sin m\theta)
\label{eq: Vpolar_2}
\eeq
where $(a_m, b_m) $ have been absorbed into the redefined coefficients $c_m, d_m$.
The potential is symmetric about the $x$ axis or 
$ \Phi(b, 2\pi - \theta) = \Phi(b, \theta)$, so $ d_m = 0$. 
Furthermore, the anti-symmetry of the potential with respect to the vertical $y$ axis implies 
$ \Phi(b,\pi - \theta) = - \Phi(b, \theta) $. 
Imposing this requirement in Eq.(\ref{eq: Vpolar_2}), one concludes 
that $ c_0 = 0$ and $m = $ odd. 

For the potential exterior to the plates we start from the general form 
\beq
\Phi_{ex}(b\le r \le a, \theta) = c_0 + A_0 \ln r + \sum_{m= 1} [A_m r^m + B_m r^{-m}][C_m \cos m\theta + D_m \sin m\theta]
\label{eq: Vpolar_ext}
\eeq
where the coefficients $A_m, B_m, C_m, D_m$ are different from the coefficients $a_m,b_m,c_m, d_m$ of the interior solution.  
They are determined from the  boundary conditions on the exterior potential.
Requiring that this potential  vanish at $r=a$ implies
$  c_0  =  - A_0 \ln a$ , and $ B_m  =  - A_m a^{2m}$. 
Matching the interior and exterior solutions at $r = b$ yields  $   A_0 = 0 = D_m $, the index $m=1,3,5 \ldots$, and the
coefficients $A_m$ can be absorbed into $C_m$ which can be expressed in terms of the interior coefficients $c_m$ as
$C_m  [ 1 - (a/b)^{2m}]  = c_m $ where $m$ is odd.
Expressed in terms of the interior coefficients, the solutions for the interior and exterior potentials are
\begin{align}
\Phi_{in}(r, \theta) & =    V_p \sum_{m=1, 3, ...} X_m  (\fr{r}{b})^m \cos m\theta , \;\;\; & 0 \le r \le b  \label{eq: Vpolar_int_3}
\\
\Phi_{ex}(r, \theta) & =   V_p \sum_{m=1, 3, ...} \fr{X_m }{[ 1 - (a/b)^{2m}]} [(\fr{r}{b})^m - (\fr{a^2}{b r})^{m}]
\cos m\theta , \;\;\;  & b \le r \le a  
\label{eq: Vpolar_ext_3}
\end{align}
where we have introduced new scaled dimensionless coefficients $X_m = b^m c_m/V_p$.
Imposing the boundary condition on the right plate at $r=b$ and
matching the normal derivative across the gaps at $r = b$ yields respectively
\beqr
\sum_{m=1, 3, ...}  X_m   \cos m\theta  &  =  -1 , \;\;\;   &  \theta \in (P_1, P_2)  \label{eq: dip_odd_BC1} \\
\sum_{m=1, 3, ...} m  g_m(a, b) X_m   \cos m\theta & =  0  , \;\;\;  &  \theta \in (G_1, G_2) \label{eq: dip_odd_BC2}
\eeqr
where the dimensionless geometric coefficient $g_m$ is defined as
\beq
g_m(a, b) = \fr{1 }{[ 1 - (b/a)^{2m}]} \ge 1 ,  \;\;\; b <  a  \label{eq: gm_def}
\eeq
We note that $g_m$ decreases with increasing index and approaches 1 as $m \rarw \infty$. 

Integrating Eq. (\ref{eq: dip_odd_BC1}) over the angular extent of the plate at $r=b$ and 
Eq.(\ref{eq: dip_odd_BC2}) over half of the top gap $G_1$: $\theta_0 \le \theta \le \pi/2 $ (the integral over the complete gap
vanishes  because of anti-symmetry) yields the two equations 
\beqr
 \sum_{m=1, 3, ...}  X_m \fr{\sin m \theta_0}{m\theta_0} & = & -  1   \label{Int_Xm_polar_int} \\
 \sum_{m=1, 3, ...} g_m(a, b) X_m [ (-1)^{(m-1)/2} - \sin m\theta_0] & = & 0  \label{Int_Xm_polar_ext}
\eeqr
These are integral conditions which must be satisfied for a given set of geometric parameters. 
The coefficients $X_m$ must however be found from the local conditions in Eq. (\ref{eq: dip_odd_BC1}) and
Eq. (\ref{eq: dip_odd_BC2}) which are valid at every point within their respective domains. Below we discuss two methods for
determining them.

\subsubsection{Least Squares Method} \label{subsec: least_sq}

We follow the method used in \cite{Wang_78} which parallels a development by Sommerfeld in \cite{Sommerfeld} to
treat the problem of light waves reflecting off a curved mirror.
Essentially, the method consists of determining the expansion coefficients so as to minimize the quadratic residual error
on the boundary conditions.  Taking the sum of the squared difference of  Eqs. (\ref{eq: dip_odd_BC1}) and (\ref{eq: dip_odd_BC2}) and integrating over the appropriate azimuthal ranges yields an error function as
\beq
Err({\bf X}) = \int_{-\theta_0}^{\theta_0}[1 + \sum_{m=1, 3, ...} X_m \cos m \theta] ^2 d\theta +
\int_{\theta_0}^{\pi - \theta_0} [\sum_{m=1, 3, ...} m  g_m(a, b) X_m   \cos m\theta]^2 d\theta  \label{eq: dip_err_func}
\eeq
The minimum residual is obtained by setting the partial derivatives to zero 
$ \fr{\del Err({\bf X})}{\del X_j} = 0$ which yields matrix equations for the coefficients. 
 Define the vector ${\bf b}$ with components $b_n$ and matrix ${\bf A}$ with elements $A_{m n}$ as follows
    \beqr
    b_n & =  & \int_{-\theta_0}^{\theta_0} \cos n\theta d\theta = \fr{2}{n} \sin n\theta_0   \label{eq: def_bn}  \\
    A_{m  n} & = &  \int_{-\theta_0}^{\theta_0} \cos m \theta \cos n\theta \; d\theta  
\!\!\! =   [\fr{\sin (n-m)\theta_0}{n - m}  +  \fr{\sin (n + m)\theta_0}{n + m}]  , \;\;\; m \ne n \label{eq: def_Amn} \\
    A_{n n} & = & \int_{-\theta_0}^{\theta_0} \cos^2 n\theta \; d\theta = [ \theta_0 + \fr{1}{2n}\sin 2n\theta_0]  \label{eq: def_Ann}
    \eeqr
The diagonal elements follow from the off-diagonal elements on using $\lim_{m \rarw n}\sin (n-m)\theta_0/(n - m) = \theta_0$. 
This matrix ${\bf A}$ is symmetric, $A_{m n} = A_{n m}$.  These elements will arise in all the situations to be discussed in this
paper for both the dipole and quadrupole kickers. 

Collecting terms leads to the matrix equation ${\bf B} \cdot {\bf X} = - {\bf b}$ or in component form
  \beqr 
& \mbox{}    \sum_{m={\rm odd}} B_{m n}X_m  = - b_n , \;\;\; &  n = \; {\rm odd}  \label{eq: dip_lsq_matrix}  \\
  B_{n n} & = A_{n n} +  n^2  g_n(a, b)^2 \left(\fr{\pi}{2} - A_{n n} \right) , \;\;\;  & n = \; {\rm odd} \label{eq: dip_Bnn} \\
  B_{m n} & =   \left[1  -   m n  g_m(a, b) g_n(a, b) \right] A_{m n},   \;\;\;  &  m= \; {\rm odd} \; \ne n
  \eeqr
  The matrix ${\bf B}$ is a non-singular, square, symmetric matrix of dimension $N\times N$ to solve for the $N$ coefficients
  $X_1, X_3, \ldots,  X_{2N-1}$. 
  This matrix equation (\ref{eq:  dip_lsq_matrix}) must be truncated and solved numerically.

  \subsubsection{Projection Method}  \label{subsec: project}

This follows the second method investigated in \cite{Wang_78} which is referred to as the ``simple integration'' method.
We choose to call it the projection method since it is based on projecting the coefficients $X_n$ on to a basis set of harmonic
functions. 
Multiplying  Eq.(\ref{eq: dip_odd_BC1}) by $\cos n\theta$ and integrating over the plate on the right $P_1:  -\theta_0 \le \theta \le \theta_0$ leads to the set of equations
\beq
 \sum_{m={\rm odd}} A_{mn}X_m   + b_n = 0 , \;\;\; n = \; {\rm odd}
\label{eq: polar_Xm_int}
\eeq
These coefficients form a potential that only satisfies the boundary conditions on the plates but not in the gaps. The matrix ${\bf  A}$
is also singular which is a consequence of the fact that it does not specify a unique potential.

Multiplying Eq.(\ref{eq:  dip_odd_BC2}) by $\cos n\theta$ and integrating over the gap from $\theta_0$ to $\pi - \theta_0$
yields the set of equations
\beq
n g_n(a,b) [\fr{\pi}{2} - A_{nn}] X_n - \sum_{m=1,3,..., \ne n} m g_m A_{mn} X_m  =  0, \;\;\;\; n = {\rm odd}
\label{eq: polar_Xm_ext}
\eeq
We can combine Eqs. (\ref{eq: polar_Xm_int}) and (\ref{eq: polar_Xm_ext}) into a system of $N$ equations
for $N$ unknowns as
\begin{align}
  - b_n & =   \sum_{m= {\rm odd}}C_{m n}X_m  , \;\;\;\;\; & \;\;\;\;  n = {\rm odd}   \label{eq: matrix_proj_2}  \\
C_{nn} & =    [1 -  n g_n(a, b) ] A_{nn} + n\fr{\pi}{2}  g_n(a,b),  & \;\;\;\;   n = {\rm odd} \label{eq: dip_Cnn} \\
C_{m n} & =    [1 - m g_m(a,b)] A_{mn},   \;\;\;  &  m= \; {\rm odd} \; \ne n    \label{eq: dip_Cmn}
\end{align}    
The equations have been combined so that the matrix ${\bf C}$ is in general non-singular and can be used to numerically find the desired coefficients $X_m$.
In order to find an approximate expression for the lowest order coefficient, we can keep only the first term and find
\beq
X_1 \approx -\fr{b_1}{C_{11}} =  \fr{4 (1- (b/a)^2) \sin \theta_0}{ \pi - (2 -  (b/a)^2)(2\theta_0 + \sin 2 \theta_0)}
\eeq
This provides  a quick, rough estimate of the dipole field.
Appendix A shows the coefficients $X_1, X_3$ when calculated to the next order and discusses the errors with these
approximations.

\subsection{Potential for the Even Mode}

Here we consider the potential when both plates are at the same voltage. This mode is excited when the beam induces image
currents and a voltage difference with respect to the beampipe. This is the so-called even mode or common mode and its
characteristic
impedance is involved in matching to the external impedances, as is discussed later in Section \ref{subsec: Zc_dip}. 

We assume that both plates are at a voltage $V_b$. Now the potential is symmetric about
the $y$ axis (as opposed to the anti-symmetry in the odd mode) and has identical symmetry about the $x$ axis, i.e.
$\Phi(b,\pi - \theta) = \Phi(b, \theta), \;\;\;\;  \Phi(b, 2\pi - \theta) = \Phi(b, \theta)$.
We start with the form for the interior potential in Eq.(\ref{eq: Vpolar_2}).
Symmetry about the $x$ axis requires that $d_m = 0$, while symmetry about the $y$ axis requires $m$ is  even.
For the exterior solution  we start with the general form in Eq.(\ref{eq: Vpolar_ext}).
 Requiring that the external potential vanishes at $r= a$ and matching the exterior and interior potentials at $r=b$
 $\forall \theta$ yields its form.
 We introduce the scaled dimensionless  coefficients  $X_0 = c_0/V_b$,  $X_m  = b^m c_m/V_b$. The two even mode
 potentials  can be written      as
 \beqr
 \Phi_{in}(r, \theta) & = & V_b[ X_0 + \sum_{m= {\rm even}} X_m (\fr{r}{b})^m \cos m\theta , \;\;\; \hspace{3em}  0 \le r \le b \\
   \Phi_{ex}(r, \theta) & = & V_b \left[ X_0 \fr{\ln(r/a)}{\ln(b/a)} +\!\!\!\!\!\! \sum_{m= {\rm even}} \fr{ X_m}{[1 - (\fr{a}{b})^{2m}] }
   [  (\fr{r}{b} )^m - (\fr{a^2}{b r})^m] \cos m\theta \right], \;\;\;     b \le r \le a  \non \\
   \eeqr
   Matching the two potentials on the plates and their radial derivatives in the gaps leads to the boundary conditions
    \begin{align}
    1 & =   X_0 + \sum_{m={\rm even}}X_m \cos m \theta ,  \;\;\;  &  \theta \in (P_1, P_2)    \label{eq: dip_even_BC1} \\
    X_0 & =  2 \ln(b/a)\sum_{m={\rm even}}X_m m g_m \cos m \theta , \;\;\; &  \theta \in (G_1, G_2) \label{eq: dip_even_BC2}
    \end{align}
    The integral conditions obtained by integrating Eq.(\ref{eq: dip_even_BC1}) over a plate and Eq.(\ref{eq: dip_even_BC2})
      over either gap  are
      \beqr
      1 & = &  X_0 + \sum_{m={\rm even}}X_m \fr{\sin m \theta_0 }{m \theta_0}   \label{eq: dip_global_even_BC1} \\
      (\pi/2 - \theta_0) X_0 & = &   - 2\ln(b/a)\sum_{m={\rm even}} g_m X_m \sin m\theta_0, 
      \label{eq: dip_global_even_BC2} 
      \eeqr
  
For the sake of brevity, we consider only the projection method to determine the potential in this mode. 
Proceeding as before, i.e. multiplying Eq. (\ref{eq: dip_even_BC1}) and Eq. (\ref{eq: dip_even_BC2}) by $\cos n\theta$
and integrating over the appropriate range of $\theta$, we obtain
    \begin{align*}
    b_n &  =  X_0 b_n + \sum_{m={\rm even}, \ne n}X_m A_{m, n} + X_n A_{n, n}   \;\;\; & n = {\rm even}  \\
    -X_0 b_n & =  2 \ln(b/a)\left[ -\sum_{m={\rm even}, \ne n}X_m m g_m A_{m,n} + X_n n g_n (\pi/2 - A_{n, n}) \right]  \;\;\;
&   n = {\rm even} 
  \end{align*}
  These equations can be combined to yield the matrix system
  \begin{align}
     b_n & =     \sum_{m={\rm even}} D_{m n}X_m , \;\;\;\;\;\;\;\;\; & n = {\rm even} \\
     D_{n n} & =   [(1 + 2\ln(b/a) n g_n) A_{n, n} - n \pi g_n \ln(b/a) ],   & \;\; n = {\rm even}  \label{eq: dip_Dnn} \\
      D_{m n} & =   (1 +   2\ln(b/a) m g_{m, n}) A_{m, n} \;\;\;  &  m = {\rm even} \ne n  \label{eq: dip_Dmn} 
      \end{align}
      Once the $X_n, n > 0$ are found, $X_0$ can be found from either of the integral conditions in
      Eq. (\ref{eq: dip_global_even_BC1}) or Eq. (\ref{eq: dip_global_even_BC2}). 

      To lowest order, keeping only the first term in the matrix equation, we have the coefficients
      \beqr
      X_2 & \approx & \fr{b_2}{D_{2 2}} = -\fr{2 (1 - (b/a)^4) \sin 4 \theta_0}{ -4 (1 - (b/a)^4) \theta_0 +  8 (\pi - 2 \theta_0) \ln (b/a)
-  (1 - (b/a)^4 + 4 \ln (b/a)) \sin 4 \theta_0} \\
     X_0 & \approx & 1 - X_2 \fr{\sin 2\theta_0}{2\theta_0}
    \eeqr

\subsection{Electric and Magnetic fields in the odd mode}
From the interior potential, it follows that the electric fields in the interior along the Cartesian axes
acting on a particle with polar coordinates $(r < b, \theta)$ are
 \beqr
 E_x & =  &    - \fr{V_p}{b}\sum_{m={\rm odd}} m X_m (\fr{r}{b})^{m-1} \cos (m-1)\theta 
= - \fr{V_p}{b} \left[  X_1  + 3 X_3 (\fr{r}{b})^2\cos 2\theta  +  \ldots  \right] \\
 E_y & =  &   \fr{V_p}{b}\sum_{m={\rm odd}} m X_m  (\fr{r}{b})^{m-1}\sin (m-1)\theta = 
\fr{V_p}{b}\left[  3 X_3 (\fr{r}{b})^2\sin 2\theta  + 5 X_5 (\fr{r}{b})^{3}\sin 4\theta  + \ldots  \right]  \non \\
 \eeqr
We expect (and verify in Section \ref{sec: Numerics}  )  that the coefficient magnitudes $|X_m|$ decrease with increasing order. 
Along the $x$ axis, the horizontal field $E_x$ has its maximum value  while $E_y$ vanishes along both the horizontal and
vertical axes.
The first term in $E_y$ has a maximum at $\theta = \pi/4$, the second term along $\pi/8$ etc. 
Thus for small enough beam size ($\sg_{\perp}\ll b$, $\sg_{\perp}$ is the transverse rms size) the field in this kicker approaches
that of a pure dipole, while for larger beam sizes the beam experiences nonlinear kicks  in both directions.

In the analysis to follow here and later, we will assume that the relevant modes are matched so that there are no reflections
from either end i.e. we have only a pure TEM wave propagating from the power source.  For such a wave propagating
along $+\hat{z}$, the electric $\vec{E}$ and magnetic $\vec{B}$ fields obey
\beq
 \hat{z} \times \vec{E} = c \vec{B}, \;\;\; \Rarw c B_x = - E_y , \;\;\;  c B_y =  E_x , \;\;\; E_z = 0 = B_z
\label{eq: fields_TEM}
\eeq
A particle with charge $q$ propagating along $-\hat{z}$ or in a  direction opposite to that of the EM wave, will experience a
force with horizontal component
\beq
F_x = q[E_x + (\vec{v}\times \vec{B})_x ] =  q (1 + \bt) E_x(t)  \label{eq: Fx}
\eeq
and a similar expression for $F_y$. 
The change in momentum $\Dl p_x$ due to this force from a kicker of length $L_k$ is found from
$\Dl p_x   =  \fr{L_k}{\bt c} F_x$ while the angular kick $\Dl x'$ is given by $\Dl p_x= m_0\gm  \bt c   \Dl x'$,
where $m_0$ is the rest mass.  Hence the total horizontal angular kick is
  \beqr
  \Dl x' & = &   \fr{q(1 + \bt)}{\bt^2  \gm m_0 c^2}E_x L_k   
  \eeqr
  which is the sum of  kicks from the electric and magnetic fields.
  If the particle propagates in the same direction as the wave, the two forces oppose each other leading to a near
  cancellation for relativistic particles.  At low energies, for example the 2 MeV proton beam in IOTA
has $\bt = 0.07$, the magnetic kick is a small fraction of the kick from the electric field, so the relative direction of
propagation of the wave and particles does not matter much. 
We can estimate the dipole kick by using the approximate analytic solution $X_1^{(2)}$ for $X_1$ using a 2x2 matrix and shown in equation
\ref{eq: X1_anal_2nd}  in Appendix A. Using the geometry of the existing injection kickers in IOTA \cite{Antipov_strip}, we have
$\theta_0 = 32.5^{\circ}$ and
$b/a = 0.8$, we obtain $X_1= -0.65$ from \ref{eq: X1_anal_2nd}. The right plot in Fig. \ref{fig: X1sol-error} shows that at
this (half)
coverage angle, this value underestimates the correct value by about 20\%. Including this correction, a more precise value is $X_1 = -0.78$.
Assuming a plate voltage of
  1 kV, a plate radius of 20 mm, a compact kicker length of  20 cm, the kick on the IOTA beam is $\Dl x' = 1.64$ mrad. In terms
  of the average beam size at a location with $\bt_x = \bt_{av}= 1.2$ m and average beam size $\sg_{av} = 2.2$ mm, this
  amounts to a kick $\bt_x \Dl x' \simeq 0.94\sg_{av}$. If instead we apply the above estimate to the existing injection stripline
  kicker  where $V_p = 25$kV, $L_k=0.635$m, $b=0.02$m, we have a beam kick
  $\bt_x \Dl x' \simeq 72\sg_{av}$. This is considerably larger than required ($\sim 5-10 \sg_x$) in order to explore the
  nonlinear aspects of the dipole kick on echoes \cite{Sen_18}.

\subsection{Characteristic Impedance} \label{subsec: Zc_dip}

An arrangement of $n$ deflecting plates enclosed by a conducting beam pipe forms a set
of $n$  coupled transmission lines.  For a TEM wave, in the frequency domain the voltage and current amplitudes $V_i$  and
$I_i$ associated with each one of the plates are locally related to each other through the two relations  
\begin{equation}
  k \, {\bf V} = \omega {\bf L}_M' \,  {\bf I},   \;\;\;\;\;   k \, {\bf I} = \omega {\bf C}_M' \,  {\bf V}
 \label{impedance_matrix}
 \end{equation}
  where ${\bf V}$ and ${\bf I}$ are vectors of dimension $n$ while $k$, $\omega$ are the spatial and angular 
   frequencies and  ${\bf L'}_M$, ${\bf C}_M'$ are the distributed Maxwell inductance and capacitance matrices. 
   The equations express the fact that current in any one of the conductors induces a proportional voltage
     in all the others and vice-versa.
  Although the elements  $\ell_{ij}'$ and $c_{ij}'$ of  ${\bf L'}_M$, ${\bf C}_M'$ respectively have units of [H/m] and [F/m], 
   they do not represent conventional circuit elements.  
   In what follows, we shall reserve the notation  $L'_i, L'_{ij}, C'_i, C'_{ij}$  for  such elements.

   Combining both equations yields the dispersion relation
   \begin{equation}
     c^2 \;\;{\bf L}'_M {\bf C}'_M = {\bf I}_{unit}
  \label{dispersion_relation}
  \end{equation}
   where we used the fact that the wave velocity $c = \omega/k$ and ${\bf I}_{unit}$ is the unit diagonal matrix.  Note that by reciprocity, the matrices ${\bf L}'_M$, ${\bf C}'_M$ 
   as well as their product are symmetric for any arrangement of the plates (symmetric or not).
   Using equation (\ref{dispersion_relation}) to substitute for $\omega/k = c$ in  the first part of Eq. (\ref{impedance_matrix})
   relating ${\bf V}$ to ${\bf I}$ yields 
  \begin{equation}
  {\bf V} = \left[ ({\bf C}'_M)^{-1} {\bf L}'_M \right]^{1/2} \; {\bf  I }  = \frac{1}{c}\; ({\bf C}_M')^{-1}  {\bf I} = \; {\bf Z}_c {\bf  I }
 \label{impedance_matrix1}
 \end{equation}
 where ${\bf Z}_c$
 is known as the characteristic impedance matrix. 
  
 For an $n$-fold symmetric plate arrangement, the number of independent elements of ${\bf C}'_M$ (or ${\bf L}'_M$) is
 reduced and all the $c'_{ii} = c'_{11} $ while the off diagonal  $c'_{ij}$ depend only on the angular distance
 between the  electrodes $i$ and $j$.
  For a dipole kicker, $n=2$, one can verify that the eigenvectors of ${\bf C}'_M$ are  ${\bf u}_o = (1,-1)$ and ${\bf u}_e =(1,1)$.  These eigenvectors, which are shared 
 by the capacitance and characteristic impedance matrices, define the so-called coupled modes. An arbitrary excitation can be 
  expressed as a linear combination of these modes. Using the dispersion relation, the eigenvalues of the characteristic impedance matrix  $Z_{c,  odd}$ and $Z_{c, even}$
  may be expressed in terms of those of the ${\bf L}'_M$ and ${\bf C}'_M$ matrices to define the effective capacitance
  and inductance of the even and odd modes $C'_e, L'_e, C'_o, L'_o$.    
  \begin{eqnarray}
  Z_{c, odd} & =  &  \frac{1}{c}\frac{1}{c'_{11} -   c'_{12} } = \frac{1}{cC'_o}  = \sqrt{ L'_o/C'_o} \label{eq: Zc_2_1}  \\
  Z_{c, even} & = &  = \frac{1}{c}\frac{1}{c'_{11} +   c'_{12} } =  \frac{1}{cC'_e} = \sqrt{ L'_e/C'_e} \label{eq: Zc_2_2}
   \end{eqnarray}
  where the $c_{ij}$  (called ``coefficients of induction'' in \cite{Jackson}, Chapter 1) are elements of the Maxwell capacitance matrix while the $C'_k$ are the modal capacitances. 
  In  general, $c'_{ij} \le 0, \;\; i \ne j$; hence,  $Z_{c, even} \ge Z_{c, odd}$. 
  A derivation of these results is presented in Appendix B where it is also shown how the $c'_{ij}$ are related to the mutual capacitances $C'_{ij}$. 
  
  When each stripline  is terminated by an impedance to ground equal to the characteristic impedance of a given mode,
  there is no reflection of that mode. Assuming that all terminations have impedance $Z_L$,  both modes are 
  perfectly matched when  $Z_{c,  even} =  Z_{c, odd} = Z_L$. However, this condition is restrictive
  since it implies $C'_{12}/C'_{11} \rightarrow 0 $ which happens with   increasing angular plate separation or  alternatively by
  increasing $b/a $.  
  Appendix C shows that in general a load termination network can be devised to match all modes with any number of plates.
  For a dipole configuration, this requires an additional resistor between the two electrodes, a scheme that was also proposed in  \cite{Belver-Aguilar_14}. 
   
  Another often used alternative follows from these weaker requirements: (1) No injected power is reflected back to the generat
or  (2) power deposited in the even (or common) mode is coupled out of the striplines.
With the line extremities terminated by a load $Z_L$  both these conditions are satisfied  when''  
  \begin{equation}
  Z_{c, odd} Z_{c, even} = Z^2_L
  \label{geometric_mean}
  \end{equation}
  is fulfilled.  In fact, when \ref{geometric_mean} holds, the plate arrangement is a directional coupler. However, for high
  beam  current  applications as mentioned previously, it may sometimes be preferable and simpler to match the even mode to $Z_L$ and to 
  tolerate some odd mode mismatch. 
   
   We now calculate the frequency independent (low frequency)  part of the characteristic impedance.
   By definition, a mode characteristic impedance is the ratio of its voltage $V_p$ and current $I_p$ mode amplitudes : 
   $ Z_c = V_p/I_p$.
  The current $I_p$ can be expressed in terms of the surface current density, i.e. the current per unit length
  normal to the direction of current flow. Let $\vec{K_p}$ be the current density on a plate
  \beq
  I_p = \int_{L_p} \lvert \vec{K_p} \rvert dl 
  \eeq
where $dl$ is an element of length and $L_p$ defines the contour of the plate. 
At the interface between two media (vacuum in our case), the
discontinuity between the tangential components of the magnetic field on either side of the interface is given by \cite{Jackson}
  \beq 
\vec{K} =  \hat{n} \times [\vec{H}_{1} -  \vec{H}_{2} ] = \hat{n} \times [ \fr{1}{\mu_1}\vec{B}_{1} -  \fr{1}{\mu_2} \vec{B}_2 ]
  \eeq
  where   $\hat{n}$ is the unit normal from media 1 (region interior to the plates)  towards media 2 (region exterior
  to the plates).  In the second equality we have assumed the media are linear so that $\vec{B} = \mu \vec{H}$, $\mu$ is 
  the magnetic permeability. The normal component of the $\vec{B}$ field is continuous across the plate.
 For a TEM wave propagating  in the stripline, the $\vec{E}$ and $\vec{B}$ fields
  are orthogonal everywhere to the direction of propagation.   We have $ c \vec{B} = \hat{z}\times \vec{E}$, use the relation
  $\hat{n}\times \hat{z} \times \vec{E} = (\hat{n}\cdot \vec{E}) \hat{z}$ and
  let  $\mu_1 = \mu_2 = \mu_0$ (vacuum permeability) to  write the surface current density on the plate in terms of the
  discontinuity in the normal (or radial) components of the electric field across the plate.
\beq
 K_p = \fr{1}{Z_0}[ E_{in, r}(r=b) -  E_{ex, r}(r=b) ]
\eeq
where $Z_0= \mu_0 c$ is the impedance of free space.

Next we calculate the characteristic impedance of the odd and even modes. The transverse and longitudinal beam coupling
impedances are proportional to the characteristic impedances of the odd and even modes respectively
\cite{Gold_Lamb, Blednykh_13}.

\subsubsection{Odd Mode Characteristic Impedance} 
Using the potential forms in Eqs.(\ref{eq: Vpolar_int_3}) and (\ref{eq: Vpolar_ext_3}), we have for the surface current density
\beqr
K_p & = &  \fr{V_p}{b Z_0} \vert \sum_{m={\rm odd}} m X_{m}\left [ 1 - \fr{1 + (a/b)^{2m}}{1 - (a/b)^{2m}} \right]  \cos m\theta \rvert
 =   \fr{2V_p}{b Z_0} \lvert \sum_{m={\rm odd}} m X_{m} g_m(a, b) \cos m\theta \rvert  \non \\
\eeqr
The current on either plate is
\beqr
I_p & = & b \lvert\int_{-\theta_0}^{\theta_0} K_p(\theta) \; d\theta \; \rvert  = 4\fr{V_p}{Z_0} \lvert \sum_{m={\rm odd}}  X_{m} g_m(a, b) \sin m\theta_0 \rvert
\label{eq: Ip_dip_polar}
\eeqr
Hence the characteristic impedance is
\beqr
Z_{c, odd} & = &  \fr{Z_0}{4 \lvert \sum_{m={\rm odd}}  X_{m} g_m(a, b) \sin m\theta_0 \rvert} =
\fr{Z_0}{ 4 \lvert\sum_{m={\rm odd}}  (-1)^{(m-1)/2} X_{m} g_m(a, b)  \rvert}  \label{eq: Zc_dip}
\eeqr
where we used the integral condition in Eq.(\ref{Int_Xm_polar_ext}) in the second equality above. 
Eq. (\ref{eq: Zc_dip}) shows for example that $Z_{c, odd}$ is determined entirely by the half coverage angle $\theta_0$ and the ratio
$b/a$ [see Eq.(\ref{eq: gm_def})] and not by the specific values of $a, b$. As the ratio $b/a$ increases, $Z_{c, odd}$ decreases and $Z_{c, odd} \to 0$ when $b/a \to 1$.

\subsubsection{Even Mode Characteristic Impedance} 

The surface current density defined in terms of the discontinuity in the radial electric field across a plate is
\beqrs
K_p  & = & \fr{V_b}{b Z_0}\left[ \lvert \fr{X_0}{\ln(b/a)} - 2\sum_{m={\rm even}} m X_{m} g_m(a, b) \cos m\theta \rvert \right]
\eeqrs
The current on a plate is
\beqr
I_p & = &  2\fr{V_b}{Z_0} \lvert \fr{X_0 \theta_0}{\ln(b/a)} - 
2\sum_{m={\rm even}}  X_{m} g_m(a, b) \sin m\theta_ 0 \rvert  
=  \pi \fr{V_b}{Z_0}\fr{\lvert  X_0\rvert}{\ln(a/b)}   \label{eq: Ip_dip_polar_even}
\eeqr
where in the last step we used the integral condition in Eq.(\ref{eq: dip_global_even_BC2} ). 
Hence the characteristic impedance of the even mode is
\beq
Z_{c, even} = \fr{Z_0}{\pi} \fr{\ln(a/b)}{ \lvert X_0 \rvert}    \label{eq: Zc_even}
 \eeq
 This expression resembles the characteristic impedance of a coaxial line with a single cable, $Z_c = Z_0\ln(a/b)/(2\pi)$
 and differs by a factor of two from a similar expression for the characteristic impedance of a single stripwire
 kicker \cite{Barry_Liu_90}.

\section{Quadrupole Kicker with circular symmetry} \label{sec: Quad}

  For a four-fold symmetric quadrupole kicker, the capacitance matrix has three independent elements and the 
  four characteristic impedances are
  \begin{eqnarray} 
  Z_{c1} &=& Z_{c2} =  \frac{1}{c} \frac{1}{c_{11}'-c_{13}'} = \frac{1}{cC_1'}=
  \sqrt{L_1'/C_1'}  \label{eq: Zc_4_1}   \\  
  Z_{c3} &=&  \frac{1}{c} \frac{1}{c_{11}'+ c_{13}' - 2 c_{12}'}= \frac{1}{cC_3'} =
  \sqrt{L_3'/C_3'}  \label{eq: Zc_4_2}  \\
  Z_{c4} &=& \frac{1}{c} \frac{1}{c_{11}'+ c_{13}' + 2c_{12}'} = \frac{1}{cC_4'} = \sqrt{L_4'/C_4'}  \label{eq: Zc_4_3}
\end{eqnarray}
As in the dipole case, the $c'_{ij}$ are elements of the Maxwell capacitance matrix. A derivation can also be found in Appendix B.
Modes 1 and 2 are known as dipole modes, 
mode 3 will be referred as the  quadrupole mode (the mode which applies a  quadrupolar kick) and
mode 4 as the sum mode (a beam induced mode where all plates are at the same potential). 
From the above definitions of the modes and the property of the Maxwell capacitances $c_{i j}' \le 0, \;\; i \ne j$, it follows that in general
$Z_{c, sum} \equiv Z_{c4} \ge Z_{c, quad} \equiv Z_{c3}$. 
In analogy with the dipole kicker case, one may either choose a suitable load network to match all modes (Appendix C) or alternatively, a directional coupler configuration  with individual lines terminated with loads $Z_L$. In the latter case, power sent on any one of the lines will not be reflected provided that the conditions "
  \begin{equation}
  Z_{c,quad} Z_{c,sum} = Z_{c,dipole}^2 = Z_L^2 
  \label{quad_condition}
  \end{equation}
are fulfilled. In this section we again solve for the potential and determine the quadrupole and sum modes characteristic
  impedances following the method of the previous section.

\subsection{Potential solution for the quadrupole mode}

The plates are numbered in anti-clockwise order starting from the right, a sketch is shown  in  Fig. \ref{fig: quad_polar}. 
\bfig
\centering
\includegraphics[scale=0.4]{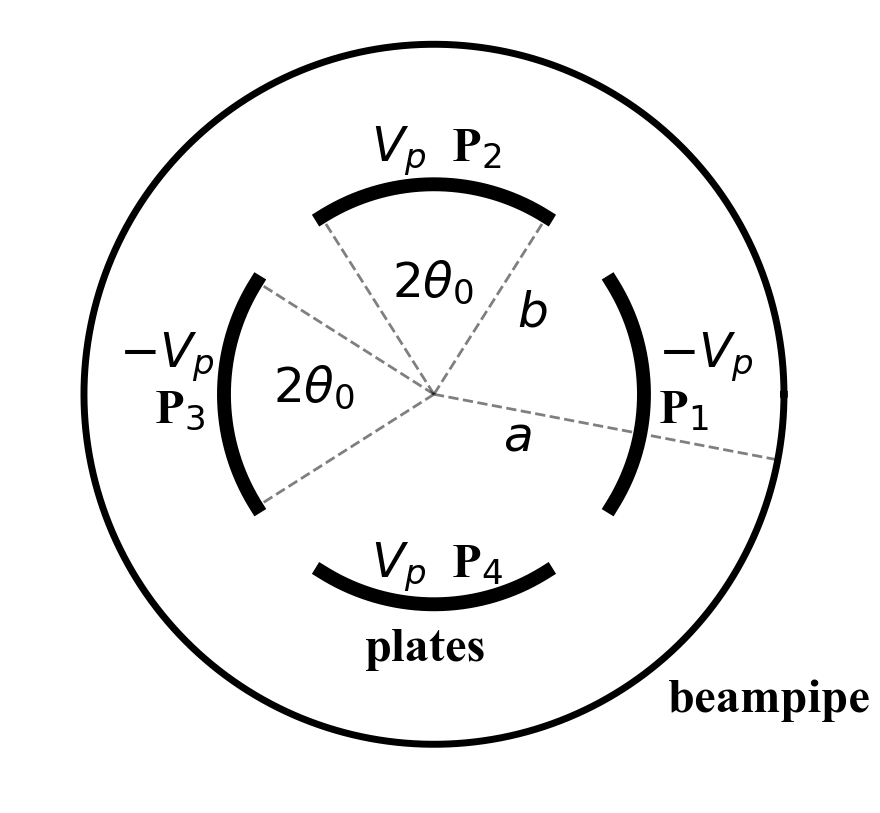}
\caption{Quadrupole kicker plates along the arc of a circle.}
\label{fig: quad_polar}
\efig
In this configuration, the potentials on the plates alternate in sign on adjacent plates with $\Phi(r = b, \theta) = -V_p$
on plate $P_1$ on the right which extends over the angles $ - \theta_0 \le \theta \le \theta_0$,
$\Phi(r = b, \theta) = V_p$ over plate $P_2: \pi/2 - \theta_0 \le \pi/2 + \theta_0$ etc. 
The gaps $G_1, G_2, G_3, G_4$ extend over the angles not covered by the plates, e.g.
$G_1: \theta_0 \le \theta \le  \pi/2 - \theta_0$, $G_2: \pi/2 + \theta_0 \le \theta \le  \pi - \theta_0$ etc. 
The general expressions for the potentials $\Phi_{in}(r,\theta)$ interior and $\Phi_{ex}(r,\theta)$  external to the plates
are respectively,
\beqr
\Phi_{in}(0 \le r \le b, \theta) & = & c_0  +  \sum_{m=1} r^m (c_m \cos m \theta + d_m \sin m\theta)  \\
\Phi_{ex}(b\le r \le a, \theta) & = & B_0 + A_0 \ln r + \sum_{m= 1} [A_m r^m + B_m r^{-m}][C_m \cos m\theta + D_m \sin m\theta]
\eeqr
Symmetry about the x-axis, $\Phi_{in}(b,2\pi - \theta) = \Phi_{in}(b, \theta)$ implies that $ d_m = 0$. 
Symmetry about the y-axis, $\Phi_{in}(b,\pi - \theta) = \Phi_{in}(b, \theta) $  implies that $ m = $ even. 
Since the plates are symmetric,   the potential is
anti-symmetric about the lines midway between adjacent plates (at opposite voltages) along $\theta = \pi/4, 3\pi/4$.
These anti-symmetries can be written as
 $  \Phi_{in}(b, \pi/2 - \theta) = -\Phi_{in}(b, \theta), \;\;\;    \Phi_{in}(b, 3\pi/2 - \theta) = -\Phi_{in}(b, \theta) $.
  These imply $c_0 = 0$ and
$ \cos m\pi/2 = -1, \;\;\; \Rarw m = 2, 6, 10, \dots = 2(2n - 1)$.
Applying the above symmetries and all the matching conditions, the potentials are
\begin{align}
\Phi_{in}(r, \theta) & =  V_p \sum_{m=2, 6, 10,...} X_m (\fr{r}{b})^m \cos m\theta  ,  \;\;\;\;\;\;  & 0 \le r \le b   \label{eq: Phi_in_quad} \\
 \Phi_{ex}(r, \theta) & =  V_p \sum_{m=2, 6, 10, ...} \fr{X_m }{[ 1 - (a/b)^{2m}]} [(\fr{r}{b})^m - (\fr{a^2}{b r})^{m}] \cos m\theta , \;\;\;   &  b \le r \le a  
 \label{eq: phi_ex_quad}
 \end{align}
The  boundary conditions satisfied by these coefficients have the same form as those for the dipole odd mode in
Eqs. (\ref{eq: dip_odd_BC1}) and (\ref{eq: dip_odd_BC2}) except that $m = 2, 6, \ldots$. 
Integrating these equations over any plate  and any gap leads to the two integral conditions
\beqr
\sum_{m=2, 6, 10,...} X_m \fr{\sin m\theta_0}{m\theta_0}  =  - 1   \label{eq: cm_1_quad} \\
\sum_{m=2, 6, ...} g_m(a, b) X_m [(-1)^{(m-2)/4} - \sin m\theta_0] & = & 0  \label{eq: cm_2_quad}
\eeqr
In Eq. (\ref{eq: cm_2_quad}) the first term in square brackets results in alternating signs for successive values of $m$, hence this
condition has also the same form as Eq.(\ref{Int_Xm_polar_ext})  except for the different values of $m$.

We now write down the matrix equations for the two methods discussed previously. 
The error function to minimize in the least squares method has the same form as in Eq.(\ref{eq: dip_err_func}), except that
the second integral (over the gap) runs from $\theta_0$ to $\pi/2 - \theta_0$ and the index $m$ runs over the values 2, 6, 10, \ldots. 
Minimizing  the error function yields the matrix equation ${\bf \bar{B}} \cdot {\bf X} = - \bf{b}$ where the off diagonal
elements of the matrix ${\bf \bar{B}}$ are the same as for the matrix ${\bf  B}$ for the odd mode in the dipole case, while
the diagonal elements are
\beqr
  \bar{B}_{n n} & = &  A_{n, n} +  n^2  g_n(a, b)^2 \left(\fr{\pi}{4} - A_{n, n} \right) , \;\;\; n= 2, 6, 10, \ldots 
  \eeqr
The $\pi/2$ term in $B_{n n}$ (see Eq.(\ref{eq: dip_Bnn})) is replaced by $\pi/4$ in $ \bar{B}_{n n}$ and the indices have different values.

To apply the projection method, multiplying the first boundary condition
by $\cos n\theta$ and integrating over a plate  leads to
exactly the same as Eq.(\ref{eq: polar_Xm_int}) for the odd mode in the dipole kicker, except for the values of the
index $m$.
 Multiplying the second boundary condition by $\cos n\theta$ and integrating over a gap, we have
 \beq
  n g_n X_n [\fr{\pi}{4} -  A_{n, n}]  -  \sum_{m=2, 6, ...} m  g_m(a, b) X_m A_{m, n}  = 0   \label{eq: quad_IC2}
  \eeq
  Hence the matrix equation is ${\bf \bar{C}} \cdot {\bf X} = - \bf{b}$ where ${\bf \bar{C}}$ is similarly
  related to the matrix ${\bf C}$ for the dipole odd mode defined in Eq. (\ref{eq: dip_Cnn}) and (\ref{eq: dip_Cmn})
  as the matrices ${\bf \bar{B}}$ and ${\bf B}$ above.

\subsection{Potential for the sum mode}
  In this mode, all plates are at the same voltage. Now, we have symmetry about
  the axes at $\pm 45^{\circ}$ in addition to the symmetries about the $(x, y)$ axes. As with the quadrupole mode, the
  symmetries about the $x, y$ axes lead to $ d_m = 0, \;\;\;\;\;  m={\rm even}$.
  The symmetries about the other axes along the $\pm 45^{\circ}$ angles imply  $ m = 4, 8, ...$
  Hence, the potential in the interior and exterior can be written as
   \beqr
   \Phi_{in}(r, \theta) & =  V_b[ X_0 + \sum_{m= 4, 8, ...} X_m (\fr{r}{b})^m \cos m\theta]  , & \;\;\; 0 \le r \le b \non  \\
\mbox{} \\
   \Phi_{ex}(r, \theta) & =  V_b \left[ X_0 \fr{\ln(r/a)}{\ln(b/a)} + \sum_{m= 4, 8, ...} \fr{ X_m}{[1 - (\fr{a}{b})^{2m}] }
   [  (\fr{r}{b} )^m - (\fr{a^2}{b r})^m] \cos m\theta \right], \;\;\;  & b \le r \le a  \non \\
   \eeqr
   These are of the same form as for the even mode in the dipole kicker, except  for for the indices.
   Hence the boundary conditions are the same as in Eqs. (\ref{eq: dip_even_BC1}) and (\ref{eq: dip_even_BC2})
   and the integral conditions are the same as in Eqs. (\ref{eq: dip_global_even_BC1}) and (\ref{eq: dip_global_even_BC2})
   except for the replacement $\pi/2$ by $\pi/4$ in the latter equation and  the index $m=4, 8, \ldots$. 
   Using the projection method, the matrix equation is ${\bf \bar{D}} \cdot {\bf X} =  \bf{b}$ where ${\bf \bar{D}}$ is
   similarly related to the ${\bf D}$ matrix defined in Eqs.(\ref{eq: dip_Dnn}) and (\ref{eq: dip_Dmn}) as
    ${\bf \bar{B}}$ is related to ${\bf B}$. When the coefficients $X_m, m \ge 4$ are found from this matrix equation, 
      $X_0$ can be found  subsequently  by using either of the integral conditions.

\subsection{Electric and Magnetic fields, Characteristic Impedance}

We consider first the fields in the quadrupole mode. The electric fields in Cartesian coordinates are
 \beqr
 E_x  \!\! & \!\! = \!\!& \!\!\! - \fr{V_p}{b}\!\!\! \sum_{m=2, 6, \ldots}\!\!\!\!\!\!\!\!\! m X_m  (\fr{r}{b})^{m-1} \cos (m-1)\theta \!\! = \!\!\!  -  \fr{V_p}{b}[2 X_2 \fr{x}{b} +
   6 X_6 (\fr{r}{b})^5 \cos 5\theta + \ldots ] \\
 E_y    & = &   \fr{V_p}{b} \sum_{m=2, 6, \ldots} \!\!\!\!\!\! m X_m (\fr{r}{b})^{m-1}\sin (m-1)\theta   =    \fr{V_p}{b} [2 X_2\fr{y}{b} +
   2 X_6 (\fr{r}{b})^5 \sin 5\theta + \ldots ]  \non \\
 \eeqr
 Keeping only the first term gives us the quadrupole fields
 \beq
 E_{x, quad}  = - 2\fr{V_p X_2}{b^2} x,  \;\;\;\;   E_{y, quad} =  2\fr{V_p X_2}{b^2}  y
 \eeq
Using the expressions for the forces derived above in Eq.(\ref{eq: Fx}), we have for the quadrupole  kicks from a kicker of length $L_k$
  \beqr
  \Dl x' & = &  - 2 \fr{q(1 + \bt)}{\bt^2 m_0c^2 \gm} \fr{V_p X_2 L_k}{b^2} x  , \;\;\;\; 
  \Dl y'   =  2 \fr{q(1 + \bt)}{\bt^2 m_0c^2 \gm} \fr{V_p X_2 L_k}{b^2} y
  \eeqr
  Hence the integrated quadrupole gradient or inverse focal length defined from $\Dl x' = - K_q x$ is
  \beq
  K_q \equiv \fr{1}{f_q} = - \fr{q(1 + \bt)}{\bt^2 \gm m_0c^2 } \fr{2 V_p X_2 L_k}{b^2} = 
\fr{q(1 + \bt) }{\bt^2  \gm m_0c^2 } \fr{\del E_x}{\del x}_{| x = 0} L_k
\eeq
We now estimate the quadrupole kick for the 2 MeV IOTA proton beam referred to in Section \ref{sec: Dipole}. 
We assume a plate voltage $V_p = 1$ kV, the kicker length to be $L_k = 0.2$ m, $b/a=0.8$ and
$\theta_0 = 30^{\circ}$. We use Eq.(\ref{eq: X2q_2}) in Appendix A and the correction of 17\% at these values of
$b/a, \theta_0$ from Fig. \ref{fig: X2sol-error} to estimate 
$X_2 = -1.23$. This yields $K_q = 0.26$ m$^{-1}$, or the dimensionless quadrupole strength  $q= \bt_x K_q = 0.32$.
According to the theory of nonlinear echoes \cite{Sen_18}, this value of the quadrupole kicker strength $q$ will
suffice for nonlinear effects of the quadrupoles on the echoes to be observable. 

  The characteristic impedance can be calculated similarly as for the dipole kicker. We have for the quadrupole mode,
\beqr
Z_{c,quad} & = &  \fr{Z_0}{4 \lvert \sum_{m=2, 6, ...}  X_{m} g_m(a, b) \sin m\theta_0 \rvert }  = \fr{Z_0}{4 \lvert \sum_{m=2, 6, ...}  (-1)^{(m-2)/4}
  X_{m} g_m(a, b)  \rvert} \non \\
\eeqr
For the sum mode, the characteristic impedance is
\beq
Z_{c, sum} = 2 \fr{Z_0}{\pi} \fr{\ln(a/b)}{ \lvert X_0 \rvert}
\eeq
As discussed earlier,  the geometric mean of the quadrupole mode impedance and
the sum mode impedance is matched to that of the external load $Z_L$, or
\beq
Z_{c, geom} \equiv \fr{Z_0}{\sqrt{2\pi}} [\ln(a/b) \fr{1}{ \lvert X_0 } \fr{1}{\sum_{m=2, 6, ...}  (-1)^{(m-2)/4}  X_{m} g_m(a, b)  \rvert}]^{1/2}
  = Z_L    \label{eq: quad_Zcgeom}
  \eeq
This seems to be the adopted matching scheme in some quadrupole kicker designs \cite{Iriso_09, Wu_12}.

\section{Numerical solutions and comparisons with FEMM}  \label{sec: Numerics}

In this section we compare the fields found using the two methods described in Sections \ref{subsec: least_sq},
\ref{subsec: project} with  those obtained from the 2D electrostatic (and magnetostatic) code
FEMM  \cite{FEMM} which uses the finite element method.
 The domain of interest is subdivided into triangular regions; the program uses quadratic Lagrangian interpolation 
 over these regions and solves for the potential at the nodal locations.

The  Fourier series converges poorly near the tips of the plates due to discontinuities in the 
 derivative of the potential (Gibbs phenomena). A Gaussian filter is applied to improve convergence by smoothing them out.
After filtering, the series $\sum_m X_m \cos m \theta$ becomes $\sum_m \exp[-m^2/(2 \sg^2)]X_m \cos m\theta $
 where $\sg$ is the smoothing parameter (we used $\sg= 100/(2\pi)$). 
For both methods, we found that after 100 terms in the
matrix equations the solutions change by less than 1\% with the addition of more terms. The convergence rate with
both methods is about the same, the rate with the projection method is marginally faster. We used 200 terms 
 in the results discussed here to ensure that the convergence errors are at less than the 1\% level. 

\subsection{Potential and fields in a dipole kicker}

In this section we first compare the potential obtained with two series expansions to the FEMM results.
 Next we discuss some insights they provide on the influence of the geometrical parameters. 
\bfig
\centering
\includegraphics[scale=0.25]{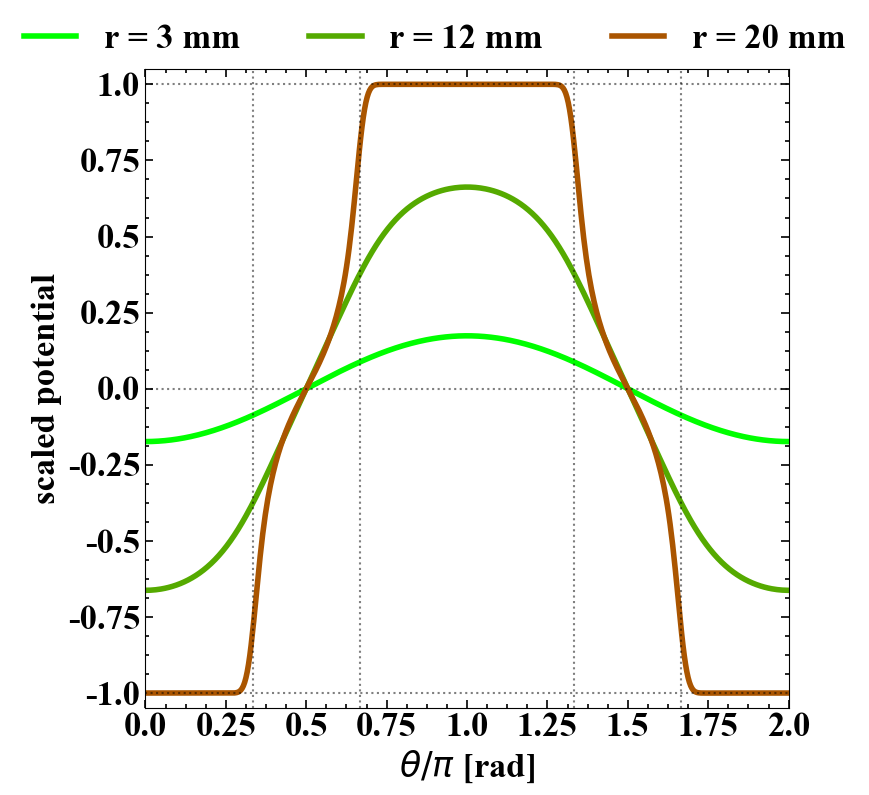} 
\includegraphics[scale=0.25]{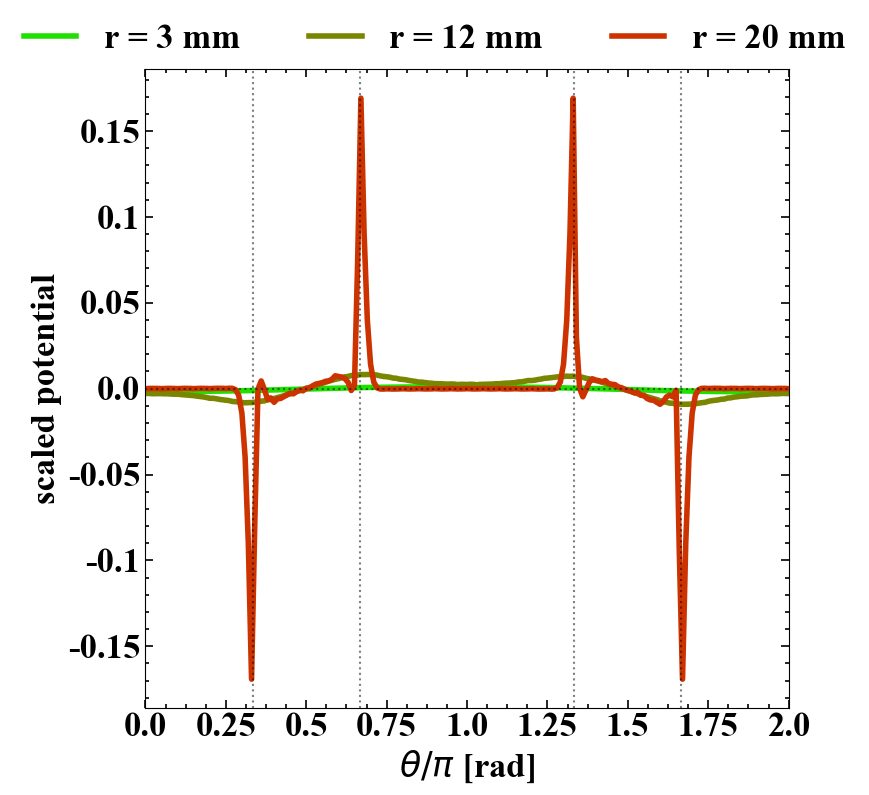}
\includegraphics[scale=0.25]{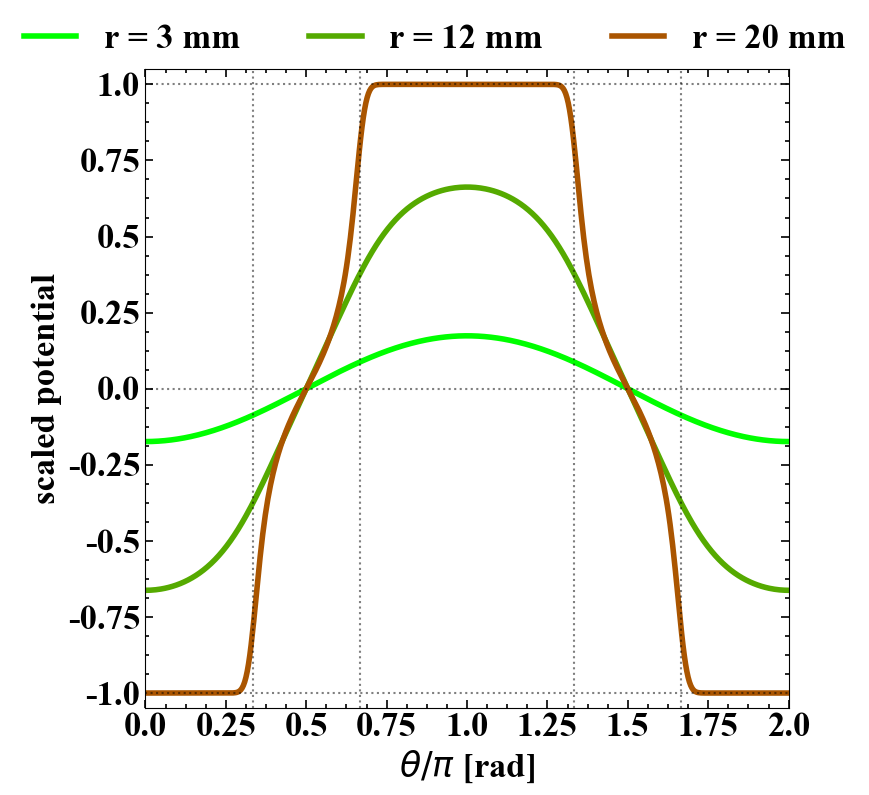}
\includegraphics[scale=0.25]{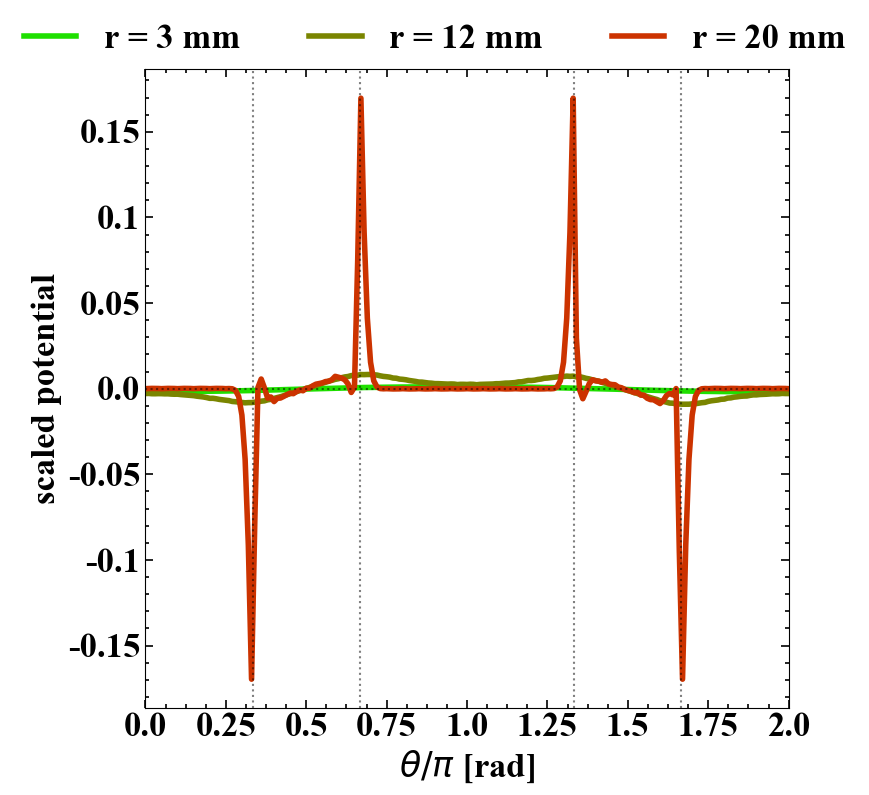}
\caption{Top Left: The scaled potential $\Phi/V_p$ using the least square method as a function of the angle $\theta$ for
  different values of the radial distance $r$ for plates with $b=20$ mm, $a=25$ mm,  $\theta_0= \pi/3$. 
 Top Right: The absolute relative difference between the solutions found using the least squares method 
  and the program FEMM. The difference is well below 5\% everywhere except at the
  tips of the plates.
  Bottom Left and Right: Same as in the top row, except that the projection method is used. The
  two series expansions give nearly the same result.}
\label{fig: dip_pot_anal_FEMM_033}
\efig
Figure \ref{fig: dip_pot_anal_FEMM_033} compares the potential solutions obtained using the
two series expansions with the FEMM results. The potential (scaled by $V_p$) is shown as  a function of $\theta$
 for three values of $r\le b$ with the kicker's geometric parameters fixed at $a= 25$ mm, $b = 20$ mm and
$\theta_0=\pi/3$. Both series solutions  yield nearly identical results. 
At the plate radius  $r=b$, both methods lead to the
scaled potential equal to -1 and +1 on the right  and the left plates respectively. The  plots on the right in
Fig.\ref{fig: dip_pot_anal_FEMM_033} show the absolute difference with FEMM results: $\le 5\%$ everywhere 
except near the tips of the plates where it is $\sim$20\%.  This is likely due to a combination of two factors:
(1) the filtering  applied to the analytic solutions  increases the difference by $\sim 5\%$ at the tips and (2) 
 the FEMM polynomial basis functions can not model 
 the singular variation of the potential in the vicinity of a sharp edge without resorting to an extremely dense mesh. 
 In the central region of the beampipe within 
 the area occupied by the beam,  the differences are negligible. We verified that the above difference bounds are valid for
 values of $\theta_0$ in the range $0.2\pi \le \theta_0 \le 0.45\pi$; this should cover most cases of practical interest.

It is instructive to consider the behavior of the coefficients $X_m$ in the Fourier expansion of the potential.
Fig. \ref{fig: dip_odd_X1_X3} shows the dependence of the two lowest order coefficients (in the odd mode) on the
geometric parameters $\theta_0$ and $b/a$. We observe that $X_1$ is always negative and its magnitude increases
monotonically with $\theta_0$. When $X_1 = -1$, all the other coefficients sum to zero and the potential is nearly
linear in the $x$ coordinate, resulting in a uniform horizontal electric field. This is true for different values of $b/a$.
These values will need to be checked against the
requirement of matching the characteristic impedance, to be discussed later. The top right plot in
Fig. \ref{fig: dip_odd_X1_X3} shows that $X_1$ is mostly constant for $b/a \le 0.6$ and then decreases in magnitude
for $b/a > 0.6$ where the change decreases as $\theta_0$ increases. For example, the decrease in $X_1$
is $< 5\%$ for $b/a > 0.6$. To a good approximation  $X_1$ is independent of $b/a$ for $\pi/3 \le \theta_0 < \pi/2$.
The bottom plot shows that $X_3$   is an oscillatory function of $\theta_0$,  crossing zero in the range
$0.25 \pi \le \theta_0 < 0.3 \pi$ for the different $b/a$.
The dependence on $b/a$ is similar to that of $X_1$. The behavior of higher order coefficients $X_m, m > 3$ is
similar to that of $X_3$, with their absolute values decreasing with order $m$. These lead to the expected conclusion
that the potential interior to the plates is mostly independent of the beampipe radius for $b/a \le 0.6$, the presence
of the beampipe perturbs this potential significantly only  for small coverage angles and as the distance between the plates 
 and the beampipe wall decreases.
\bfig
\centering
\includegraphics[scale=0.25]{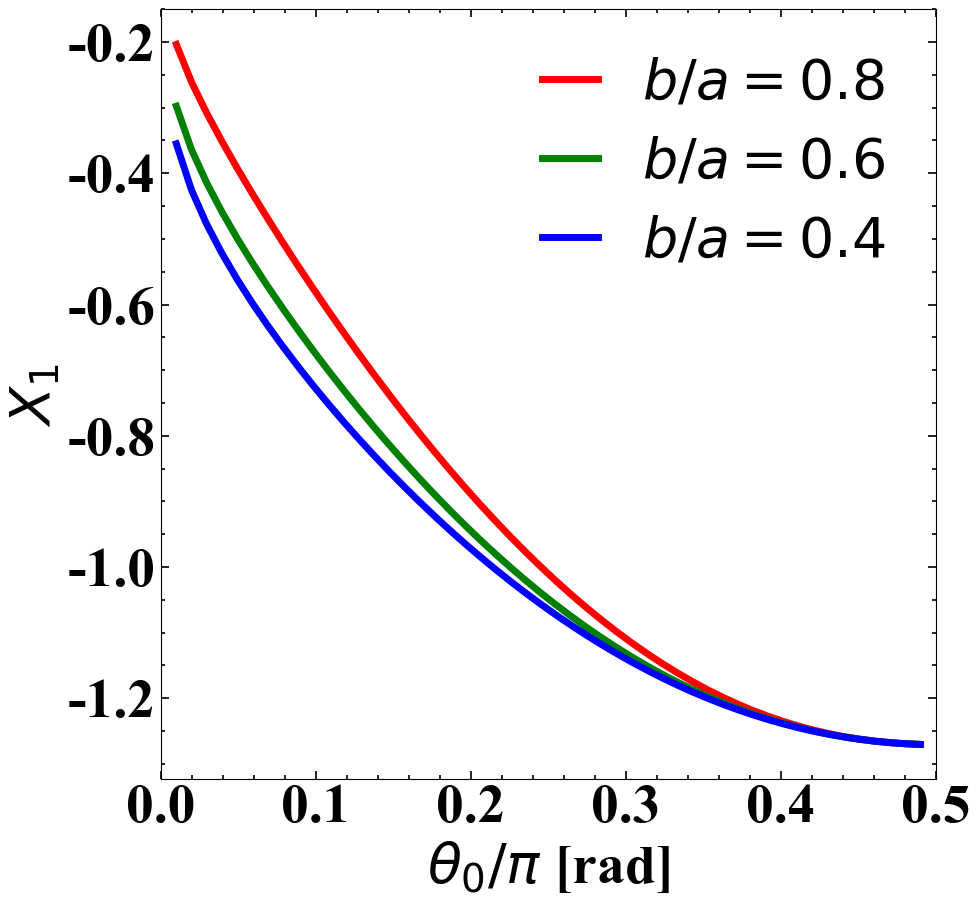} 
\includegraphics[scale=0.25]{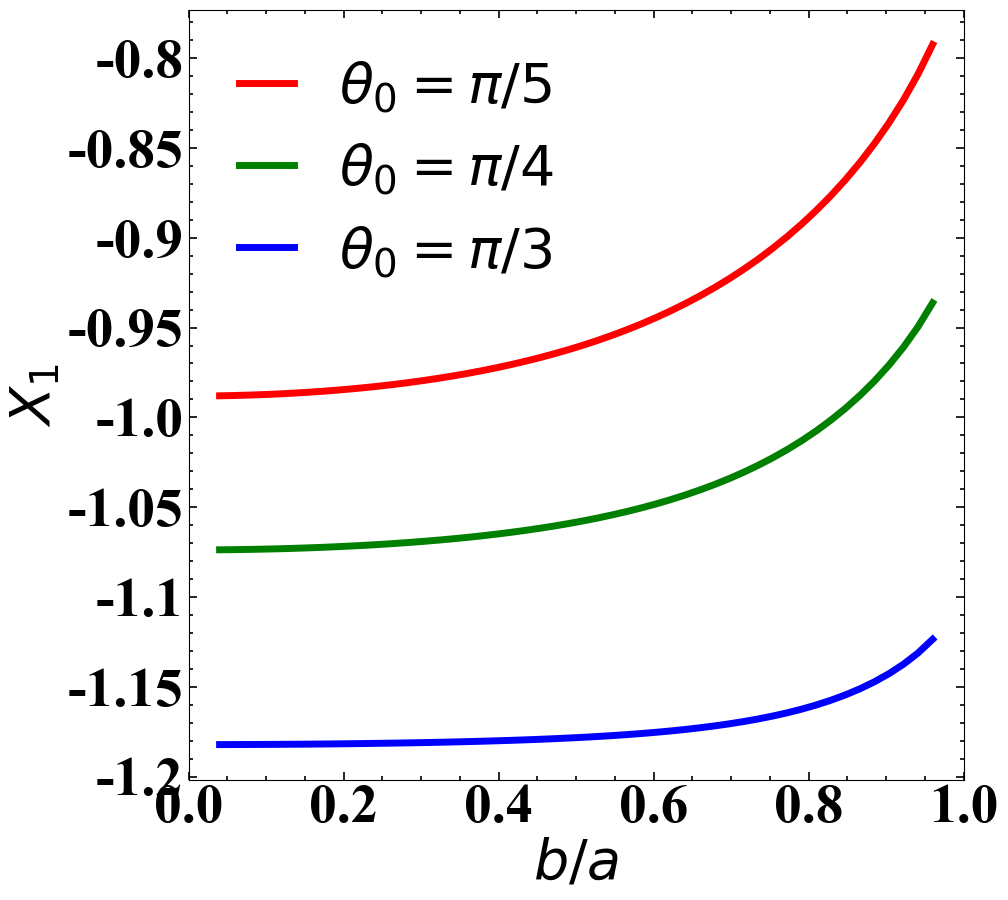} 
\includegraphics[scale=0.25]{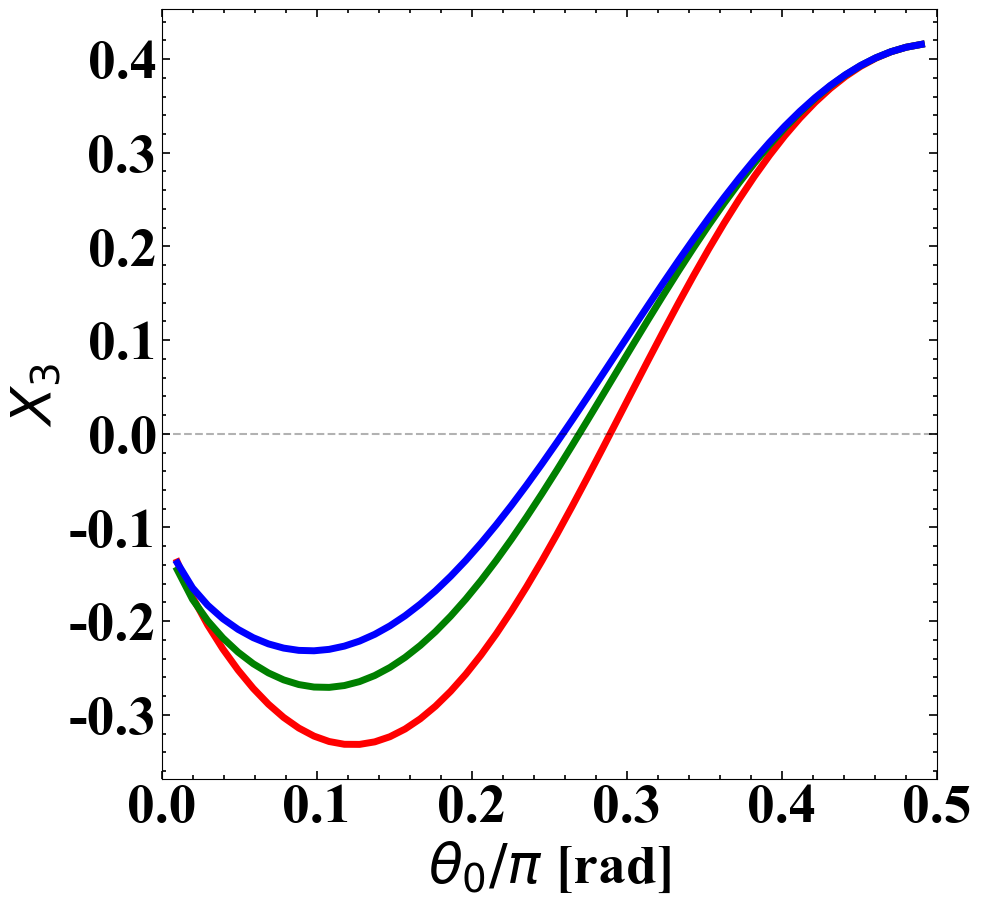}
\includegraphics[scale=0.25]{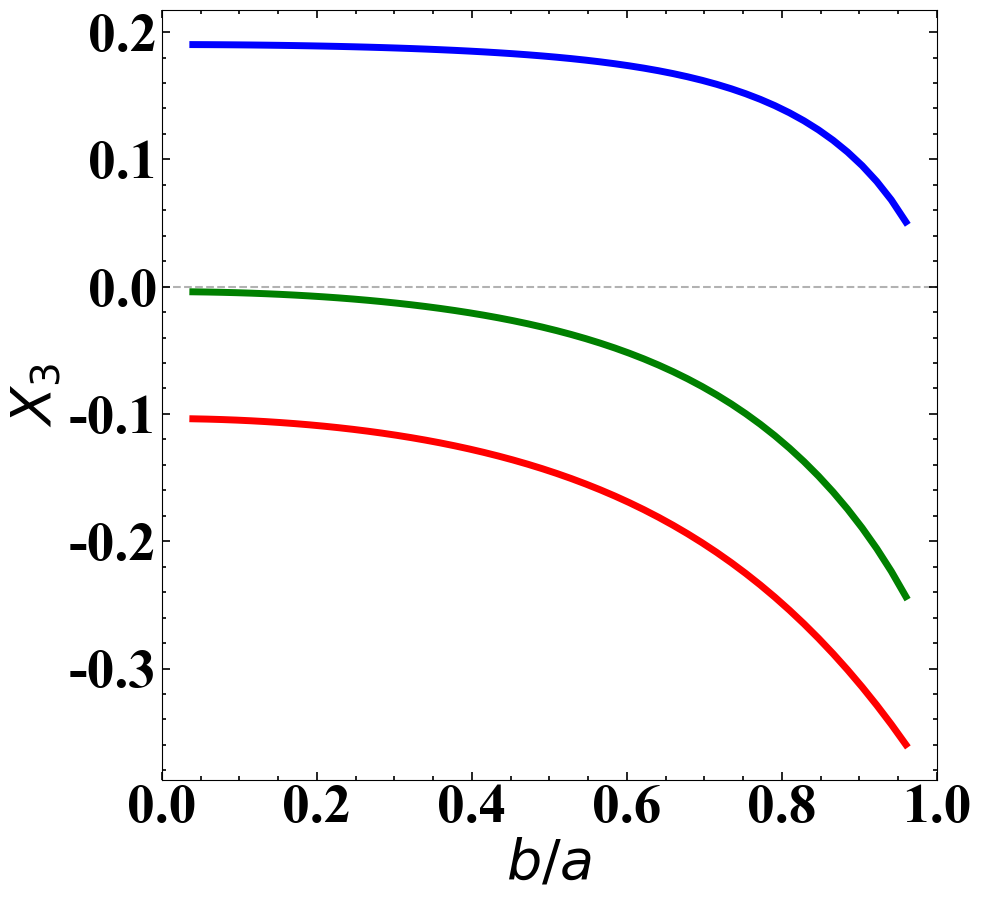} 
\caption{Odd mode:  Variation of the two lowest order coefficients with half coverage angle $\theta_0$ for different values of $b$ (left) and  with $b/a$ (right) for different values of $\theta_0$.  Top row: $X_1$, Bottom row: $X_3$. }
\label{fig: dip_odd_X1_X3}
\efig
\bfig
\centering
\includegraphics[scale=0.2]{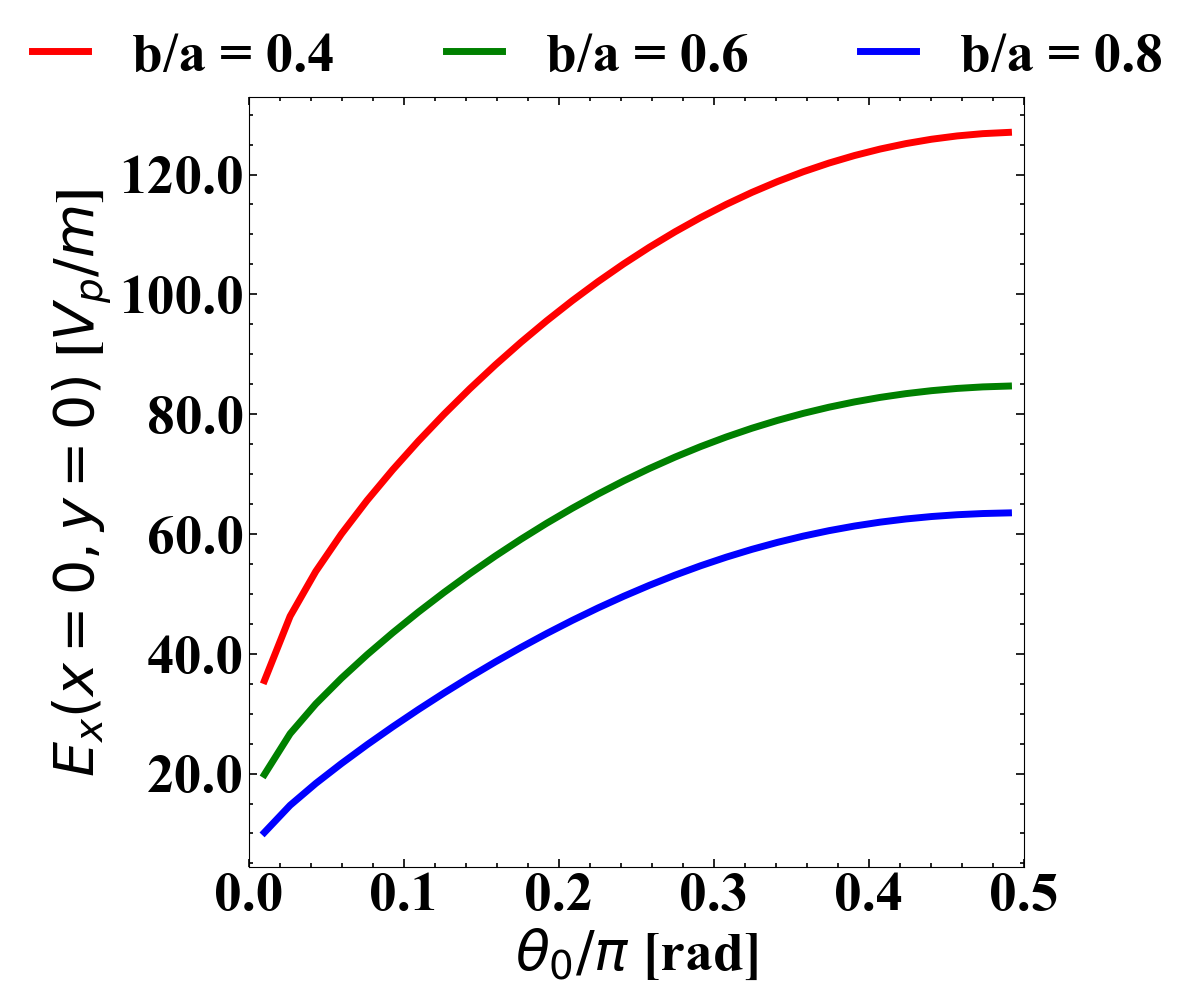}  
\includegraphics[scale=0.2]{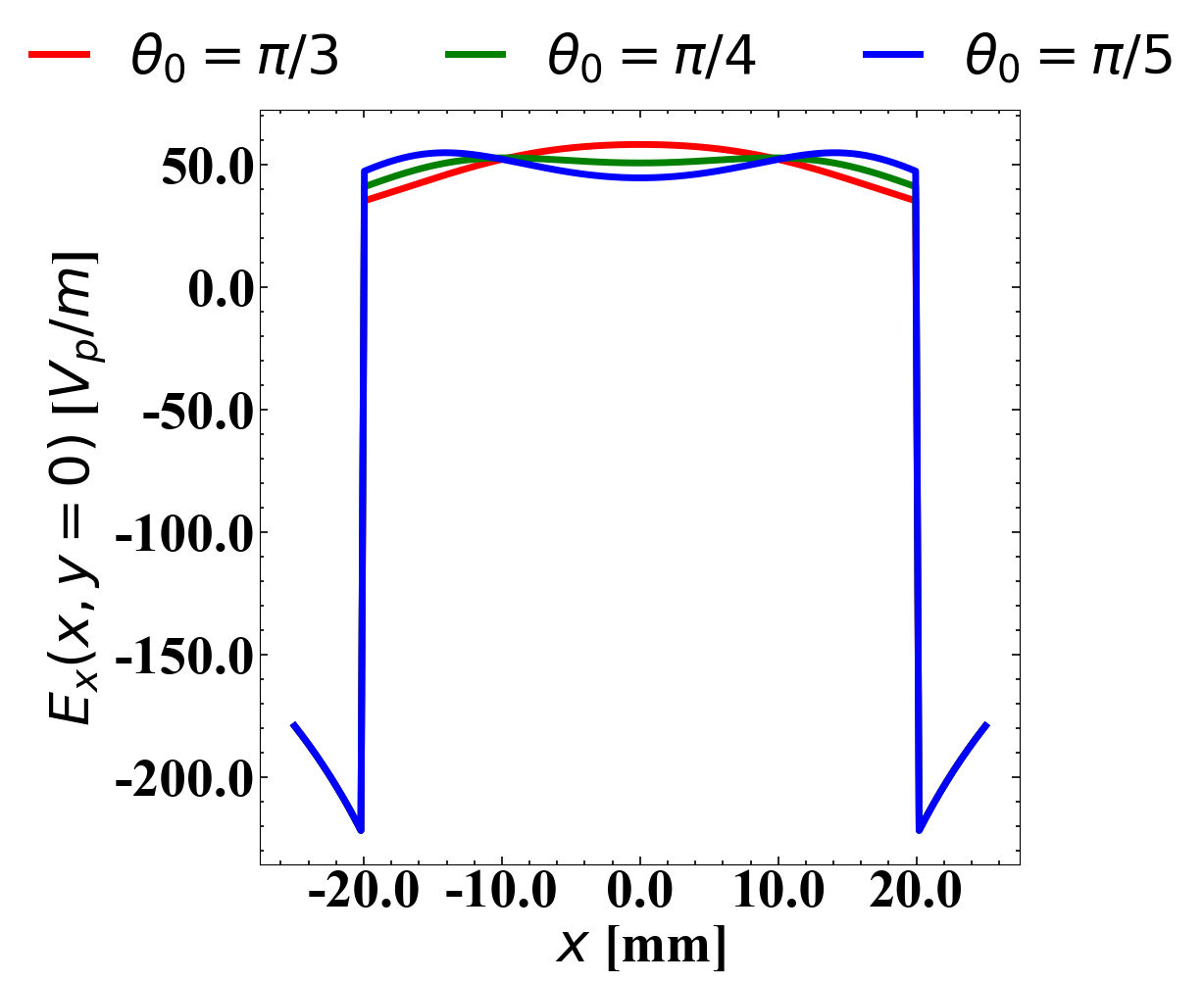} 
\caption{Left:  Horizontal electric field $E_x(0, 0)$ at the center of the beam pipe as a function of $\theta_0$ for different
  values of  $b/a$.  
  Right: Horizontal electric field $E_x(x, 0)$ along the $x$ axis between the plates at $x= \pm 20$mm with $a= 25$ mm
  for different values of $\theta_0$.  }
\label{fig: dip_Ex}
\efig
The left plot in Fig. \ref{fig: dip_Ex} shows the horizontal electric field  $E_x(0,0)$ at the origin as a function of $\theta_0$
for different values of $b/a$. The important observation here is that while $E_x(0,0)$ initially increases with $\theta_0$, it
eventually saturates around $\theta_0 \simeq 0.4\pi$, so further increase in the coverage angle does not increase the
electric field by much. The right plot in Fig. \ref{fig: dip_Ex} shows the horizontal electric field $E_x(x, 0)$
along the horizontal axis out to the plates at $x = \pm 20$ mm with $a= 25$ mm. Of the three profiles for different
$\theta_0$,
the flattest profile is obtained for $\theta_0 = 0.25\pi$, which is expected from the above discussion on the variation
of $X_1$ on $\theta_0$. 

For the even mode where the lowest order coefficients are $X_0, X_2, ...$, we observe qualitatively the same behavior, 
shown in Fig. \ref{fig: dip_even_X0_X2}, as for $X_1, X_3, ...$ in the odd mode.
 The difference is that $X_0$ is always positive, increases monotonically with
$\theta_0$ and reaches a maximum value of +1 as $\theta_0 \to \pi/2$. The higher order coefficients are again oscillatory
 functions of $\theta_0$. In comparison to the odd mode case, the coefficients depend more strongly on $b/a$.
\bfig
\centering
\includegraphics[scale=0.25]{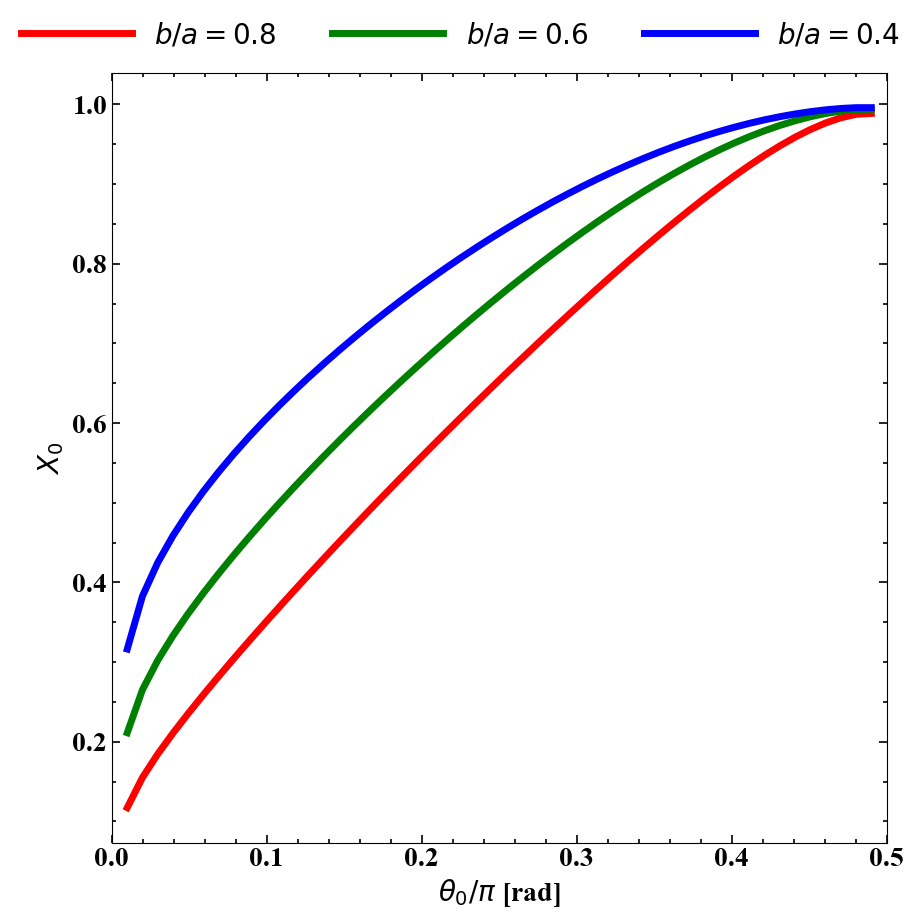}
\includegraphics[scale=0.25]{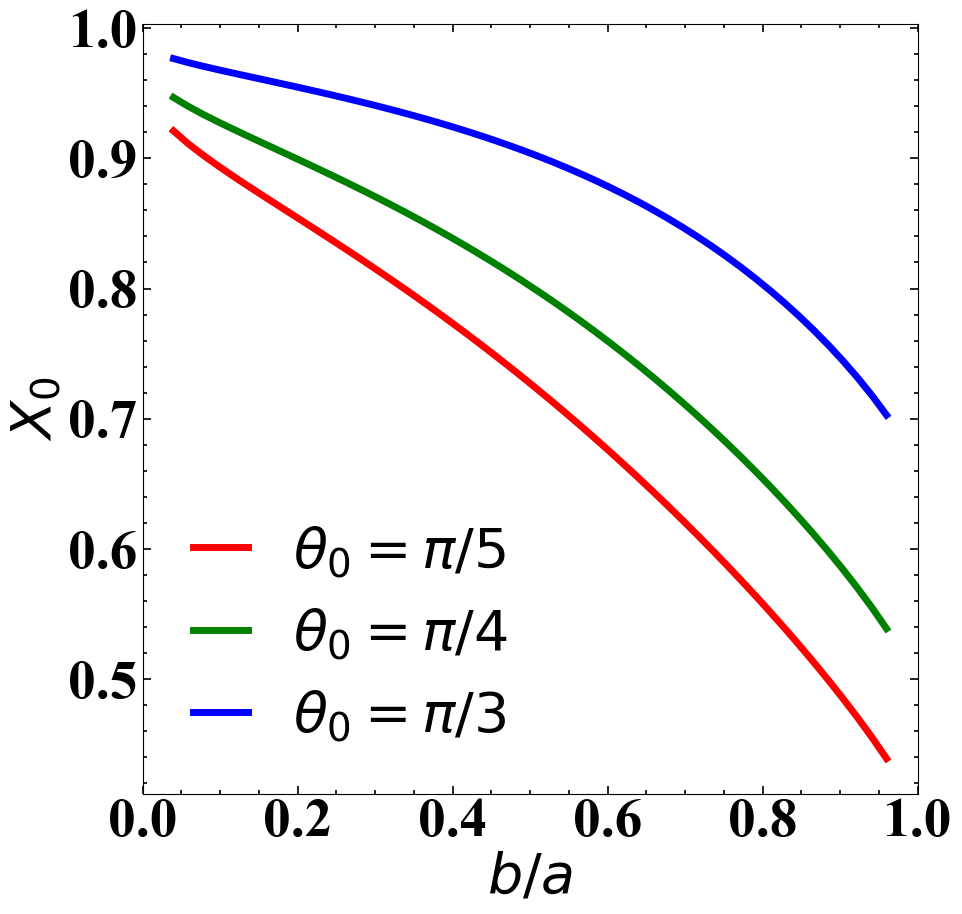}
\includegraphics[scale=0.25]{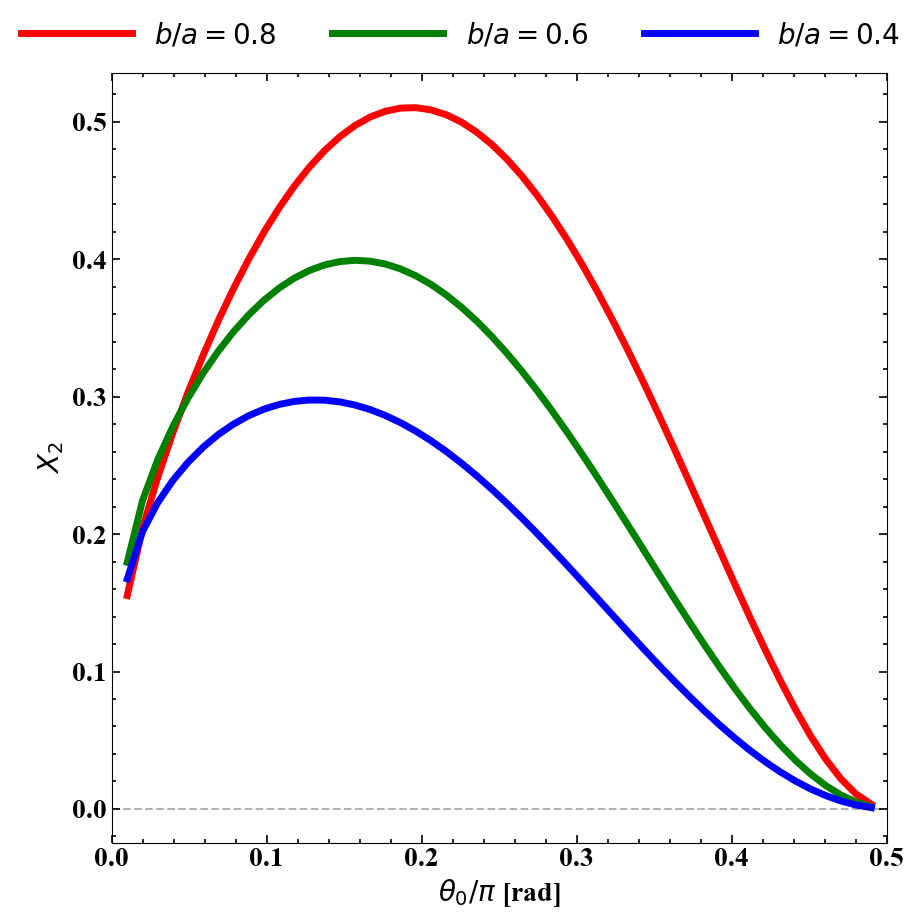}
\includegraphics[scale=0.25]{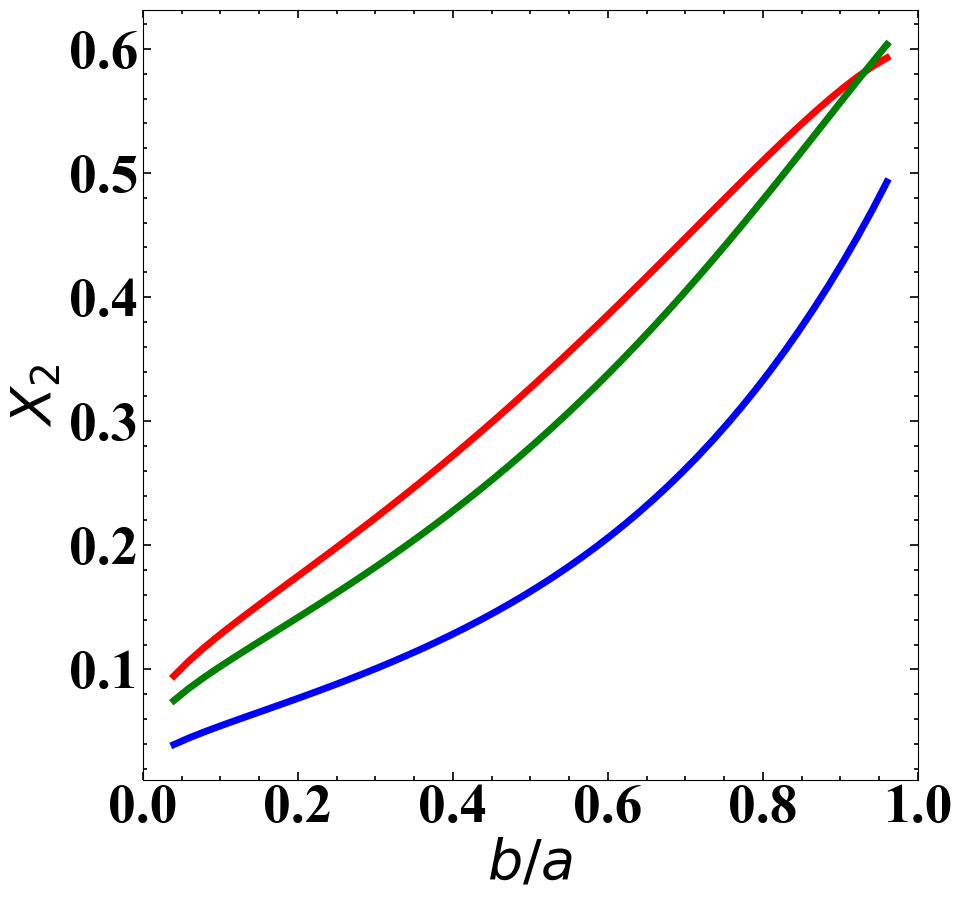}
\caption{Even mode:  Variation of the two lowest order coefficients with half coverage angle $\theta_0$ for different values of $b$ (left) and  with $b/a$ (right) for different values of $\theta_0$.  Top row: $X_0$, Bottom row: $X_2$. }
\label{fig: dip_even_X0_X2}
\efig

\bfig
\centering
\includegraphics[scale=0.5]{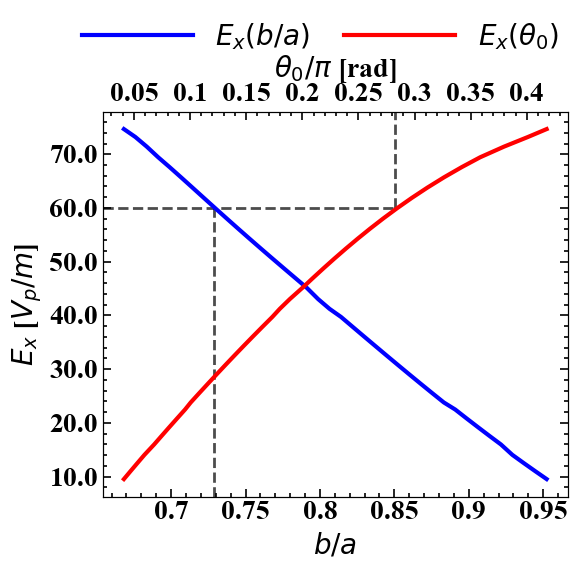}
\caption{The odd mode electrical field $E_x$ at the center as a function of $(b/a, \theta_0)$. The axes
  are labeled as follows; bottom: $b/a$, top: $\theta_0$, vertical: $E_x$ in units of $V_p/m$ where $V_p$ is the voltage
  on the plates. The beam pipe radius $a= 25$ mm. The even mode characteristic impedance   $Z_c = 50 \Om$ is constant along each  curve.
}
\label{fig: dip_Ex_b_theta}
\efig
We saw previously that the characteristic impedance in either mode is determined entirely by $(b/a, \theta_0)$. Assuming
that $Z_{c, even}$ is matched, we can determine the choice of parameters that result in the desired electric field.
Fig. \ref{fig: dip_Ex_b_theta} shows the horizontal electric field at the origin as a function of  $b/a$ and $\theta_0$ 
under the constraint that  $Z_{c, even} = 50 \Om$. 
The values of $(b/a, \theta_0)$  that yield a desired field $E_x$ are 
obtained by taking the intersections of the horizontal line at this $E_x$ with the two curves and reading
  the corresponding values. As an example, the dashed lines show that for a desired $E_x= 60 V_p$/m
  of the field in the odd mode  requires $b/a \simeq 0.73, \theta_0 \simeq 0.28 \pi$ in a beampipe of radius 25 mm.
  For any other beampipe radius, say $a_2$, the field needs to be scaled by the ratio $25/a_2[{\rm mm}]$. 

\subsection{Potential and fields in a quadrupole kicker}

We now compare the quadrupole kicker solutions obtained  using the series expansions with FEMM. 
Once again the least squares method and the projection method give very close results; therefore, only the projection
method's  results will be discussed. 
\bfig
\centering
\includegraphics[scale=0.25]{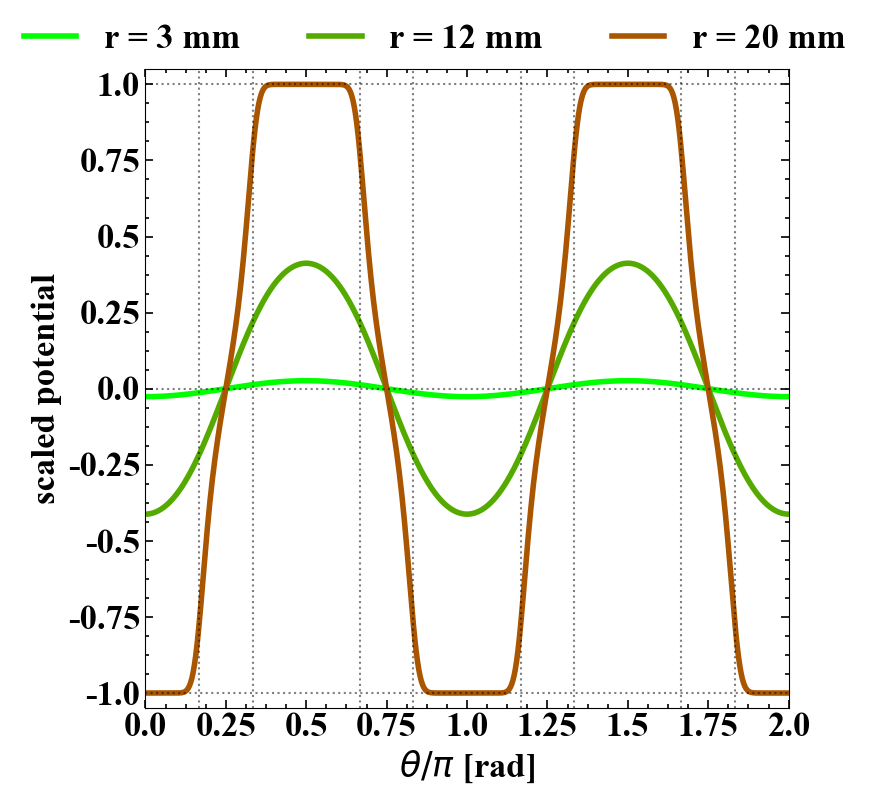}  
\includegraphics[scale=0.25]{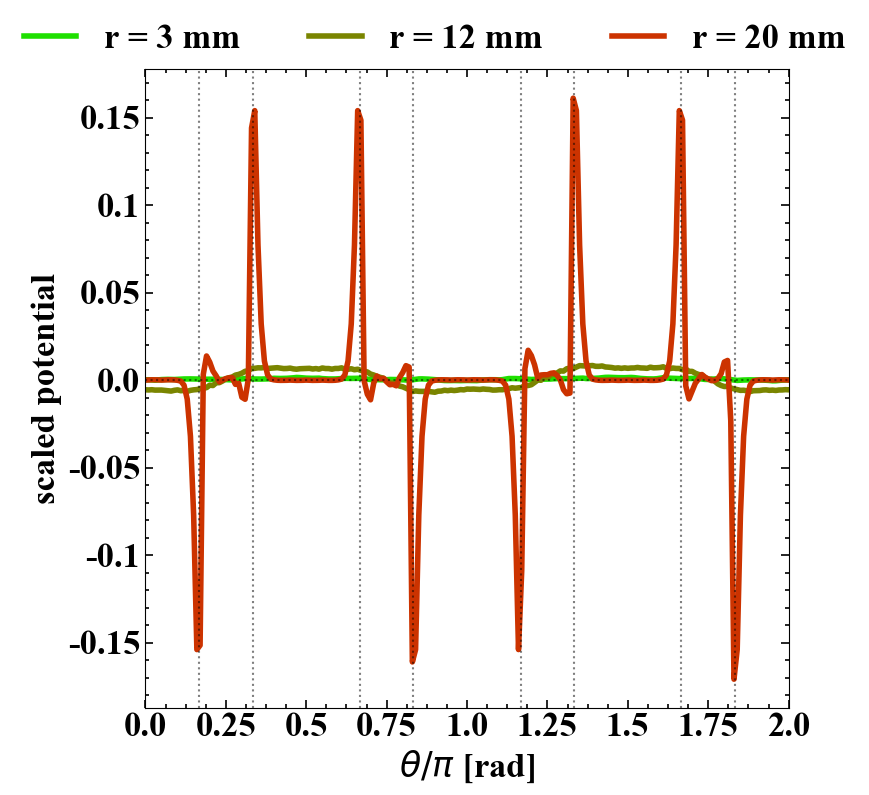} 
\caption{Left: Potential (using the projection method) in the quadrupole mode as a function of $\theta$ for 3 values of $r$, the
  largest $r= b$ with
  $\theta_0 = \pi/6$ and $a= 25$ mm, $b = 20$ mm. Note that for the quadrupole, $0 < \theta_0 < \pi/4$,  Right: Difference
  between the projection method and FEMM values for the potential. The difference reaches 15\% at the tips of the plates
but is less than 5\% everywhere else. }  
\label{fig: pot_quad}
\centering
\includegraphics[scale=0.225]{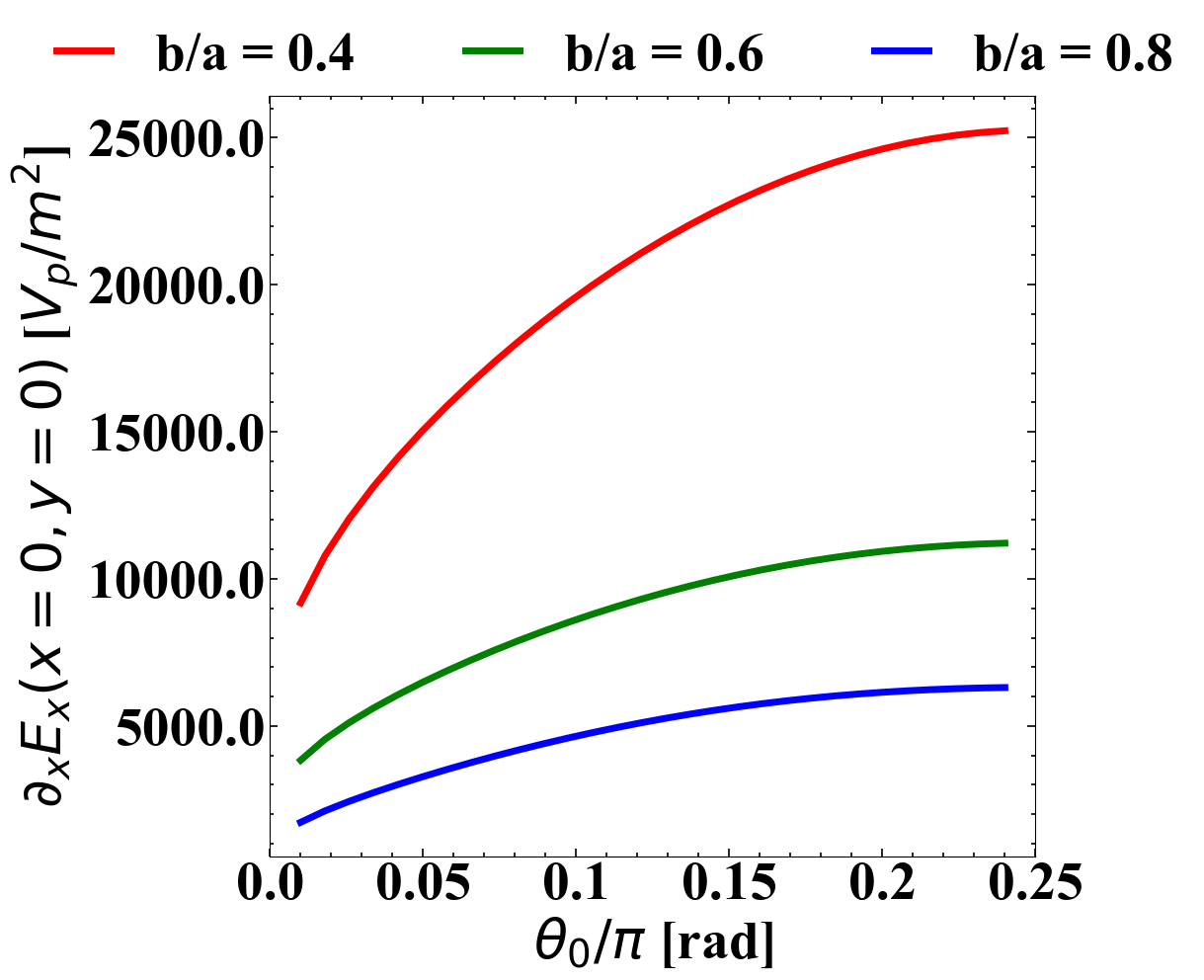} 
\includegraphics[scale=0.225]{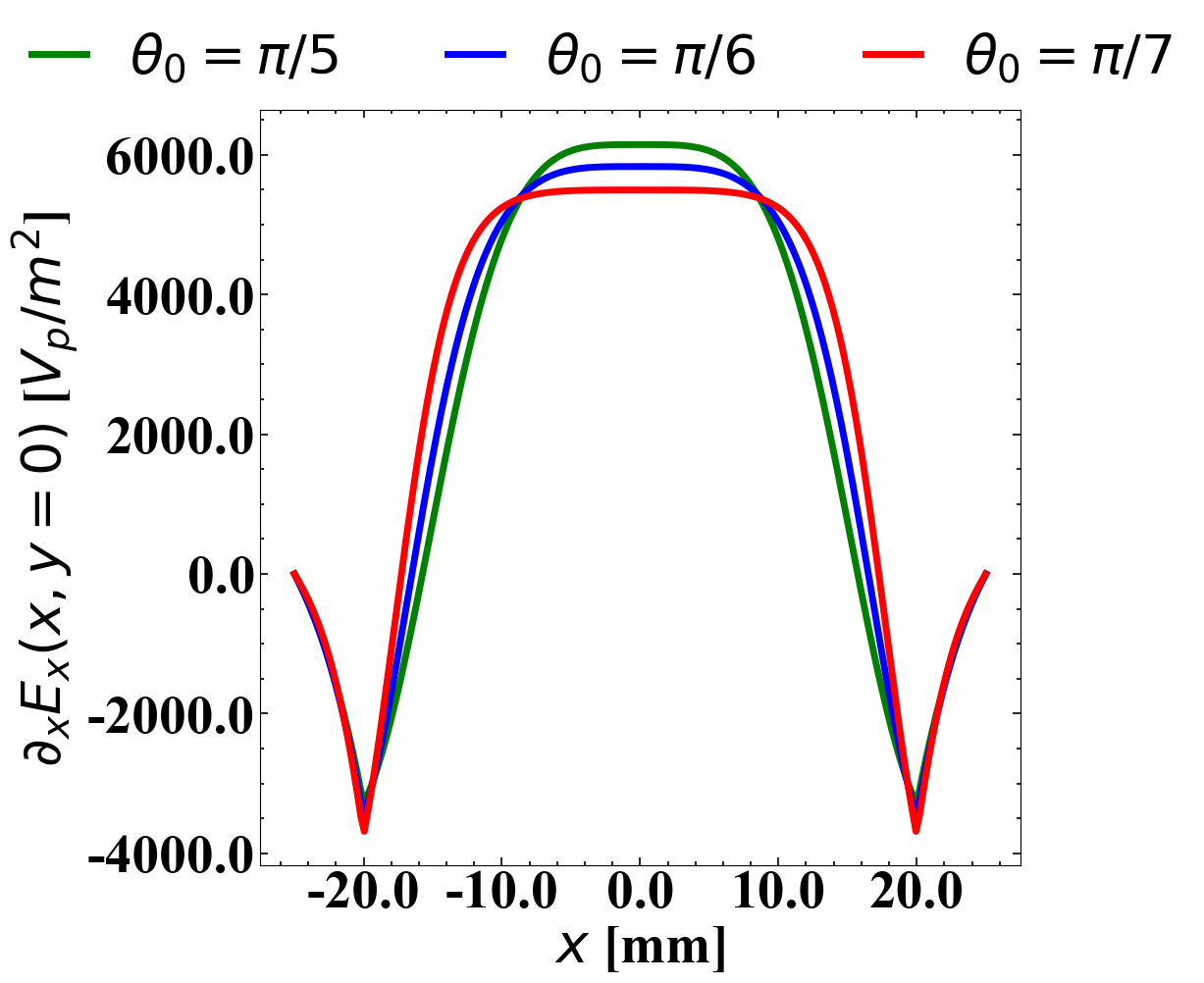}   
\caption{Scaled horizontal gradient of the electric field $\del E_x/\del x$ at the origin as a function of $\theta_0$ (left) and along
  the x-axis between the plates at $x = \pm 25 $ mm (right). Here $b/a = 0.8$  }
\label{fig: dEx_quad}
\efig
Fig. \ref{fig: pot_quad} shows as an example, the potential in the quadrupole mode calculated using the projection method
and its difference with the FEMM result.
The differences reach 15 - 20\%  at the tips of the plates, depending on the half coverage angle $\theta_0$,
but are less than 5\% everywhere else. The dependence of the coefficients $X_2, X_6, X_{10}, ....$ on $\theta_0, b/a$ in this
quadrupole mode mirrors the dependence of $X_1, X_3, X_5,...$ on these parameters in the dipole odd mode case. 
Similarly the coefficients $X_0, X_4, X_8, ...$ in the sum mode have a similar dependence on the same parameters
as do $X_0, X_2, X_4, ...$  in the dipole even mode case.
Fig. \ref{fig: dEx_quad} shows the gradient of the horizontal electric field scaled by the
potential in the quadrupole mode. The left plot shows the gradient at the origin while the right plot shows the gradient along
the horizontal axis.
The gradient increases with $\theta_0$ but the range over which the gradient stays constant
around the origin decreases with increasing $\theta_0$. 

  \bfig
\centering
\includegraphics[scale=0.5]{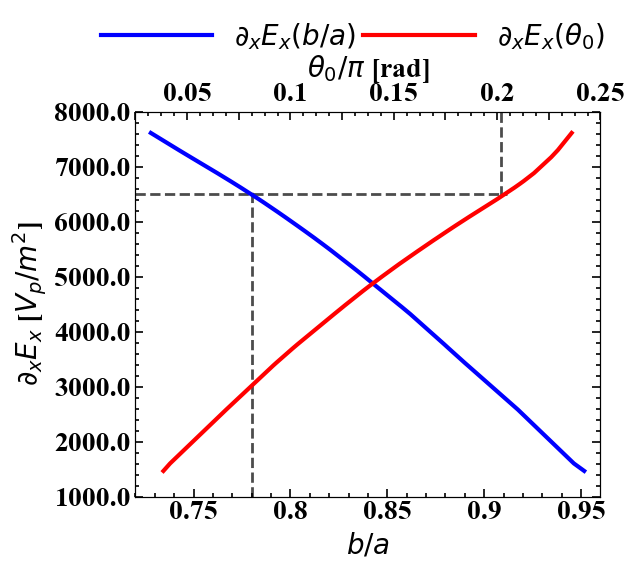}
\caption{The horizontal gradient of the electrical field $\del E_x/\del x$ in the quadrupole mode at the center as a function of
  $(b/a, \theta_0)$. The horizontal axes are labeled as in Fig.\ref{fig: dip_Ex_b_theta}, the vertical axes shows the gradient in
  units of  $V_p/m^2$ where $V_p$ is the voltage  on the plates. The beam pipe radius $a= 25$ mm. The geometric mean of the
  characteristic impedance   $Z_{c, geom} = 50 \Om$ is constant along each  curve.}
\label{fig: quad_gradx_b_theta}  
\efig
Fig. \ref{fig: quad_gradx_b_theta} shows the horizontal gradient of the horizontal electric field at the origin in the
quadrupole mode as functions of
the parameters $(b/a, \theta_0)$ with the constraint that the geometric mean of the characteristic impedances
$Z_{c, geom} = 50 \Om$.  The dotted lines show that an electric field gradient of 6500 $V_p/m^2$ requires
$b/a = 0.78$, and $\theta_0 \simeq 0.2 \pi$ in a beampipe of radius 25 mm.
For any other beampipe radius, say $a_2$, the field gradient needs to be scaled by the ratio $(25/a_2[{\rm mm}])^2$.


\subsection{Characteristic impedance}

The characteristic impedance of a mode can be found from FEMM using  the  relationship between $Z_c$ and
the capacitance per unit length $C'$, namely
\beq
Z_c = \fr{1}{c C'}  \label{eq: Zc_capacity}
\eeq
which was written  down in Section \ref{subsec: Zc_dip}. This assumes, as mentioned earlier in Section \ref{subsec: Zc_dip},
that the characteristic impedance is independent of frequency and is valid at low frequencies. 
FEMM is used to calculate the characteristic impedance of a
single mode at a time for which this relation is easily derived. 
in a TEM mode, the electric ($E_E$) and magnetic ($E_B$) field energies  per unit length  are equal. This follows by
 integrating the volume energy densities $({\cal E}_E, {\cal E}_B)$  over the entire cross-sectional area:
\beqr
 E_E & = &  \int dA \;  {\cal E}_E = \half \eps_0 \int dA \; E^2,  \;\; \; \; 
 E_B  =  \int dA\; {\cal E}_B = \fr{1}{2 \mu_0} \int dA \; B^2 = \fr{1}{2 \mu_0 c^2}\int dA \; E^2 = E_E  \non \\
\eeqr
 and
where we used $c \lvert B \rvert = \lvert E \rvert $  in a TEM mode. Using 
 $E_E = (1/2) C' V^2$, $E_B = (1/2)L' I^2$ ($L'$ is the mode inductance per unit length)  and $E_E = E_B$,   we have  
$ Z_c = V/I = \sqrt{L'/C'}$. 
Using the relation for the phase velocity $c = 1/\sqrt{L' C'}$, Eq.(\ref{eq: Zc_capacity})  follows. 
 FEMM calculates the stored energy $E_E$; the capacitance and  the characteristic impedance are found
 using the above relations.
\bfig
\centering
\includegraphics[scale=0.2]{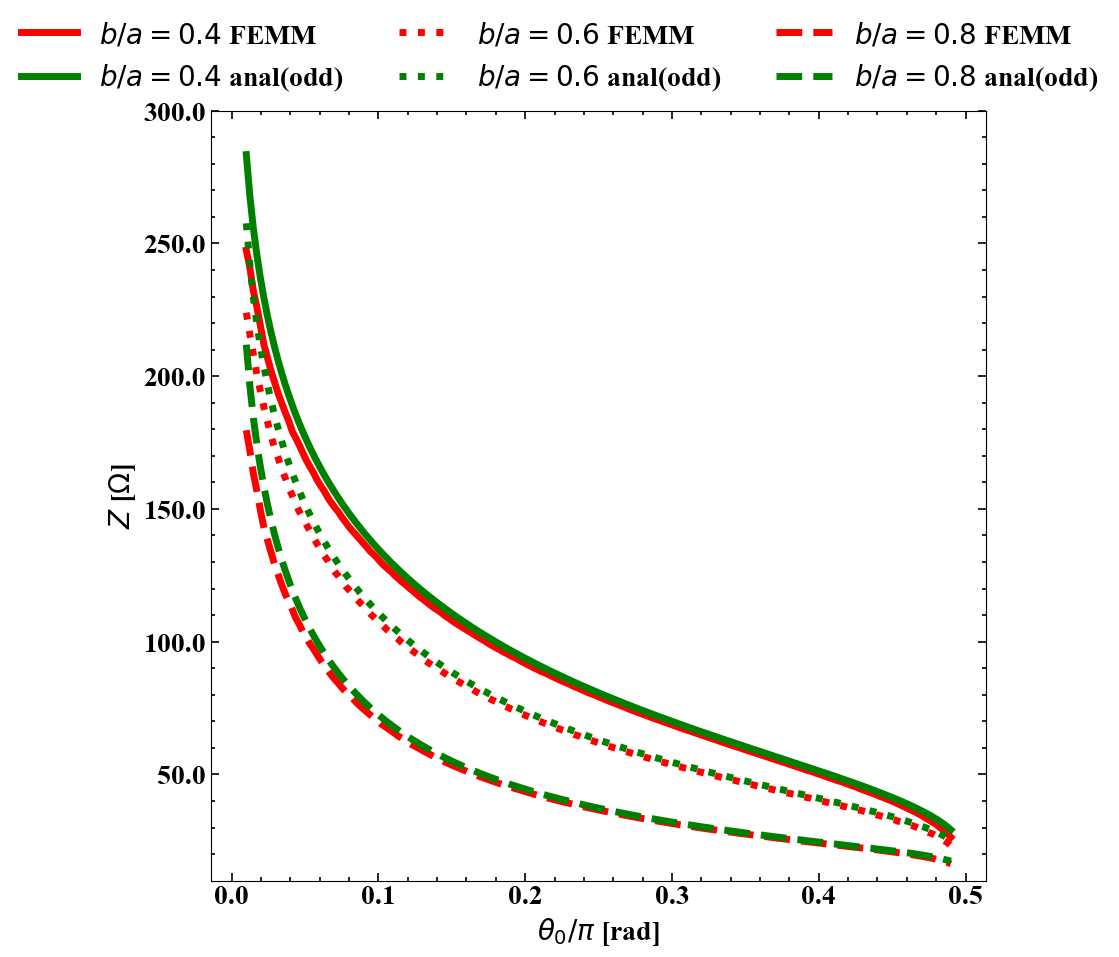}
\includegraphics[scale=0.2]{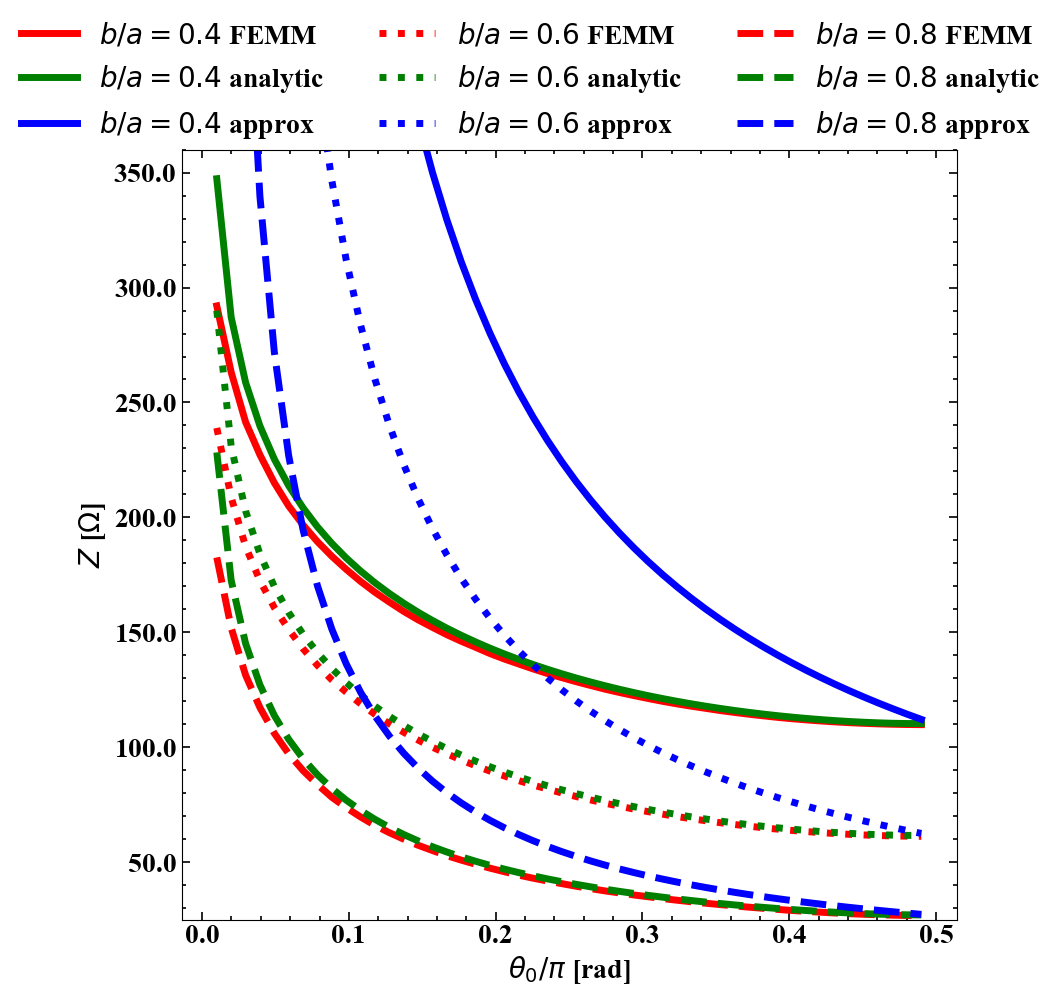}
\caption{Dipole kicker: Dependence of $Z_c$ in the even mode on $\theta_0$ calculated using FEMM, the analytic expression in
  Eq.(\ref{eq: Zc_even}) and also the approximate form in Eq.(\ref{eq: Zc_approx}) for different values of $b/a$}
\label{fig: Zc_even_theta0}
\centering
\includegraphics[scale=0.3]{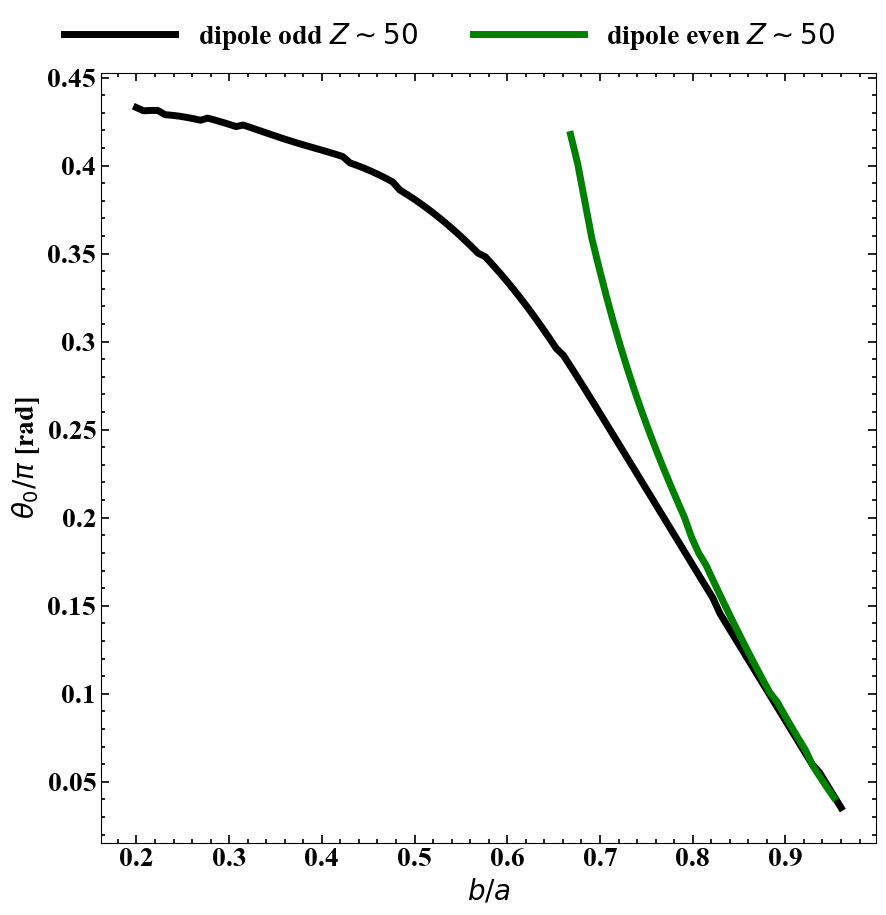}
\caption{The angle $\theta_0$ vs $b/a$ at constant characteristic impedance $Z_c$ in the two modes of the dipole kicker.
  The black curve has $Z_c = 50 \Om$ in the odd mode while the green curve has $Z_c = 50 \Om$ in the even mode.
}
\label{fig: theta0_b_constZc}
\efig
Fig. \ref{fig: Zc_even_theta0} shows the characteristic impedance in the odd and even modes of a dipole kicker as a function
of the half  coverage angle
$\theta_0$ for different values of $b/a$. For each value of $b/a$, two (odd mode) or three (even) curves are shown. One curve
represents the theoretical result in Eq.(\ref{eq: Zc_even}), another the result from FEMM and the third curve in the even mode
is the result from the approximation
\beq
Z_c^{approx} = \fr{Z_0}{2\theta_0} \ln(a/b)
\label{eq: Zc_approx}
\eeq
We observe that the theoretical and FEMM results are in close agreement for both modes except at very small $\theta_0 \to 0$,
a region not of
practical interest. We also observe that the approximate expression for the even mode agrees with the exact results in a small
range close to $\theta_0 = \pi/2$, this range increases with increasing $b/a$. 
Fig. \ref{fig: theta0_b_constZc} shows the allowed values of $\theta_0$ as a function of $b/a$ under the constraints of keeping
the characteristic impedance in the odd or the even mode constant at 50 $\Om$. 
This plot also shows that for $b/a < 0.8$, the allowed values of $\theta_0$ are larger in the even mode at a fixed $b/a$. This in
turn implies that at fixed  $(b/a, \theta_0)$ the characteristic impedance is always greater in the even mode
  while for $0.8 < b/a < 1$, the two mode impedances are nearly the same. Choosing $b/a$ in
  this range and the corresponding $\theta_0$ will allow us to very nearly match both modes to the external impedance.
  This requires that $\theta_0 < 0.15 \pi$  which as can be seen in Fig. \ref{fig: dip_Ex_b_theta}, lowers the electric field
 for a given plate voltage. Hence, the same electric field in a dipole kicker with both modes nearly matched requires a higher
 voltage compared to that  in a kicker with only one of the modes matched to the external lines.

Next we discuss the calculation of the quadrupole kicker's characteristic impedances of the two relevant modes.
\bfig
\centering
\includegraphics[scale=0.2]{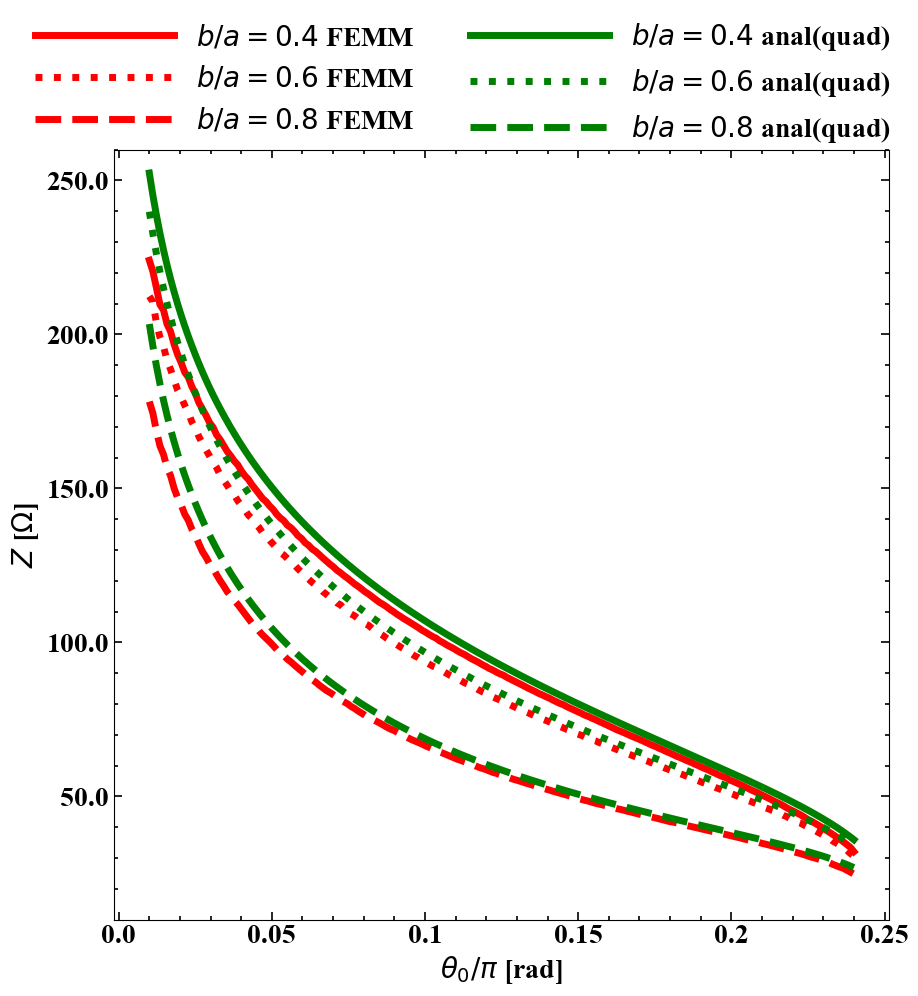}
\includegraphics[scale=0.2]{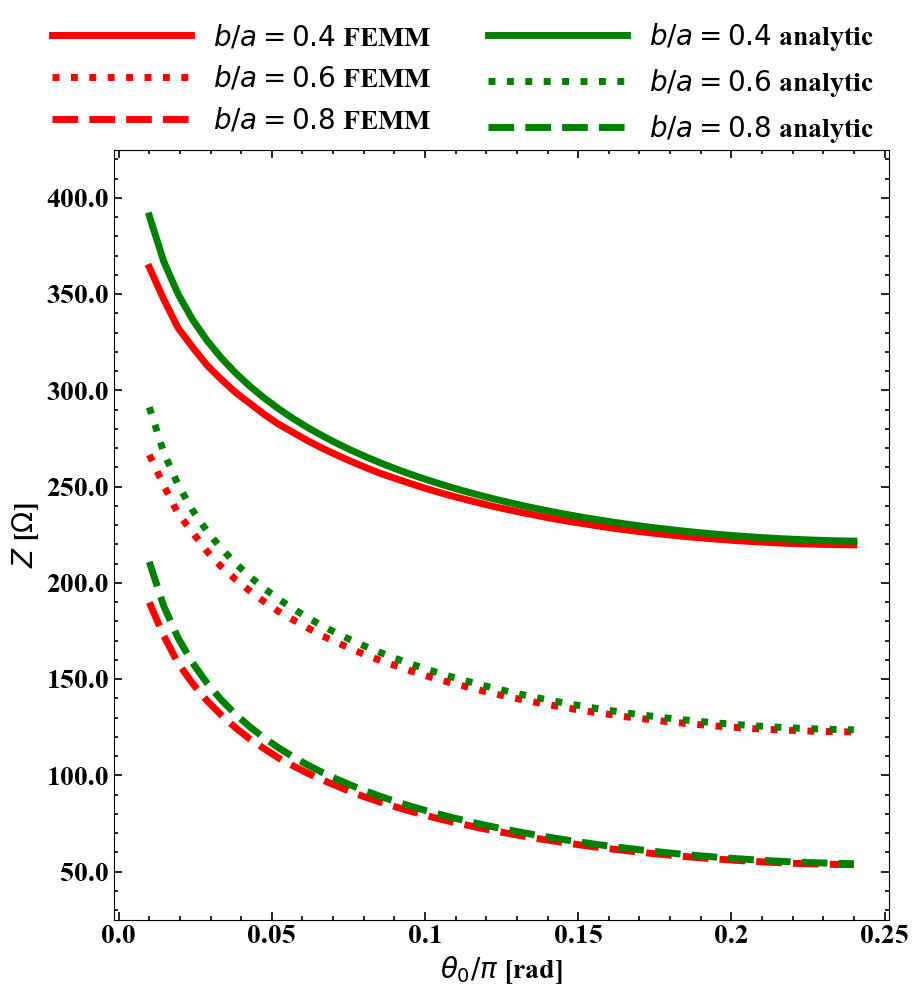}
\caption{Quadrupole mode (left) and sum mode (right) characteristic impedance  as a function of $\theta_0$ calculated with the 
  projection method and FEMM for different values of $b/a$. 
}
\label{fig: quad_Zc_sum}
\centering
\includegraphics[scale=0.35]{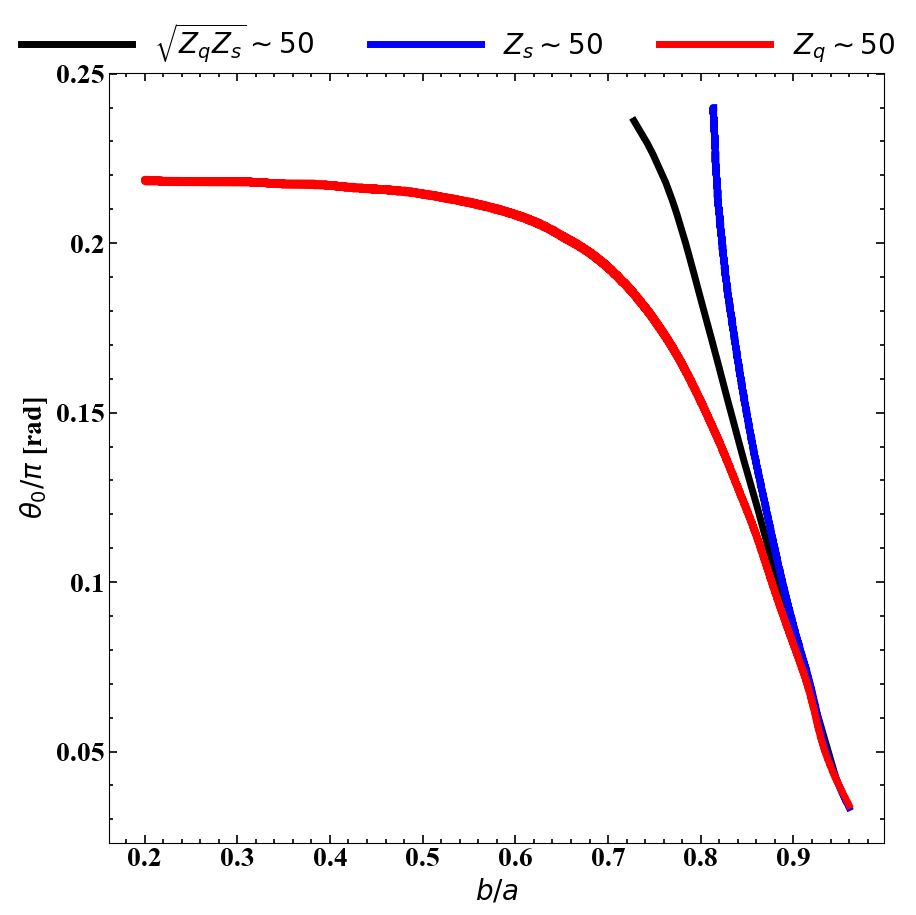}
\caption{ $\theta_0$ as a function of $b/a$ when the characteristic impedances in the quadrupole mode (red), sum mode
 (green) and  their geometric mean (black)  are each set to constant value of  50 $\Om$. 
}
\label{fig: quad_theta0_vs_b_constZc}  
\efig
Fig. \ref{fig: quad_Zc_sum} shows $Z_{c, quad}$ (left plot) and $Z_{c, sum}$ (right plot) using both the  projection method
and FEMM. As with the dipole kicker, the agreement between the two results is very good except for small values of
$\theta_0 < 0.05 \pi$. As expected,  they always obey  $Z_{c, quad} \le Z_{c, sum}$ with equality at $\theta_0 \to 0$. As functions of $b/a$, we find that $Z_{c, quad}$ varies slowly with  $b/a$ while $Z_{c, sum}$ decreases  more rapidly as $b/a$ increases and $Z_{c, quad} = Z_{c, sum} \to 0 $
as $b/a \to 1$. 
Fig. \ref{fig: quad_theta0_vs_b_constZc} shows the allowed values of $\theta_0$ as a function of $b/a$
when the characteristic impedance of either  the quadrupole or sum mode or their geometric mean  is kept constant. 
 If the geometric mean is matched to the external load, the black curve in the middle
 shows the minimum value of $(b/a)_{min} \sim 0.73$. This is slightly larger than the minimum $b/a$ for dipoles.
 We also observe that for $b/a \gtrsim 0.85$, all three impedances nearly merge so in this range, both modes will be
 nearly matched assuming the plates have zero thickness. 

 We note here that python codes which solve for the potentials and characteristic impedances for the dipole and quadrupole
 modes with applied external potentials are available here \cite{code}. 


\section{Characteristic impedance with finite plate thickness}  \label{sec: thickness}
In the analytical treatment we made the assumption that the plates have zero thickness but for a practical device
this assumption has to be dropped. 
In this section we discuss how to extend the above results to plates with finite thickness. Since most of the charge on the
plates will move to the edges, we do not expect the thickness to strongly affect the capacitance and therefore the
characteristic impedance. 
However fringe field effects around the plates do have some impact on the field configuration close to the plates.
In addition, the curvature of the tips determines the maximum field near the plates. 
These effects are stronger as the plates get closer to the beampipe (i.e. larger $b/a$) and also when the number of plates
increase causing the fringe fields to overlap. 
Here we will ignore the effects due to the frequency dependence of the plates' inductance  (primarily due to the skin effect)
so that the characteristic impedance is independent of  frequency.
This approximation is mostly valid at low frequencies (up to a few kHz) while at high frequencies (above several hundred kHz)
this impedance can be reduced by about 10-15\% \cite{Belver-Aguilar_16}. 
These effects are typically smaller than the geometric effects discussed below, especially for plate
thicknesses above $\sim 3$mm. 

We consider first the dipole kicker's even mode characteristic impedance. We can argue that the finite thickness
reduces the effective radius of the beampipe from $a$ to a value $a - t/k_f$ where $t$ is the thickness and $k_f$ is a
fit factor to be determined. By this argument, the characteristic impedance for the even mode with finite thickness can be
obtained from that with zero thickness given by Eq.(\ref{eq: Zc_even}) using
\beq
Z_{c, sc}^{even}(t) = Z_{c, even}(t=0) \fr{\ln[(a - t/k_f)/b]}{\ln(a/b)}    \label{eq: Zc_scaled}
\eeq
For each value of $b/a$, there is a single value of $\theta_0$ for which the characteristic impedance matches the external load
impedance $Z_L$, e.g. see Fig.\ref{fig: theta0_b_constZc} for the zero thickness results. Therefore we find the fit parameter
$k_f$ for each $b/a$ by minimizing the difference between $Z_{c, sc}^{even}(t)$ and $Z_L$ at the value of $\theta_0$ for which
$Z_{c, {\rm FEMM}}^{even}(t) = Z_L$. The fit parameter was determined for several values of $b/a \ge 0.7$, since the minimum
value of $b/a = 0.67$ from Fig. \ref{fig: theta0_b_constZc}. This was done for thicknesses in the range $1 \le t \le 6$ mm.
\bfig
\centering
\includegraphics[scale=0.55]{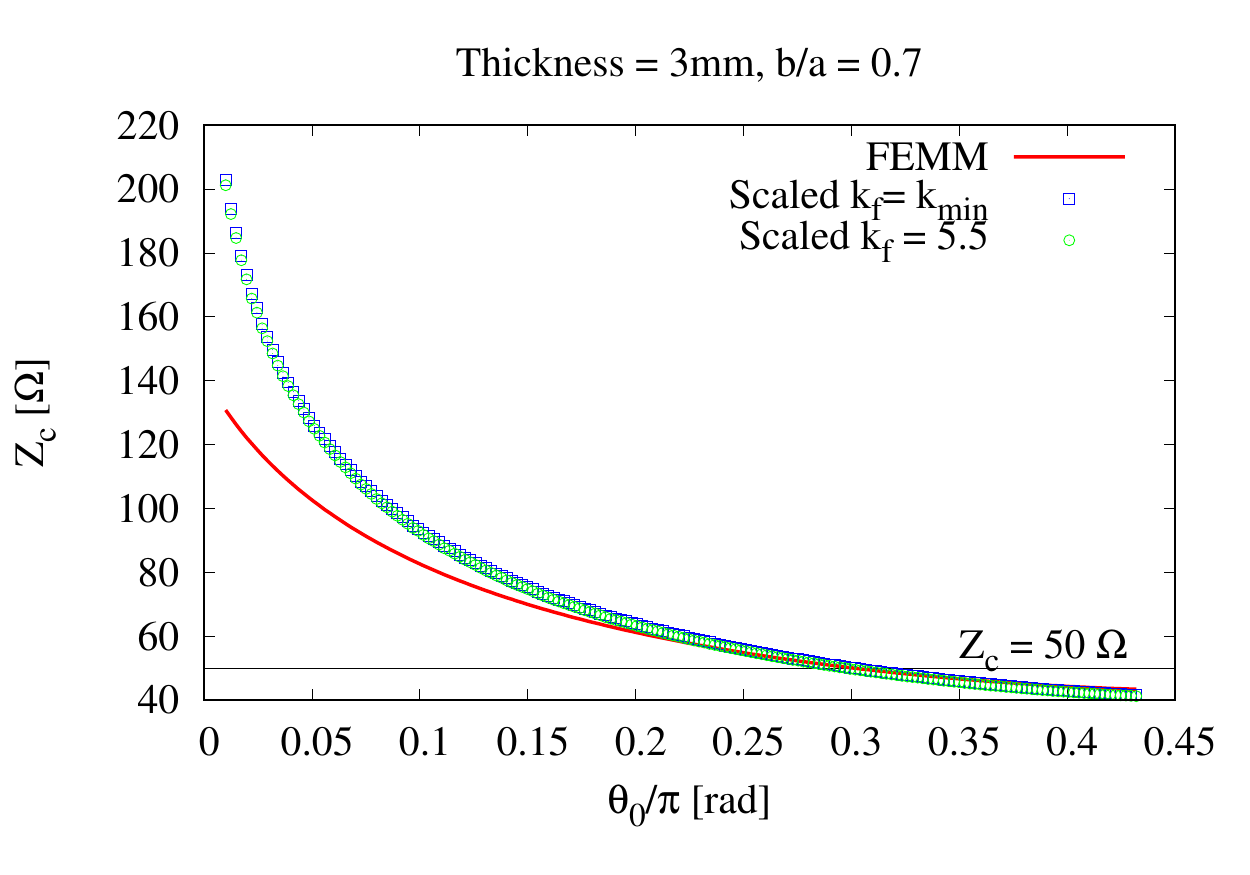} 
\includegraphics[scale=0.55]{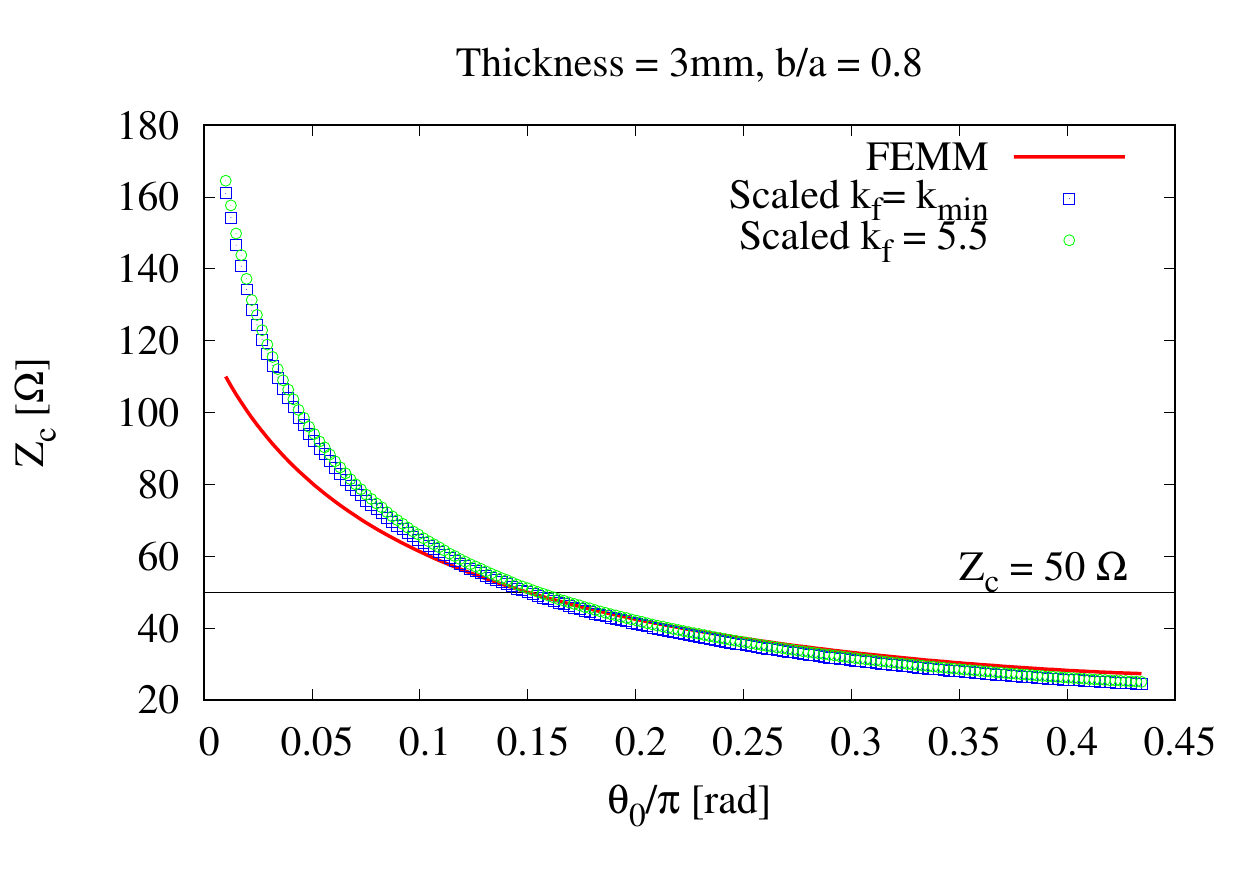} 
\includegraphics[scale=0.55]{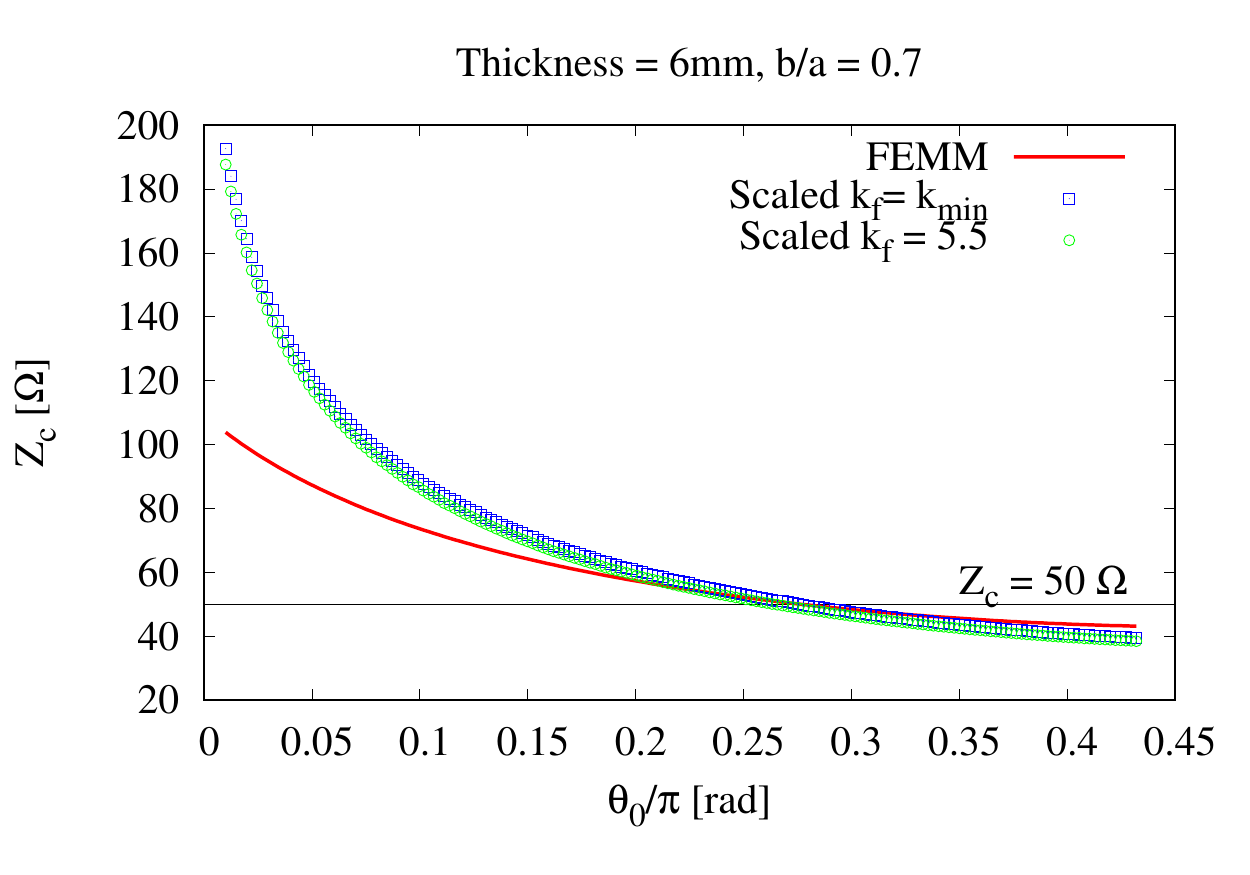} 
\includegraphics[scale=0.55]{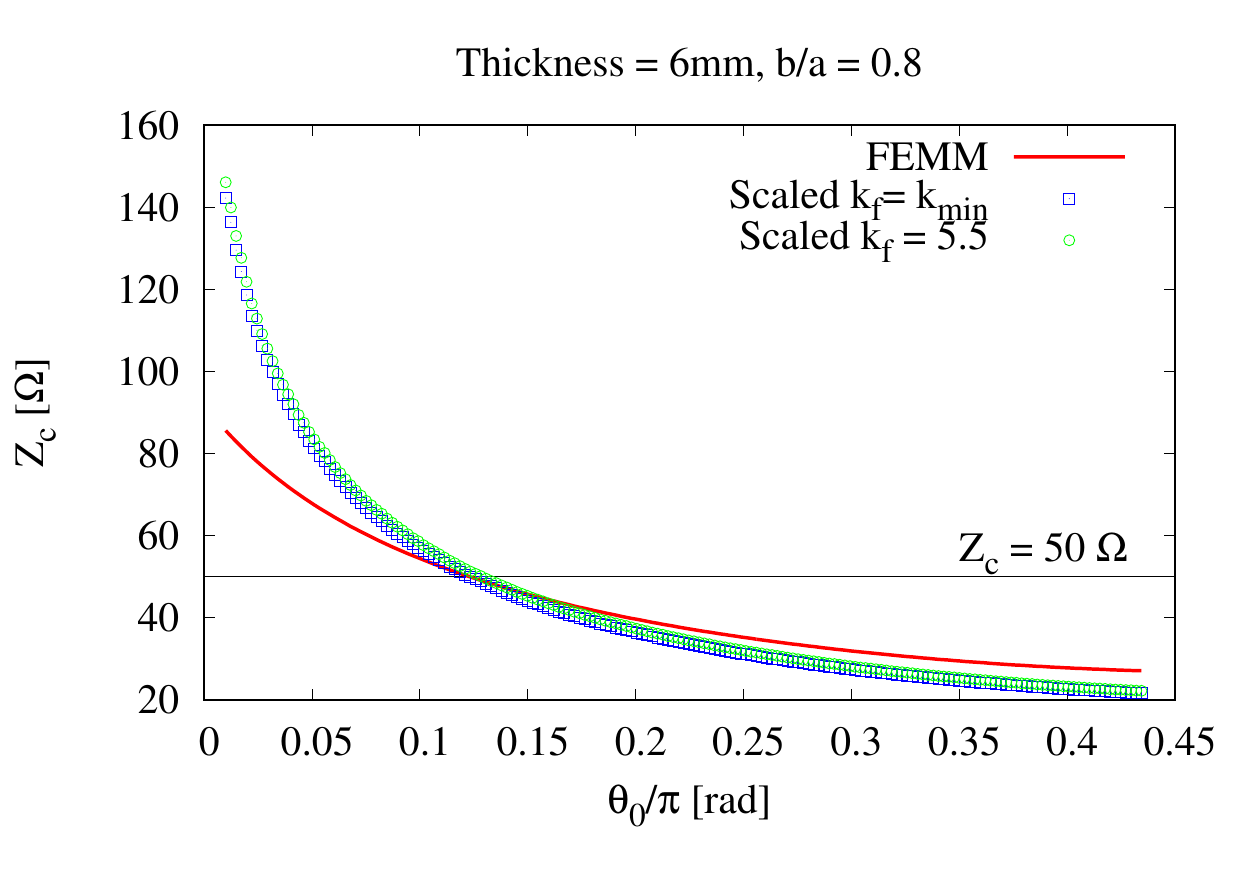}  
\caption{Dipoles: Scaled impedance with thickness for two value of the fit parameter $k_f$ in Eq.\ref{eq: Zc_scaled} compared with the
  value using FEMM as a function of $\theta_0$.
Top row with $t = 3$ mm; left plot $b/a = 0.7$,  right plot  $b/a = 0.8$.
Bottom row with $t = 6 $ mm: left plot, $b/a = 0.7$, right plot $b/a = 0.8$.
}
\label{fig: Dip_Zc_scal}
\efig
Fig. \ref{fig: Dip_Zc_scal} shows the scaled even mode impedance compared with the FEMM values for thickness $t= 3, 6$ mm
and
$b/a = 0.7, 0.8$ using two values of the fit parameter: 1) $k_f = k_{min}$, the exact fit parameter from the minimization and
2)  $k_f = 5.5$. We set $Z_L=50\Om$. While the exact fit parameter varies over the range $4.4 < k_{min} < 6.7$, we find
that using the approximate value $k = 5.5$ results in a reasonably good fit
(difference $\lvert Z_{c, {\rm FEMM}} - Z_{c, sc}(t, k_f = 5.5 )\rvert < 1.3 \Om$)
over the range of values $0.7 \le b/a \le 0.9$ and $1 \le t \le 6$ mm.
 As seen in Fig. \ref{fig: Dip_Zc_scal}, the FEMM curve and the scaled curves
are close for $\theta_0 > 0.1 \pi$ rad,  but start to diverge for smaller coverage angles.

We apply the same scaling law with thickness as in Eq,(\ref{eq: Zc_scaled}) for the geometric mean $Z_{c, geom}$ of the two
modes of interest in the quadrupoles. Strictly speaking, this scaling should be directly applicable only to the sum mode
impedance. Here however, we test the scaling on the impedance $Z_{c, geom}$ that is matched to the external impedance. 
\bfig[h]
\centering
\includegraphics[scale=0.55]{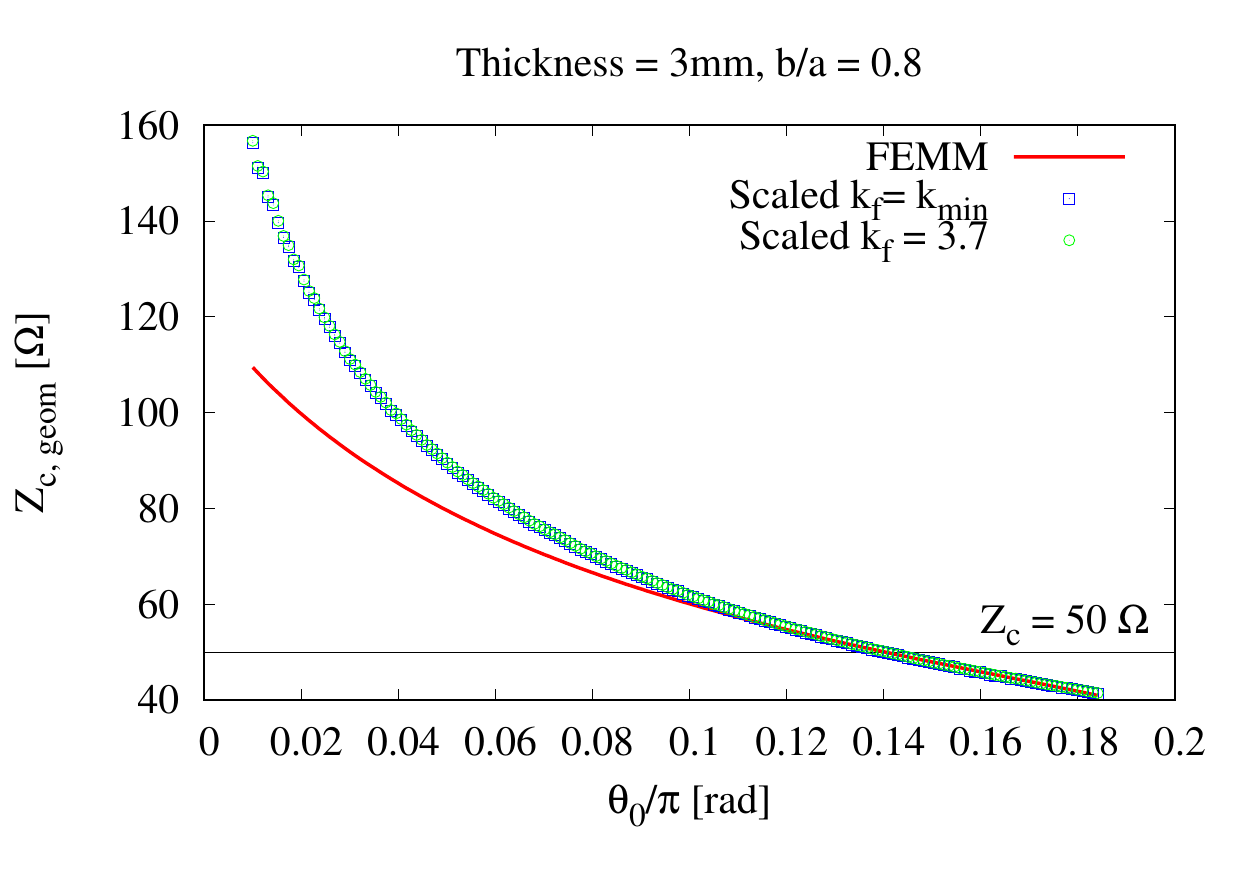}
\includegraphics[scale=0.55]{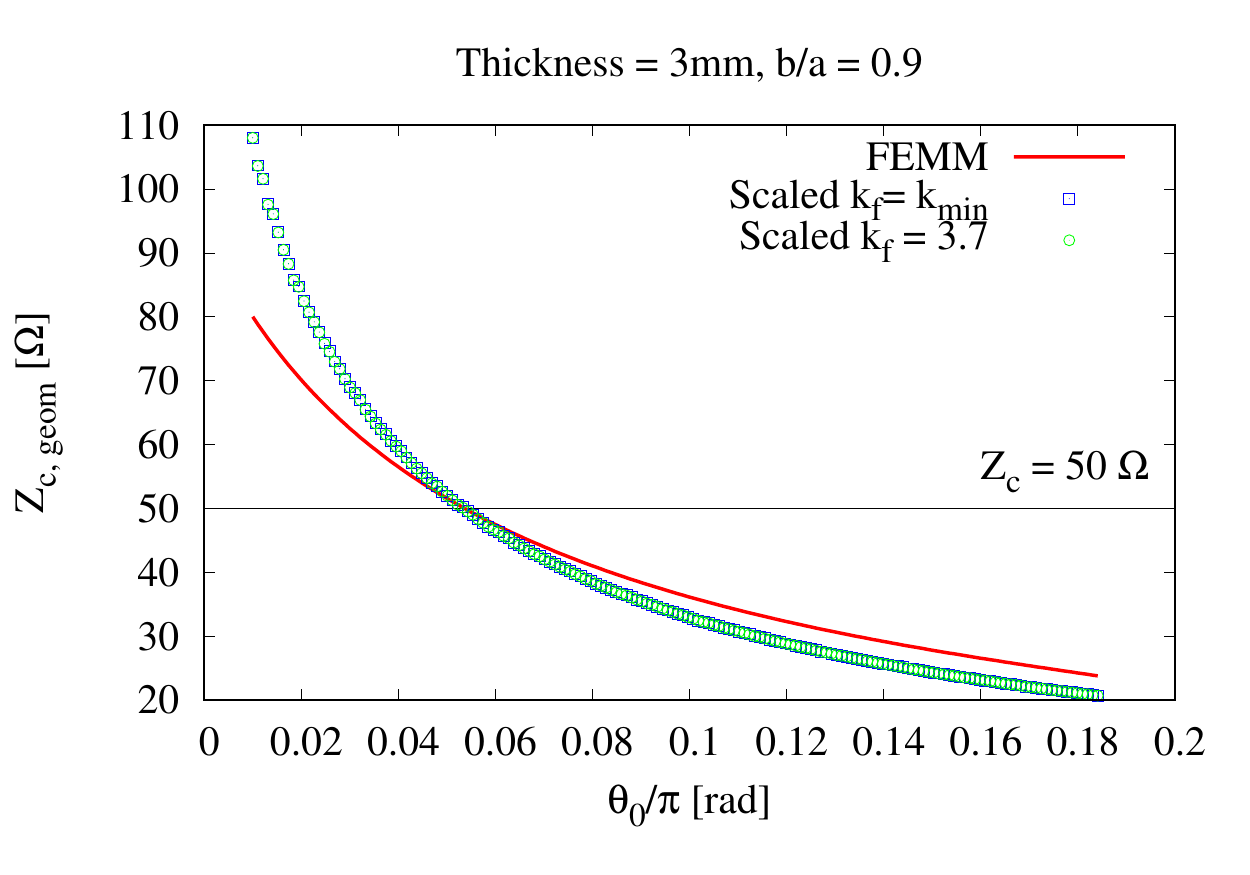}
\includegraphics[scale=0.55]{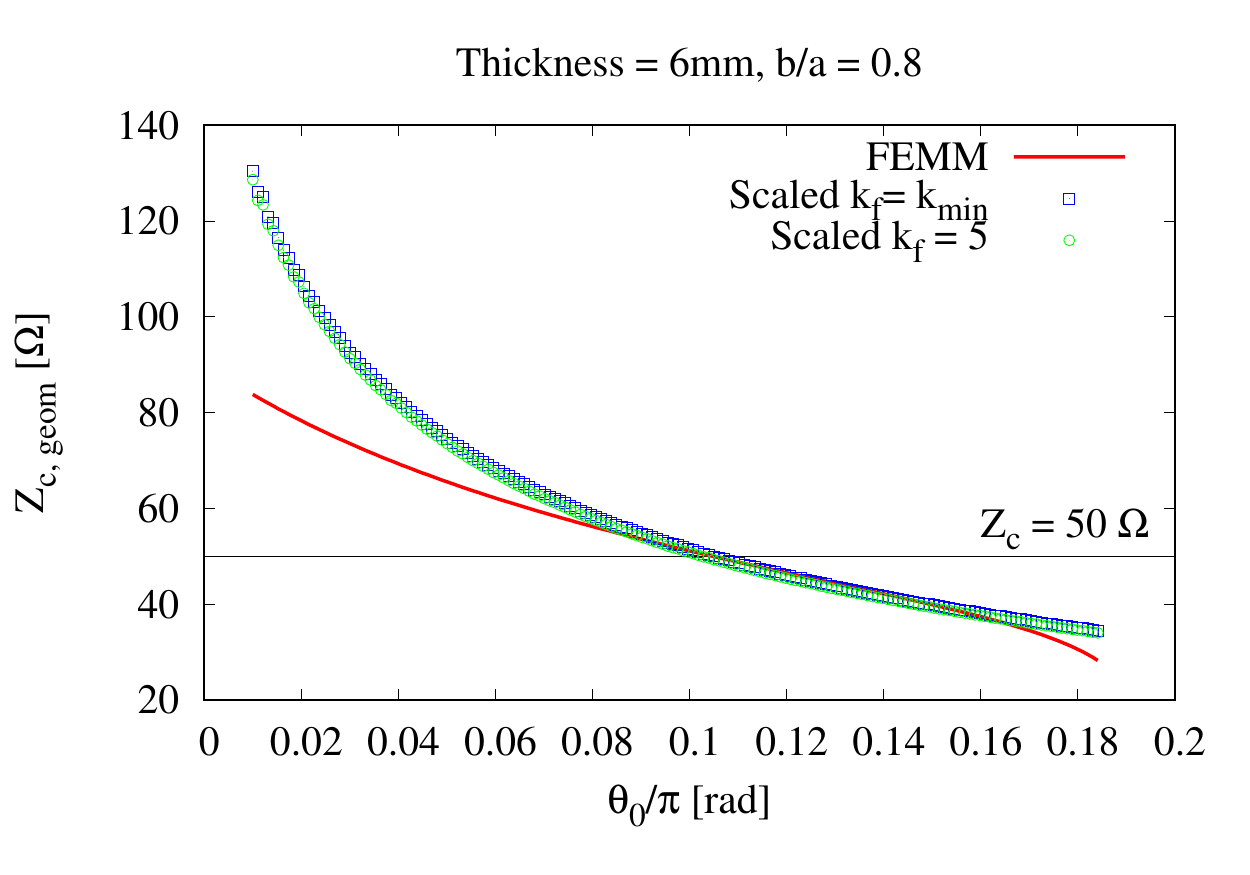}
\includegraphics[scale=0.55]{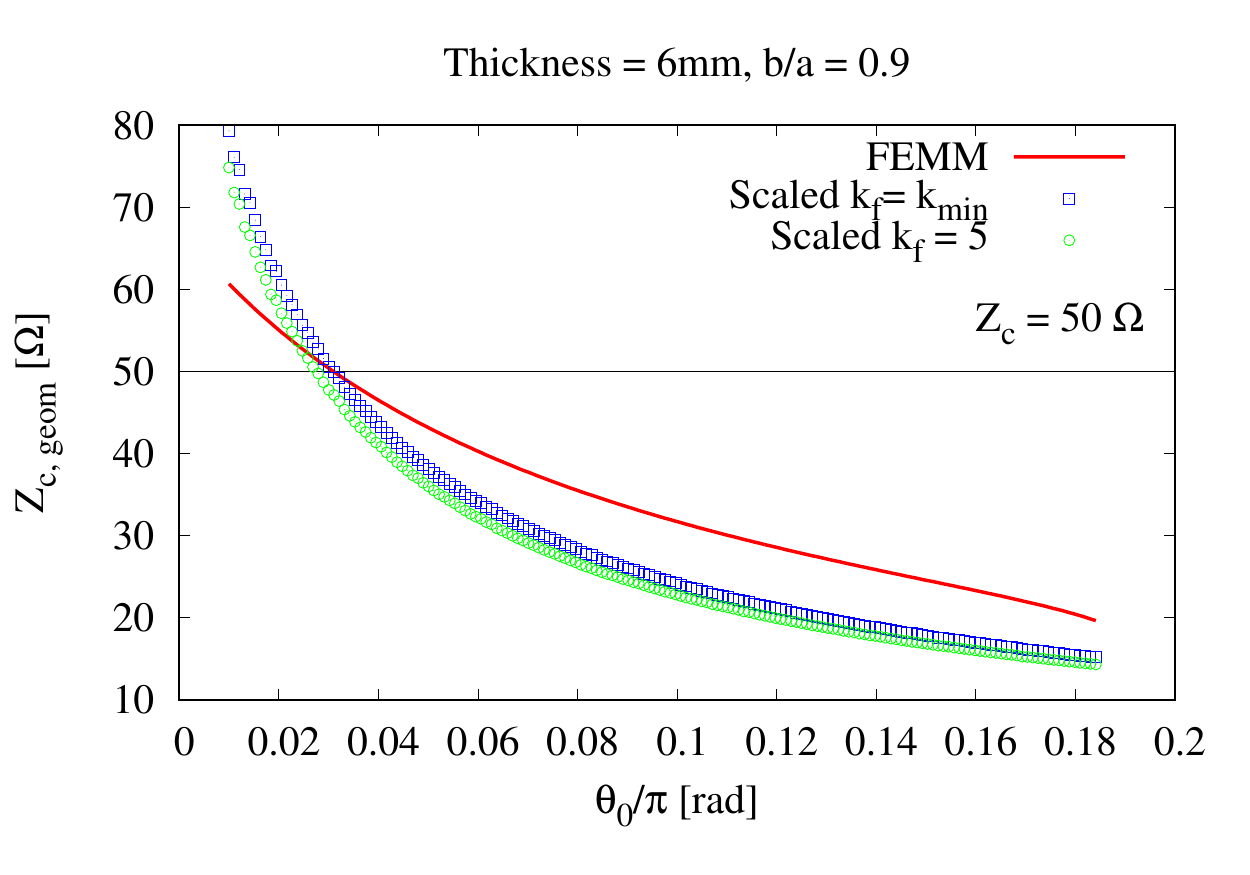}
\caption{Quadrupoles: Scaled impedance as a function of $\theta_0$ with for different thickness plates.
Top row with $t = 3$ mm; left plot $b/a = 0.8$,  right plot  $b/a = 0.9$.
Bottom row with $t = 6 $ mm: left plot, $b/a = 0.8$, right plot $b/a = 0.9$.}
\label{fig: Quad_Zc_scal}
\efig
We found in Section \ref{sec: Numerics} that the allowed range of $b/a$ for zero thickness plates in the quadrupoles at
constant $Z_{c, geom}= 50 \Om$ is $0.8 \le b/a < 1$, this range is narrower than in the dipole case.
Here we find that over the range $1 \le t \le 6$ mm with $b/a = 0.8$, the best fit
parameter $k_{f, min}$ varies from 3.6 to 3.8 while with $b/a=0.9$, $k_{f, min}$ has a different range 4.9 to  5.3.  
Fig. \ref{fig: Quad_Zc_scal} shows the scaled geometric mean characteristic impedance compared with the FEMM values for
$b/a = 0.8, 0.9$ and $t= 3, 6$ mm. 
We observe that for $b/a = 0.8$, the scaling law applies reasonably well for $\theta_0 \ge 0.1 \pi $ rad.
However for $b/a = 0.9$, the scaling starts to break down especially for the thicker plate.
The fact that there is no single value of the fit parameter
$k_f$ that can be used for different $b/a$ makes this scaling law for the quadrupoles less useful than for the dipoles.
Nevertheless for practical purposes, the scaling could be used to determine the $\theta_0$ value for which 
$Z_{geom} = 50 \Om$ even at the extreme values $b/a=0.9, t= 6$ mm. This has been confirmed with direct calculations using
FEMM.
We have verified that similar scaling behavior is observed with the sum mode impedance $Z_{sum}$ except that the fit
parameter $k_f$ are different.


\section{Conclusions}  \label{sec: conclus}

In this paper we discussed two semi-analytical methods that solve for the potentials, fields and characteristic
impedances $Z_c$ of the relevant modes in dipole and quadrupole stripline kickers. We assumed that the plates have
infinitesimal thickness and the plates and beampipe have circular symmetry. The relevant parameters are $(b/a, \theta_0)$
where $a, b$ are the beampipe and plate radius respectively and $2\theta_0$ is the angle spanned by each plate. 
Reflection symmetries or anti-symmetries as appropriate for the mode, are used and all solutions are
expressed in terms of a series of Fourier harmonics, the harmonics depend on the mode and the type of kicker.
Two methods are used to find the series coefficients: a least squares method that 
 minimizes the global error on the boundaries and a projection method  where the potential is projected onto 
 a set of basis functions. In both cases, one obtains infinite dimensional linear systems (different for each method)  
 which are then truncated and solved numerically. In both cases the series was found to converge to $\sim 1\%$  using the
first 100 terms.  
Approximate analytic expressions were derived for the two lowest order coefficients  (in Appendix A) for both kickers.
Using the second order solutions, we find that the error with the numerical solution is of the order of or less than
10\% over the range  $0.2 \pi \le \theta_0 < 0.5\pi$ (dipole) and  $0.12 \pi \le \theta_0 < 0.25\pi$ (quadrupole). So these
expressions could be used for approximate estimates of the required parameters. 
Comparisons of the numerical solutions with a finite element  code (FEMM) showed good agreement for both
types of kickers.  The deviation between the results from the series expansions and FEMM is
 $\le 5\%$ everywhere except at the tips of the plates where it can be of the order of 15-20\%,
  depending on the parameters and is generally higher in the quadrupole.
  This is likely due to a combination of the filtering used to damp the Gibbs phenomena in the analytic solutions and
 inadequate mesh density in  FEMM near the tips of the plates.
  Characteristic impedances  for the two modes of interest in each kicker (the odd and even modes in the dipole and the
  quadrupole and  sum modes in the quadrupole) were calculated and found also to be in good agreement with those
  obtained  with FEMM.
  Matching either the odd or even modes in the dipole to an external impedance (50 $\Om$) constrains the allowed values of  $(b/a, \theta_0)$. 
  Fig. \ref{fig: theta0_b_constZc} shows that this matching requires $b/a > 0.67$ and the allowed values of $\theta_0$ in this
  range. A similar plot for quadrupoles is seen in Fig. \ref{fig: quad_theta0_vs_b_constZc} which shows $\theta_0$ as a
  function of $b/a$ when
  either the quadrupole mode, sum mode or the geometric mean of the two modes is matched, In this case, matching the
  geometric mean requires $b/a \ge 0.73$. This lower bound for the quadrupole is more sensitive to the plate thickness
  (decreases with increasing thickness)  than it is for the dipole kicker. Figures \ref{fig: dip_Ex_b_theta} and
  \ref{fig: quad_gradx_b_theta} show the field
  and field gradient in a  dipole and quadrupole kicker respectively as functions of $b/a$ and $\theta_0$ under matched
  conditions. These can be used
  to determine the field and gradient for any set of parameters by scaling the beampipe  radius appropriately. 
  To account for the dependence on plate thickness, we tested a heuristic scaling law to obtain the $Z_c$ at
  a finite thickness from the value at zero thickness. In the case of the dipole kicker, this scaling law with a single value of a
  fit parameter   results in a useful approximation (to within 1 $\Om$ ) to the even mode  $Z_c$
over a range of thicknesses and $b/a$ values. For the quadrupole, the scaling law does
  not work with a single value of the fit parameter, but nonetheless can be used to find the correct value of $\theta_0$, given
  $b/a$, thickness and the external impedance.

  In Appendix B we derived the relations between the mode characteristic impedance, the elements of the Maxwell
  capacitance matrix and the mutual capacitances for the dipole and quadrupole modes.  In Appendix C, we showed how to
  choose the impedance values of a load termination network to match all modes for any configuration of  electrodes. 
  
  As mentioned in the introduction, this study was motivated by the need of
  these  kickers for beam echo generation.
  In this context, kickers are powered for the duration of a single turn; so field (or gradient) uniformity is not a primary concern. 
  For other applications where this is an issue, uniformity can be improved by shaping the electrodes
  \cite{Alesini_10, Chang_73}
. Simulations with FEMM 
   \cite{Tu_19} show that for the same applied voltage, straight parallel plates (with comparable dimensions) yield
   comparable field strengths but with better 
   field (or gradient) uniformity. The solutions presented here can be used to guide the initial design of
   either dipole or quadrupole stripline kickers  before resorting to complex software packages 
   for a detailed design to address other important issues such as minimizing the beam impedances.

\vspace{2em}

\noi {\bf \large Acknowledgment} \\
We thank the Lee Teng summer undergraduate  program at Fermilab for awarding an internship to Y.T. in 2019. 
Fermilab is operated by the Fermi Research Alliance, LLC under U.S. Department of Energy contract No. DE-AC02-07CH11359.


\appendix
\section*{Appendices}
\renewcommand{\thesubsection}{\Alph{subsection}}

\subsection{Appendix: Approximate analytic expressions for the two lowest order coefficients}
\setcounter{equation}{0}
\renewcommand{\theequation}{A.\arabic{equation}}

We saw in Section \ref{sec: Numerics} that about 100 terms were needed in the matrix equations to have
successive solutions to converge to within 1\%.  In many cases approximate solutions can be useful to obtain
rough estimates. In this appendix we will write analytic expressions for the lowest order coefficients, $X_1$ for the
dipole and $X_2$ for the quadrupole, using only one or two terms in the matrix equations. We will also show the errors
with using these expressions compared to the exact values. For both kicker types, we chose to use the matrix equations from
the projection method here. To simplify the notation, we introduce the variable  $ r_b = b/a\le 1 $.

Consider first the odd mode in the dipole kicker with a potential applied to the plates. The matrix equation is given by
Eq. (\ref{eq: matrix_proj_2}) where the matrix elements are in Eqs. (\ref{eq: dip_Cnn})-(\ref{eq: dip_Cmn}). 
Keeping only the 1st term, we have the approximate solution for $X_1$ as
\beq
X_1^{(1)} =  \fr{4 (1-r_b^2) \sin \theta_0}{ \pi- (2 -  r_b^2)(2\theta_0 + \sin 2 \theta_0)}
\eeq
Here and in the following the superscript on the $X_n$ denotes the matrix dimension, 
We note that in the limit $\theta_0 \to \pi/2$ (full coverage), $X_1^{(1)} = - 4/\pi$ which is the first coefficient in the
Fourier series expansion of a square wave - the voltage profile at the plates. 
Next  we solve the 2x2 matrix and obtain  for the dipole and sextupole coefficients $X_1^{(2)}, X_3^{(2)}$
\beqr
X_1^{(2)} \!\!\! & = &\!\!\!\!\!\!\!\!\!   \fr{- \sin\theta_0}{6(1 - r_b^6)\Dl_{dip}^{(2)} }
\left\{18 \pi  + (2 + r_b^6) (- 12\theta_0 + 3 ( \sin[2 \theta_0] + \sin[4 \theta_0]) -   \sin[6 \theta_0]) \right\}
\label{eq: X1_anal_2nd}  \\
X_3^{(2)} & = & \fr{- \sin\theta_0}{6(1 - r_b^2)\Dl_{dip}^{(2)}}
\left\{ 2 (\pi - 2 r_b^2 \theta_0) (1 +  2\cos[2 \theta_0]) +  r_b^2 (4 \sin[2 \theta_0] + \sin[4 \theta_0]) \right\} \\
\Dl_{dip}^{(2)} &  = & \fr{1}{48 (1 - r_b^2)^2 (1+r_b^2+r_b^4)} \left\{  36 \pi ( -\pi + r_b^2 (2 \theta_0  +   \sin[2 \theta_0]))
\right. \non \\
& &  +  (2 +  r_b^6) \left( \fr{15}{2} r_b^2 + 24 \pi \theta_0 - 48  r_b^2 \theta_0^2  +       6 r_b^2 (\cos[2 \theta_0] - \cos[6 \theta_0])
\right. \non \\
& & \left. \left. -  8 r_b^2 \cos[4 \theta_0] + \half r_b^2 \cos[8 \theta_0] -  24 r_b^2 \theta_0 \sin[2 \theta_0] +
(4 \pi  - 8 r_b^2 \theta_0) \sin[6 \theta_0]  \right) \right\} \non \\
\mbox{} 
\eeqr
The left plot in Fig. \ref{fig: X1sol-error} shows the values of $X_1$ calculated to 1st and 2nd order as well as the exact numerical value as a
function of $\theta_0$ with $b/a = 0.8$. The right plot in this figure shows the relative error between the analytical approximations and the exact value. It is a general feature that the lower order approximations underestimate the true value in
magnitude. At very small angles, the low order harmonics are not good approximations which is to be expected. In the limit
that $\theta_0 \to 0$, we have delta function sources which require an infinite number of harmonics. The low order
approximations improve at larger angles and we
find that with the first order solution $X_1^{(1)}$, the error is $\sim 20$\% at $\theta_0 = 0.3\pi$ and falls to $\sim 10$\% at
$\theta_0 = 0.35\pi$. The error drops more rapidly with the  solution  $X_1^{(2)}$ and the error is $\le 10$\% in the range
$\pi/4 \le \theta_0 <  \pi/2$.
This indicates that the second order expression $X_1^{(2)}$ can provide a useful estimate of the dipole field.
We note that the error with the 4x4 matrix extends the range over which the error is $< 10$\% to
$0.22 \pi \le \theta_0 < \pi/2 $ which should cover most cases of practical interest.
\bfig 
\centering
\includegraphics[scale=0.3]{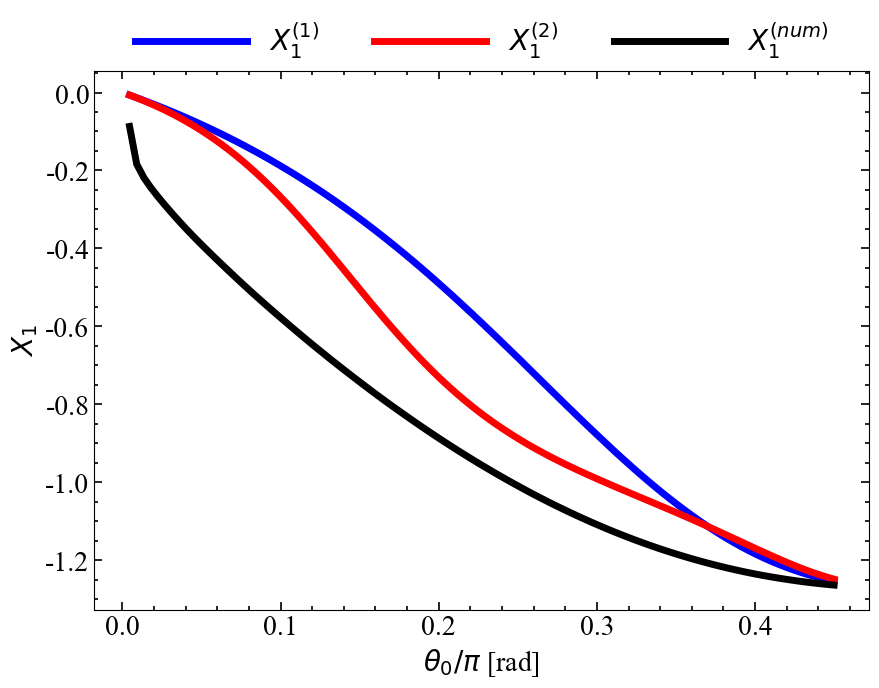}
\includegraphics[scale=0.3]{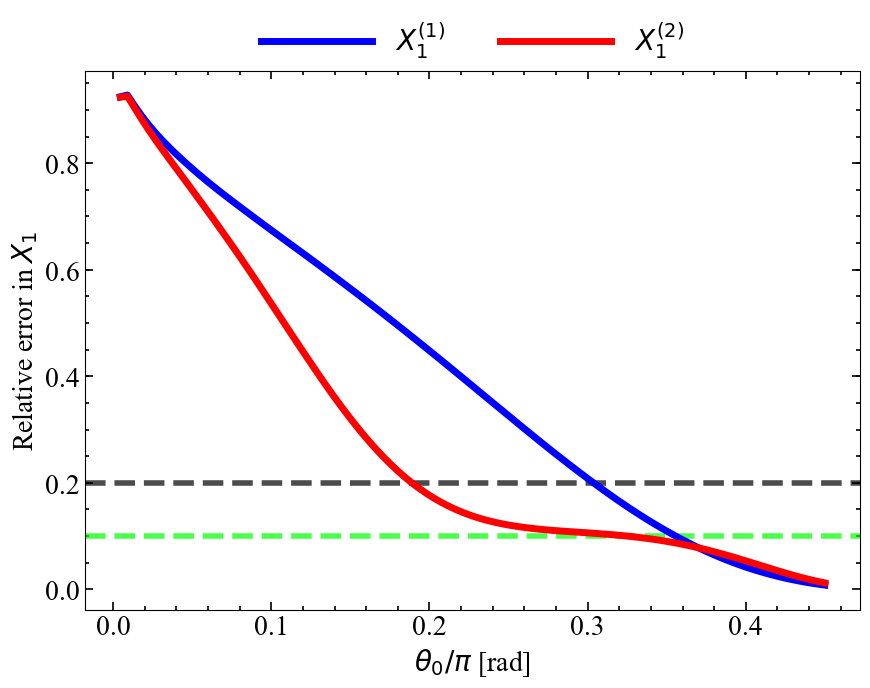}
\caption{Left: The analytical approximations to first and second order $X_1^{(1)}, X_1^{(2)}$ and the exact numerical value $X_1^{num}$  as functions of $\theta_0$ for the dipole odd mode. Right: The relative error in $X_1^{(1)}, X_1^{(2)}$ compared to
  the exact value as functions of
  of $\theta_0$. The dashed green and black lines are shown at the 10\% and 20\% error levels respectively. In both plots $b/a = 0.8$.}
\label{fig: X1sol-error}
\efig

We continue with the quadrupole mode in the  quadrupole kicker. Here the relevant matrix is $\bar{\bf C}$ defined after
Eq.(\ref{eq: quad_IC2}) in Section \ref{sec: Quad}. 
  Keeping only the 1st term,  we obtain 
  \beq
X_2^{(1)} = \fr{4 (1 - r_b^4) \sin[2 \theta_0]}{2 \pi - (1+r_b^4)(4 \theta_0+ \sin[4 \theta_0]) }
\eeq
With two terms, the solutions for the quadrupole and twelve pole coefficients are
\beqr
X_2^{(2)} \!\!\! & = & \!\!\! \fr{-\sin[2 \theta_0]}{24 (1 - r_b^{12}) \Dl_{quad}^{(2)}}
\left\{  36 \pi + (5 + r_b^{12}) \left( - 24 \theta_0 + 3  \sin[4 \theta_0] + 3  \sin[8 \theta_0] - \sin[12 \theta_0] 
\right)  \right\}   \non \\
& &  \label{eq: X2q_2}   \\
X_6^{(2)} & = &   \fr{- \sin[2 \theta_0]}{24 (1 - r_b^4) \Dl_{quad}^{(2)}} \left\{  (4 \pi (1 + 2 \cos[4 \theta_0]) \right. \non \\
&  &  \left. + (1 + r_b^4) \left( -8 \theta_0 - 16 \theta_0 \cos[4 \theta_0] + 4 \sin[4 \theta_0] + \sin[8 \theta_0] \right) \right\} \\
 \Dl_{quad}^{(2)}  & = & \fr{1}{192 (1 - r_b^4)^2 (1+r_b^4+r_b^8)} \non \\
  &  &  \left\{ 144 \pi^2 - 96 \pi \theta_0 (8 + 3 r_b^4 + r_b^{12}) -  72 \pi (1 + r_b^4)  \sin[4 \theta_0] -
  8 \pi (5 + r_b^{12}) \sin[12 \theta_0]  \right. \non  \\
  & & +  (5 + 5 r_b^4 + r_b^{12} + r_b^{16}) \left( -15/2 + 192 \theta_0^2 +   48 \theta_0 \sin[4 \theta_0] +
 16 \theta_0  \sin[12 \theta_0] \right)  \non  \\
  &  &  \left. \left. -  6 \cos[4 \theta_0] + 8 \cos[8 \theta_0] +  6  \cos[12 \theta_0] - \half \cos[16 \theta_0] \right)  \right\}
\eeqr
\bfig 
\centering
\includegraphics[scale=0.3]{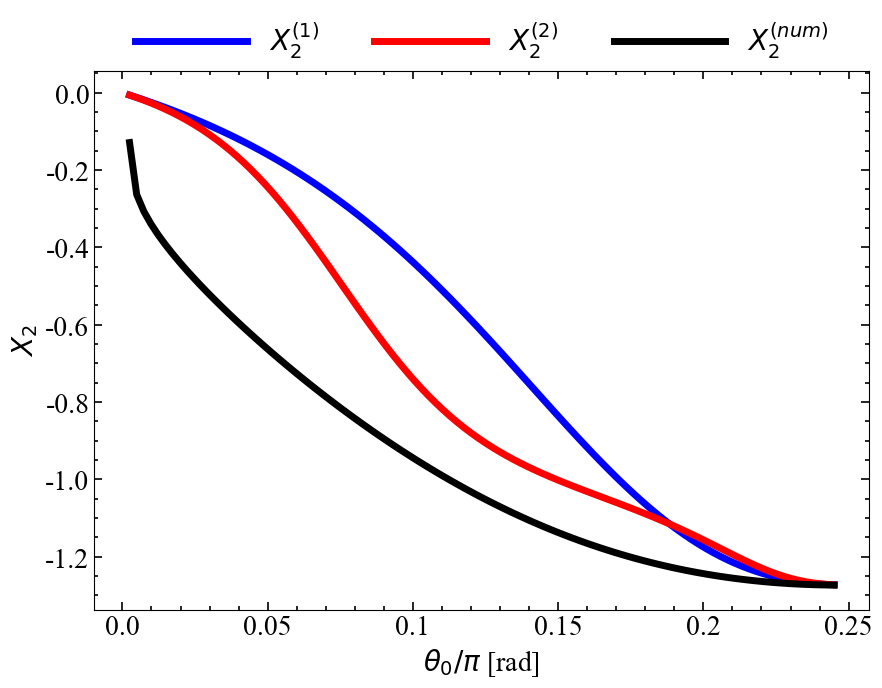}
\includegraphics[scale=0.3]{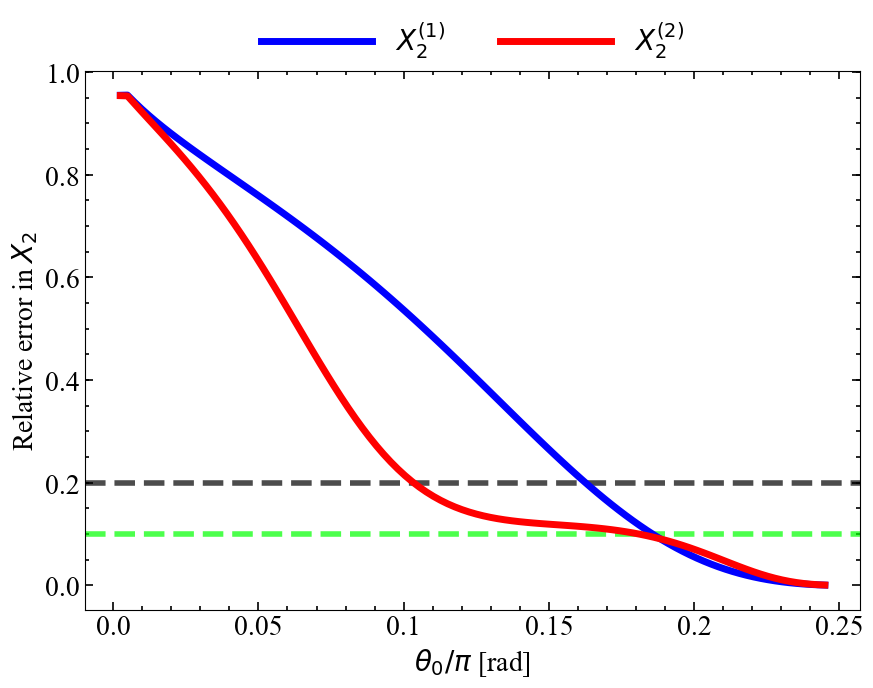}
\caption{Left: Analytical approximations to (1st, 2nd) order $X_2^{(1)}, X_2^{(2)}$ (blue, red) of the quadrupole coefficient
  and the  exact numerical value $X_2^{num}$ (black) as functions of $\theta_0$. Right: The relative error in 
  $X_2^{(1)}, X_2^{(2)}$  (blue, red) compared to the exact value as functions of
  of $\theta_0$. The dashed green and black lines are shown at the 10\% and 20\% error levels respectively. In both plots $b/a = 0.8$.}
\label{fig: X2sol-error}
\efig
The left plot in Fig. \ref{fig: X2sol-error} shows the values of $X_2$ calculated to 1st, and 2nd order as well as the exact
numerical value as a function of $\theta_0$ with $b/a = 0.8$. The right plot shows the relative error between the analytical
approximations and the exact value. The general features are the same as for the dipole case. For example, the error with
the 2nd order estimate is $\le 10\%$ for $0.16 \pi \le \theta_0 < 0.25 \pi$, We also find that for $\theta_0 \ge 0.18 \pi$,
the 1st order estimate is slightly better. The error with the 4th order estimate $X_2^{(4)}$ is $< 10$\% for
$\pi/8 \le \theta_0 < \pi/4 $.

\subsection{Appendix: Modal capacitances}
\setcounter{equation}{0}
\renewcommand{\theequation}{B.\arabic{equation}}

In this appendix, we derive the relations for the modal capacitances which are used in the expressions for the
characteristic impedances of the dipole and quadrupole kickers. 
For notational convenience we drop the prime on the capacitances in this appendix, but it should be implicitly
understood that all capacitances and inductances here  are expressed per unit length.  

We consider an arrangement of $n$ electrodes inside a closed conducting shield held at fixed
potential  $V =0 $.
The charges $Q_i$  and the potentials $V_i$  (expressed in reference to the outer shield potential). are linearly related to each other   
 \begin{equation}
V_i = \sum p_{ij} Q_j 
\end{equation}
The $p_{ij}$ are known in the literature as the  Maxwell coefficients of potential.  
This relation can be inverted to yield 
 \begin{equation}
Q_i = \sum c_{ij} V_j 
\end{equation}
The coefficients  $c_{ij}$  are known as the Maxwell coefficients of capacitance and the  matrix ${\bf C}_M$ is known as the Maxwell capacitance matrix \cite{Maxwell}.   
By reciprocity,  ${\bf C}_M$  must be  symmetric and therefore has $n$ real eigenvalues and $n$ linearly independent eigenvectors. Since the electrostatic energy is a positive definite quadratic form of the $V_i$,  none of the eigenvalues is  0 and a transformation to a frame defined by eigenvectors is always possible.  
Let  ${\bf U}$ be  an orthonormal matrix whose columns form a complete set of eigenvectors of ${\bf  C}_M$.  One has
$$
 {\bf U}^T {\bf Q }  = [{\bf U}^T {\bf C}_M {\bf U}] {\bf U}^T {\bf V}
$$
where we used the fact that ${\bf U}{\bf U}^T = {\bf I}$. It is easily verified that   ${\bf U}^T {\bf C}_M{\bf U}$ is a diagonal matrix; 
its elements, the modal capacitances (the eigenvalues) are the roots $\lm$ of the characteristic polynomial     
$  \det \left[ {\bf C}_M - \lambda {\bf I} \right] = 0 $. 
For an azimuthally  symmetric arrangement of $n$ identical conductors inside a circular conducting shield,  ${\bf C}_M$ has a special structure.  In what follows, we will mostly restrict ourselves  to the cases $n=2$  (dipole) and $n=4$ (quadrupole)  . 
By symmetry, the charge induced on conductor $i$ by a voltage applied on conductor $j$ depends only on  the (minimum) azimuthal distance between 
electrodes $i$ and $j$.
Therefore, one has 
 \begin{equation} 
             c_{ij} =  c_{1, |i-j|}   \begin{dcases*} 
                               |i-j|  \bmod n/2   &  $n$ even \\ 
                               |i-j|  \bmod (n-1)/2 & $n$ odd   
                              \end{dcases*}
\end{equation}
i.e.  ${\bf C}_M$ either has  $n/2 +1 $ independent elements for $n$ even or  $(n-1)/2$ for $n$ odd.   
Furthermore,  ${\bf C}_M$ is a circulant matrix: its columns (rows) are successive cyclic permutations of the first one. 
For such matrices,  general closed form expressions can be obtained for both eigenvectors and eigenvalues; see e.g. 
\cite{Matrix}. 

For the case $n = 2$ there are $n/2+1 = 2$ independent elements  labeled $ ( c_{11}, c_{12}) $ and 
for  $n = 4$,  there are 3 independent elements  labeled $ (c_{11} , c_{12}, c_{13}) $. 
The Maxwell capacitance matrix in these two cases are
\begin{equation}
{\bf  C}_M = \begin{pmatrix}
         c_{11}  & c_{12} \\ 
         c_{12}   & c_{11} 
        \end{pmatrix}
, \;\;\;\;
\bf {C}_M = \begin{pmatrix}
         c_{11}  & c_{12}  & c_{13} &  c_{12} \\
         c_{12}   & c_{11} & c_{12} &  c_{13} \\
         c_{13}   & c_{12} & c_{11} &  c_{12} \\
         c_{12}   & c_{13} & c_{12} &  c_{11} 
\end{pmatrix}
\label{c_matrix}
\end{equation}
Note that the diagonal elements are always positive  ($ c_{ii} > 0$)  while the off-diagonal elements are always negative i.e. $ c_{ij} < 0,\;\; i\ne j$  since setting
any  electrode $i$ to a voltage $V_i=1$ and the others to $V_j =  0, \;\; i\ne j$ induces a positive charge on electrode $i$ and negative charges on all the others. 
For $n=2$, the eigenvectors are 
\beq
{\bf u}_1 =   \frac{1}{\sqrt{2}}  (1,   -1), \;\;\; {\bf u}_2 =    \frac{1}{\sqrt{2}}  (1  , 1  ) 
\eeq
One may verify by taking the product of  the $2\times 2$ matrix in Eq. (\ref{c_matrix}) with each of these vectors that  the corresponding eigenvalues
or modal capacitances are
\beq
c_1  =    c_{11} + c_{12}, \;\;\; c_2  =    c_{11} -  c_{12}   \label{eq: modeC_2}
\eeq
For $n=4$, the eigenvectors are
\begin{eqnarray}
{\bf u}_1 & =  & \frac{1}{\sqrt{2}}  (1,   0,  -1,   0 ), \;\;\; {\bf u}_2 =    \frac{1}{\sqrt{2}}  (0  , 1,  0    -1)  \\
{\bf u}_3& =  &  \frac{1}{2} (1 , -1,  1    -1) , \;\;\;  {\bf u}_4  =   \frac{1}{2}(1,   1 , 1,   1)  
\end{eqnarray}
and using  the $4\times4$  matrix in Eq. (\ref{c_matrix}),  one easily verifies that the corresponding eigenvalues are
 \begin{eqnarray}
c_1 & =  & c_{11} - c_{13}, \;\;\;   c_2  =    c_{11} - c_{13} \\
c_3& =  &  c_{11} - 2  c_{12} + c_{13}, \;\;\; c_4  =   c_{11} + 2 c_{12} + c_{13}  \label{eq: modeC_4}
\end{eqnarray}
Using the relation for the characteristic impedance in a mode $k$ in terms of the modal capacitance
 $c_k$, i.e. $Z_{c, k} = 1/(c c_k)$, Eqs. (\ref{eq: Zc_2_1})- (\ref{eq: Zc_2_2}) for the dipole and
 Eqs. (\ref{eq: Zc_4_1}) - (\ref{eq: Zc_4_3}) for the quadrupole follow. 
\subsubsection*{Mutual Capacitance}
Even though the elements of the Maxwell capacitance matrix have the units of capacitance, they 
differ from the capacitances commonly used in circuit models, which are properly 
referred to as mutual capacitances.  While the Maxwell capacitance coefficients yield the charge induced on individual conductors, the mutual capacitances describe the buildup of 
equal and opposite charge between pairs of conductors.   The Maxwell capacitance matrix elements and the mutual capacitances can be connected to each other 
by observing that the charge on a conductor $i$ may be expressed in two equivalent forms  as
\begin{equation}
Q_i = \sum c_{ij} V_j , \;\;\; {\rm or} \;\;\;
Q_i = C_{ii}V_i + \sum_{i\ne j} C_{ij} (V_i-V_j) =  (\sum_j C_{ij}) V_i - \sum_{i\ne j} C_{ij} V_j
\end{equation}
where $C_{ij} > 0 $ is a conventional mutual (circuit) capacitance. 
 By identification, 
\beqr
 C_{ii}  & = &  \sum_j c_{ij}, \;\;\;  C_{ij}   =  -c_{ij}, \;  i \ne j  \\
 c_{ii}  & = &  \sum_j C_{ij} , \;\;\;  c_ {ij}   =  -C_{ij},  \;  i \ne j
\eeqr
where the second line follows from symmetry. 
 Expressed in terms of mutual capacitances, the modal capacitances for $n =2$ are
\begin{eqnarray}
c_1 \equiv C_1  =    C_{11}, \;\;\;\;  c_2 \equiv C_2  =    C_{11}  +2 C_{12}  
\end{eqnarray}
Similarly, for $n=4$
\begin{eqnarray}
c_1 \equiv C_1 & =  &  C_{11}  +  2C_{12}  + 2C_{13}\\
c_2 \equiv C_2 & =  &  C_{11}  +  2C_{12}  + 2C_{13}\\
c_3 \equiv C_3 & =  &  C_{11}  +  4C_{12}  \\
c_4 \equiv C_4 & =  &  C_{11}   
\end{eqnarray}

\subsection{Appendix: Characteristic Impedance Matching}
\setcounter{equation}{0}\
\renewcommand{\theequation}{C.\arabic{equation}}
The modal  characteristic impedances of an $n$-electrode arrangement are in general not equal. 
Nevertheless, it is possible  to devise a load termination network that results in a match for all modes. 
Expressed in matrix form, the relation between the electrode termination voltages and currents is        
\begin{equation}
{\bf I} = {\bf Y}{\bf V}  
\end{equation}
where ${\bf Y}$ is the nodal admittance matrix (sometimes referred to as the Laplacian matrix).  It is natural to demand that the matching network have the same 
symmetry as the as the electrodes arrangement. In that case, just as the capacitance matrix is circulant and symmetric, 
so must be ${\bf Y}$. Using $y_{ij}$  to denote an entry of ${\bf Y}$,  for $n=2$  there are two independent elements ($y_{11}$ and $y_{12}$); for 
$n=4$ there are three ($y_{11}$ , $y_{12}$ and $y_{13}$).

In direct analogy with Appendix B,  the $y_{ij}$ can be expressed in 
terms of conventional circuit admittances  $Y_{ij}$ which are defined for voltage differences $(V_i-V_j)$. 
\beq
 Y_{ii}   =  \sum_j y_{ij}, \;\;\;  Y_{ij}   =  -y_{ij}, \;  i \ne j
\eeq
 or, in a completely symmetric manner
 \beq
 y_{ii}  =   \sum_j Y_{ij} , \;\;\;  y_{ij}   =  -Y_{ij},  \;  i \ne j
\eeq
In the above equations, $Y_{ii} \; i=1,\ldots,n$ represents the admittance of a  load connected from  electrode $i$ to the ground (shield),
and $Y_{ij}\; i\ne j$  a load connected from electrode $i$ to electrode $j$.  If the loads are purely resistive
$Y_{ij} = 1/R_{ij} $. where $R_{ij}$ is the  corresponding resistance.   A mode is matched when its characteristic
impedance is equal to the inverse circuit admittance of the loading network  driven by that mode.
By inspection we observe that ${\bf Y}$ has the same eigenvectors as ${\bf C}$.  After diagonalization one can express
the matching conditions  for $n=2$
\begin{eqnarray}
Z_{c,even} & \equiv &  Z_{c,1}  =   1/Y_{11} =  R_{11} \\
Z_{c,odd} & \equiv &  Z_{c,2}  =   1/(Y_{11}  +2 Y_{12})  = \left[ \frac{R_{11}{R_{12}}}{R_{12}+2R_{11}} \right] 
\end{eqnarray}
Similarly, for $n=4$, the matching conditions are
\begin{eqnarray}
Z_{c, dip}  & \equiv & Z_{c,1}= Z_{c,2}  =   1/(Y_{11}  +  2Y_{12}  + 2Y_{13}) = \left[ \frac{R_{11}{R_{12} R_{13} }}{R_{12}R_{13} +2R_{11}R_{13} + 2R_{11}R_{12}} \right] \\
Z_{c, quad}  & \equiv & Z_{c,3}  =   1/(Y_{11}  + 4Y_{12})  = \left[ \frac{R_{11}{R_{12}}}{R_{11}+2R_{12}} \right] \\
Z_{c, sum} & \equiv &  Z_{c,4} =    1/ Y_{11} =  R_{11} 
\end{eqnarray}
The conventional termination scheme consists in  terminating each line by an impedance $Z_L$ between electrode and
ground. With either $Z_{c, even}$ or $Z_{c, sum}$ set equal to $Z_L =  R_{11}$  the even or sum mode is matched; the
remaining equation(s) can be solved for the loads  needed to match 
other modes. The result for $n=2$ is identical to that presented in section VI E of reference \cite{Belver-Aguilar_14}.  
The generalization to arbitrary $n$ is straightforward.

\end{document}